\documentclass[preprint]{elsarticle}
\usepackage{lineno,hyperref}
\usepackage{amsmath}
\usepackage{xspace}
\usepackage{siunitx}
\usepackage{graphicx}
\usepackage{booktabs}
\usepackage{makecell}
\usepackage{subcaption}
\usepackage{color}
\usepackage[utf8]{inputenc}
\modulolinenumbers[5]
\journal{Nuclear Instruments and Methods A}
\usepackage{nsw-defs}
\usepackage{geometry}
 \newcounter{daggerfootnote}
\newcommand*{\daggerfootnote}[1]{%
    \setcounter{daggerfootnote}{\value{footnote}}%
    \renewcommand*{\thefootnote}{\fnsymbol{footnote}}%
    \footnote[2]{#1}%
    \renewcommand*{\thefootnote}{\arabic{footnote}}%
}
\begin{document}

\begin{frontmatter}
\title{The large inner Micromegas modules for the Atlas Muon Spectrometer Upgrade: construction, quality control and characterization}
\author{J.~Allard, M.~Anfreville\daggerfootnote{Deceased}, N.~Andari, D.~Atti\'e, S.~Aune,  H.~Bachacou, F.~Balli, F. ~Bauer$^\dagger$, J.~Bennet, T.~Benoit, J.~Beltramelli, H.~Bervas, T.~Bey, S.~Bouaziz, M.~Boyer$^\dagger$, T. Challey, T.~Cheval\'erias,
X.~Copollani, J.~Costa,
G.~Cara, G.~Decock, F.~Deliot, D.~Denysiuk, 
D.~Desforge, G.~Disset, G.A.~Durand, R.~Durand$^\dagger$, J.~Elman; E.~Ferrer Ribas, M.~Fontaine, A.~Formica, W.~Gamache, J.~Gal\'an\footnote{Present address: Grupo de F\'isica Nuclear y Astropart\'iculas, Departamento de F\'isica Te\'orica, Universidad de Zaragoza C/ P. Cerbuna 12 50009, Zaragoza, Spain}, A.~Giganon, J.~Giraud, P.F.~Giraud, G.~Glonti,  C.~Goblin, P.~Graffin,  J.C.~Guillard, S.~Hassani, S.~Herv\'e, S.~Javello, F.~Jeanneau, D.~Jourde, S.~Jurie, M.~Kebbiri, T.~Kawamoto, C.~Lampoudis\footnote{Present address: Department of Physics, Aristotle University of Thessaloniki, University Campus, GR-54124,Thessaloniki, Greece}, J.F.~Laporte, D.~Leboeuf, M.~Lefevre, M.~Lohan, C.~Loiseau, P.~Magnier, I.~Mandjavidze, J.~Manjarr\'es\footnote{Present address: Technische Universit\"{a}t Dresden, 01062 Dresden, Germany}, P.~Mas, M.~Mur, R.~Nikolaidou, A.~Peyaud, D.~Pierrepont, Y.~Piret, P.~Ponsot, G.~Prono, M.~Riallot, F.~Rossi, P.~Schune, T.~Vacher, M.~Vandenbroucke, A.~Vigier, C.~Vuillemin, M.~Usseglio, Z.~Wu\footnote{Also affiliated to University of Science and Technology of China, No.96, JinZhai Road Baohe District, Hefei, Anhui, 230026, P.R.China.}
}
\address{ IRFU, CEA, Universit\'e Paris-Saclay, F-91191 Gif-sur-Yvette, France}

\begin{abstract}
The steadily increasing luminosity of the LHC requires an upgrade with high-rate and high-resolution detector technology for the inner end cap of the ATLAS muon spectrometer: the New Small Wheels (NSW). In order to achieve the goal of precision tracking at a hit rate of about 15 kHz/cm$^2$ at the inner radius of the NSW, large area Micromegas quadruplets with 100\,\microns spatial resolution per plane have been produced. 
IRFU, from the CEA research center of Saclay, is responsible for the production and validation of LM1 Micromegas modules. The construction, production, qualification and validation of the largest Micromegas detectors ever built are reported here. Performance results under cosmic muon characterisation will also be discussed.

\end{abstract}

\end{frontmatter}


\section{Introduction}
\label{section:introduction}
The increase of luminosity in High-Luminosity LHC, HL-LHC~\cite{HL-LHC}, implies an increase of the number of interactions per crossing of the LHC beams and then an increase of particle rate and detector irradiation. The existing forward inner parts of the ATLAS Muon Spectrometer, the  {\it Small Wheels} (SW), are expected to stand irradiations up to 15\,kHz $\mathrm{cm^{-2}}$,  too high a rate for the existing detectors. An upgrade, called the {\it New Small Wheels} (NSW)~\cite{Kawamoto:2013udg}, is foreseen,  to replace the SWs in 2021-2022.


This paper describes the construction, the assembly, the quality control and the cosmic bench validation of the “Large Modules 1” (LM1) built in IRFU/CEA that represent the largest Micromegas detectors ever built. 
In Section~\ref{section:LM1-description} and \ref{section:construction-assembly} we give a brief overview of the LM1  construction, integration and assembly steps. In Section~\ref{section:QC} we describe thoroughly the quality control measurements performed during construction. Section~\ref{section:validation} describes the validation and characterisation tests using a cosmic stand. Finally, Section~\ref{section:conclusion} is dedicated to discussion and perspectives.

\section{LM1 Micromegas quadruplets}
%
\label{section:LM1-description}
\subsection{Quadruplet specifications}
\label{subsection:mm-specifications}
The forward muon spectrometer of ATLAS consists currently of gaseous detectors namely Monitored Drift Tubes (MDT), Cathode Strip Chambers (CSC) and Thin Gap Chambers (TGC). In order to fulfil the  spatial resolution (100\,\microns per layer in the precision coordinate) requirements  and to be able to provide the required Level-1 trigger information, these detectors will be replaced by a combination of two other gaseous technologies, Micromegas (MM)~\cite{Giomataris:1995fq} and Small-Strip Thin Gap Chamber (sTGC)~\cite{Mikenberg:1988cr}, after having undergone a detailed program of design and test to assess their compatibility with HL-LHC environment~\cite{Jeanneau:2012pp,Alexopoulos:2010zz,Smakhtin:2009zz}.\\

To ensure the highest possible efficiencies,  the sTGC and MM detectors were both designed to meet the spatial resolution requirements for offline track reconstruction, and to be fast enough to participate in the L1-trigger. This allows redundancy and assures excellent performance with high efficiency. For optimal performance, the sTGC pads provide a localized trigger, and the MM stereo layers allow for a high-resolution second coordinate measurement.

\begin{figure}[!h]
    \centering
	\includegraphics[width=\textwidth]{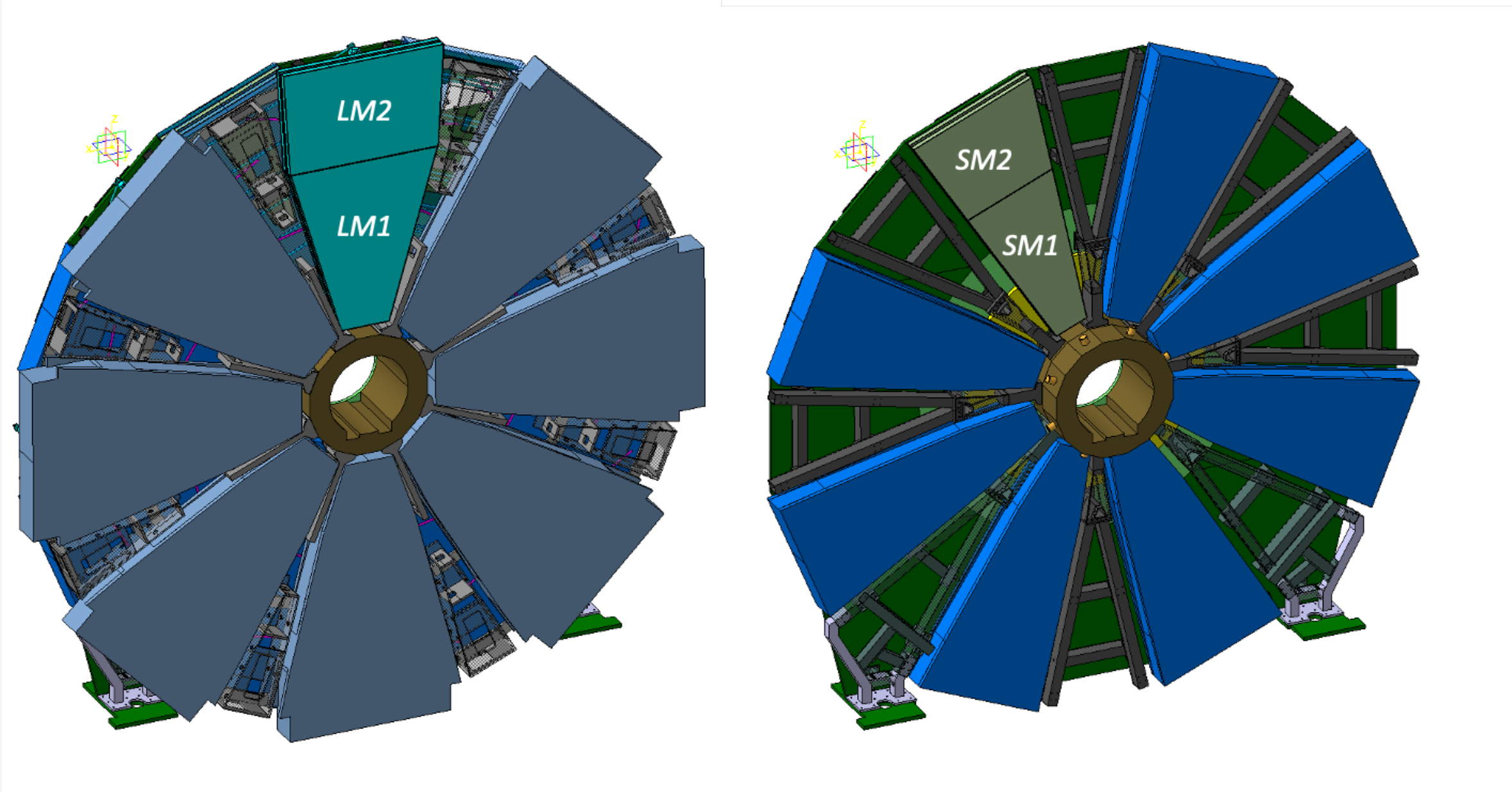}
	\caption{
	NSW detectors layout with the 8 large sectors (left) covering the 8 small sectors (right).} 
	\label{fig:nsw_layout}
\end{figure}

The NSWs are made of eight small and eight large sectors as shown in Fig.~\ref{fig:nsw_layout}.
The NSW sectors comprise a central spacer frame with MM wedges (each consisting of two quadruplet modules) rigidly attached to both faces, and all the MM services routed through the centre of the spacer frame, and with two sTGC wedges (each consisting of three quadruplet modules and all of their services) kinematically mounted outside the MM wedges. The MM wedges are assembled from two radial sections, built as separate modules. There are thus four types of MM modules: SM1 and SM2, the inner and outer quadruplets of the Small wedges, and LM1 and LM2, the inner and outer quadruplets of the Large wedges. Within the ATLAS-NSW collaboration, four consortia are in charge of the MM module construction: Italy (SM1), Germany (SM2), France (LM1) and Greece/Russia (LM2). IRFU (France) is committed to provide 32 modules LM1, each enclosing four readout planes of 3\,m$^2$.\\

In order to reconstruct the muon momentum with 15\% resolution for an energy of 1 TeV, the main specifications of MM detectors are:
\begin{itemize}
    \item spatial resolution in the bending plane ($\eta$)\footnote{The global ATLAS coordinate system is
  right-handed and has its origin in the nominal interaction point, with the
  $X$ axis pointing toward the centre of the LHC, $Y$ pointing upward, and $Z$
  along the beam line. The polar angle, $\theta$, is defined with respect to
  the positive $Z$ axis, and the azimuthal angle, $\phi$, with respect to the
  positive $X$ axis. The pseudorapidity, $\eta$, is defined as $\eta=-\ln
  \tan(\theta/2)$.}, for all track angles: 100\,\microns;
    \item spatial resolution on the second coordinate ($\Phi$), perpendicular to the bending plane : few millimeters;
    \item a pointing angular resolution of 1\,mrad when grouping track clusters from different detector layers;
    \item rate capability up to 15\,kHz/cm$^2$;
    \item no radiation aging over the projected exploitation period of 15 years.
\end{itemize}




Moreover a track sagitta resolution  of $\sim$50-70\,\microns is needed. The internal detector deformation perpendicular to its strips plane will  directly degrade the final resolution. This effect is more important for inclined tracks, i.e. for lower rapidity.
As a consequence, this implies the following constraints on the Micromegas detectors:

\begin{itemize}
    \item within a layer, strips must be straight to within 40\,\microns  along their full length;
    \item the absolute position of each strip within a layer must be known within 40\,\microns;
    \item within a module, the positions of all layers with must be known with respect to each other with a precision of 60\,microns or better
    \item a fixed reference frame, connected to the outside mechanical structure of a detector, with a known precision of 40\,microns;
    \item a module flatness over its full area of the order of or below $\sim$110\,\microns RMS.
\end{itemize}

These induce  mechanical constraints on the assembly process, in particular:
\begin{itemize}
\item the use of  granite tables with a flatness of the order of 20\,\microns RMS as absolute planarity reference planes for the assembly;
\item each individual PCB within a layer must be positioned with a precision of 30\,\microns with respect to its neighbours and to the outside mechanical reference;
\item one face of a readout panel must be positioned with an accuracy of 40\,\microns with respect to the other side;

\end{itemize}

To achieve such performance with MM modules, the precision of the components together with the construction procedures (consisting of more than 1000 steps) must be scrutinized throughout all the processes. 

Given the constraints and physics requirements on the measurement precision for these detectors, module construction requires very strict conditions of temperature, humidity and cleanliness. A dedicated clean room and precision tooling have been developed to fulfill these goals (see Section~\ref{section:tooling}).\\

\subsection{Description of LM1 Modules}
\label{subsection:mm-decription}

\begin{figure}[!h]
    \centering
	\includegraphics[width=10cm]{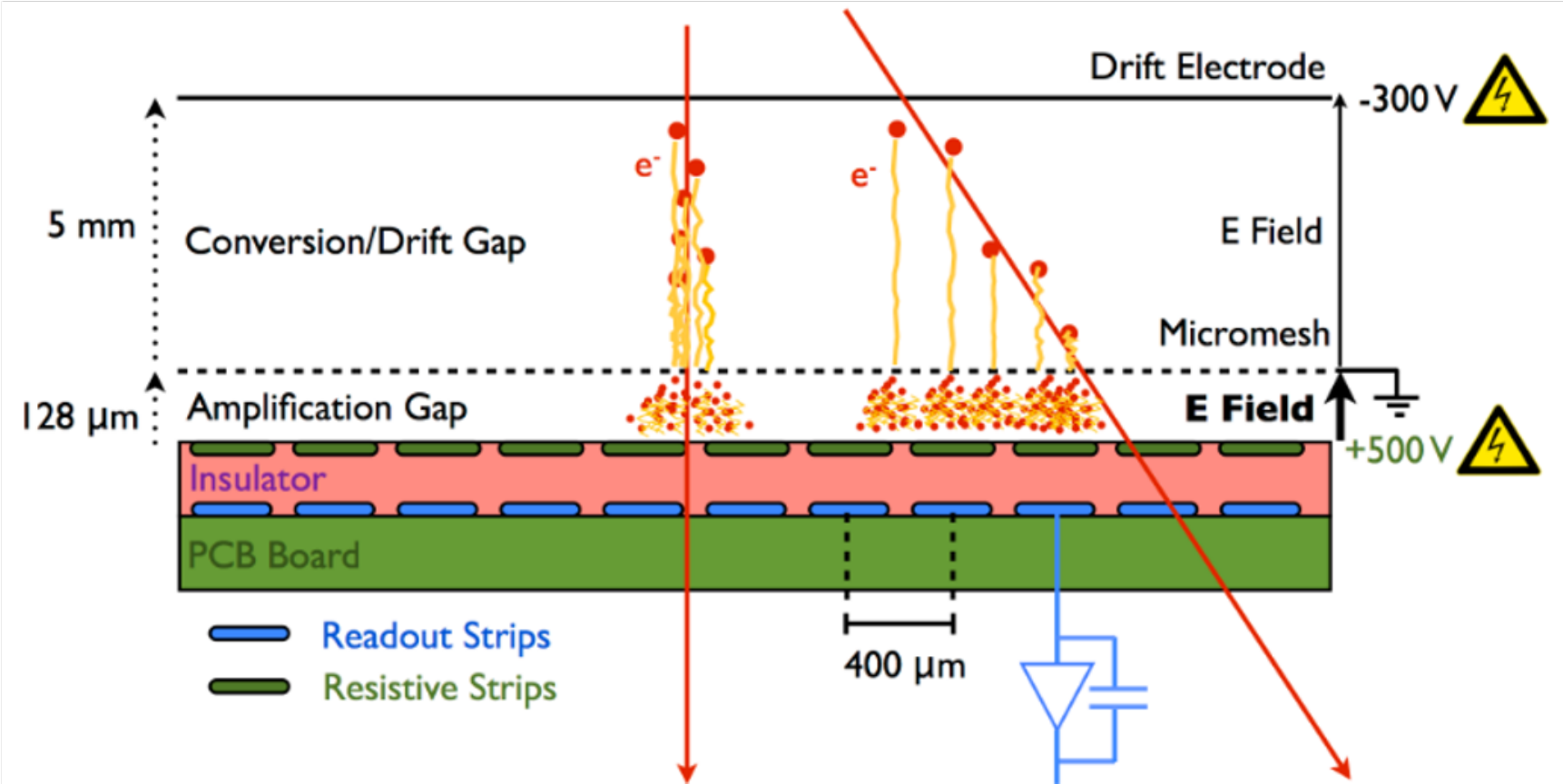}
	\caption{Design of the Micromegas architecture used in the NSW.} 
	\label{fig:mm_principle}
\end{figure}

LM1 modules are the inner MM part of large NSW sectors. They enclose four gas gaps where a Micromegas structure (see Fig.~\ref{fig:mm_principle}) is supported by five composite panels (see Section~\ref{section:construction}), three drift and two readout (RO). \\

\begin{figure}[!h]
    \centering
	\includegraphics[width=15cm]{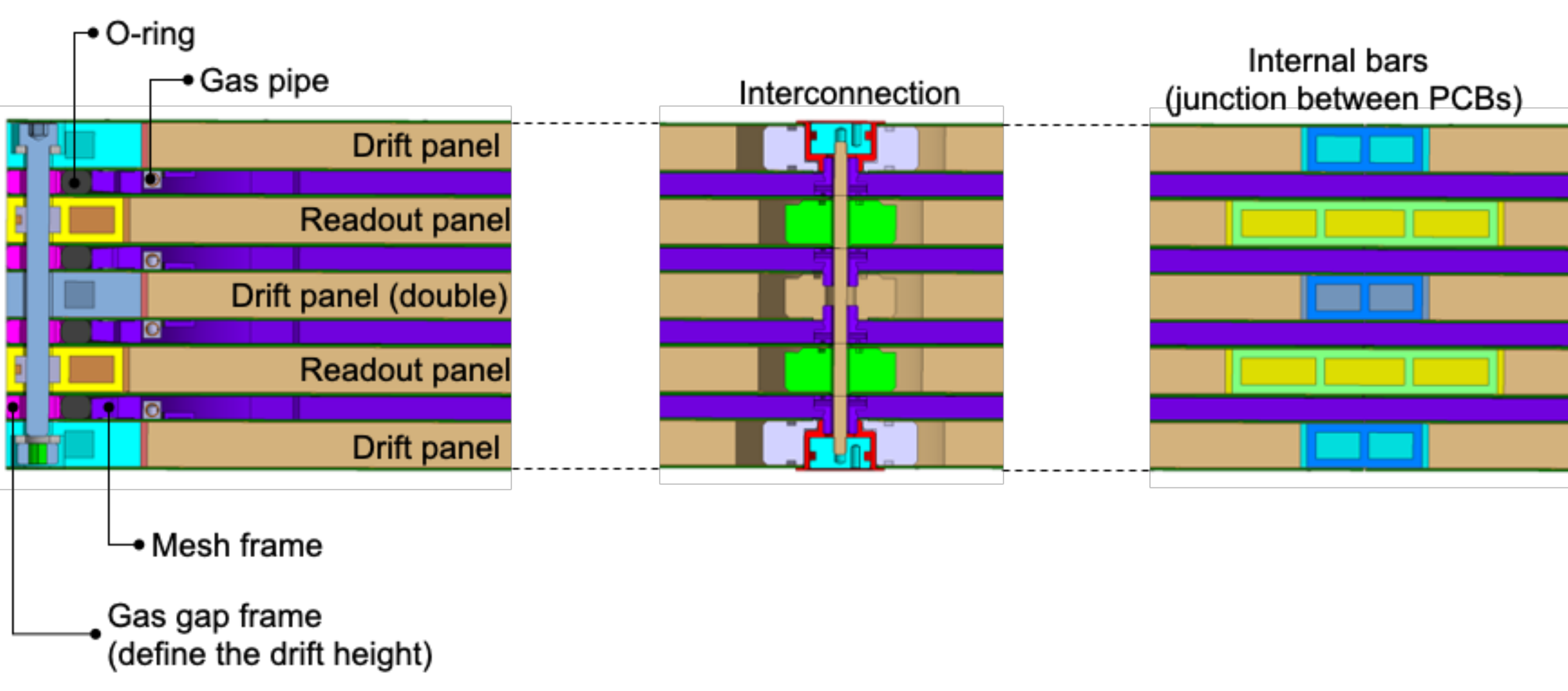}
	\caption{Section of a MM quadruplet with details of the internal components on the edges, at the interconnections and at a PCB junction. Panels are made of aluminum honeycomb (brown), external frames (cyan for drift and yellow for RO) and PCB (0.5 mm thick) where electrodes (drift cathode and anode strips) are printed. Gas-gap frames (pink) define the gas-gap thickness and the O-ring groove. Interconnections are made of two external nuts and one thread rod passing through all panels and reinforcement parts. Internal bars used to reinforce panels at PCB junctions are shown in blue for drift panels and green for RO panels (wider bars because of the angle on the PCB edge for stereo panels).
	} 
	\label{fig:lm1_section}
\end{figure}

\begin{figure}[!h]
    \centering
	\includegraphics[width=0.4\textwidth]{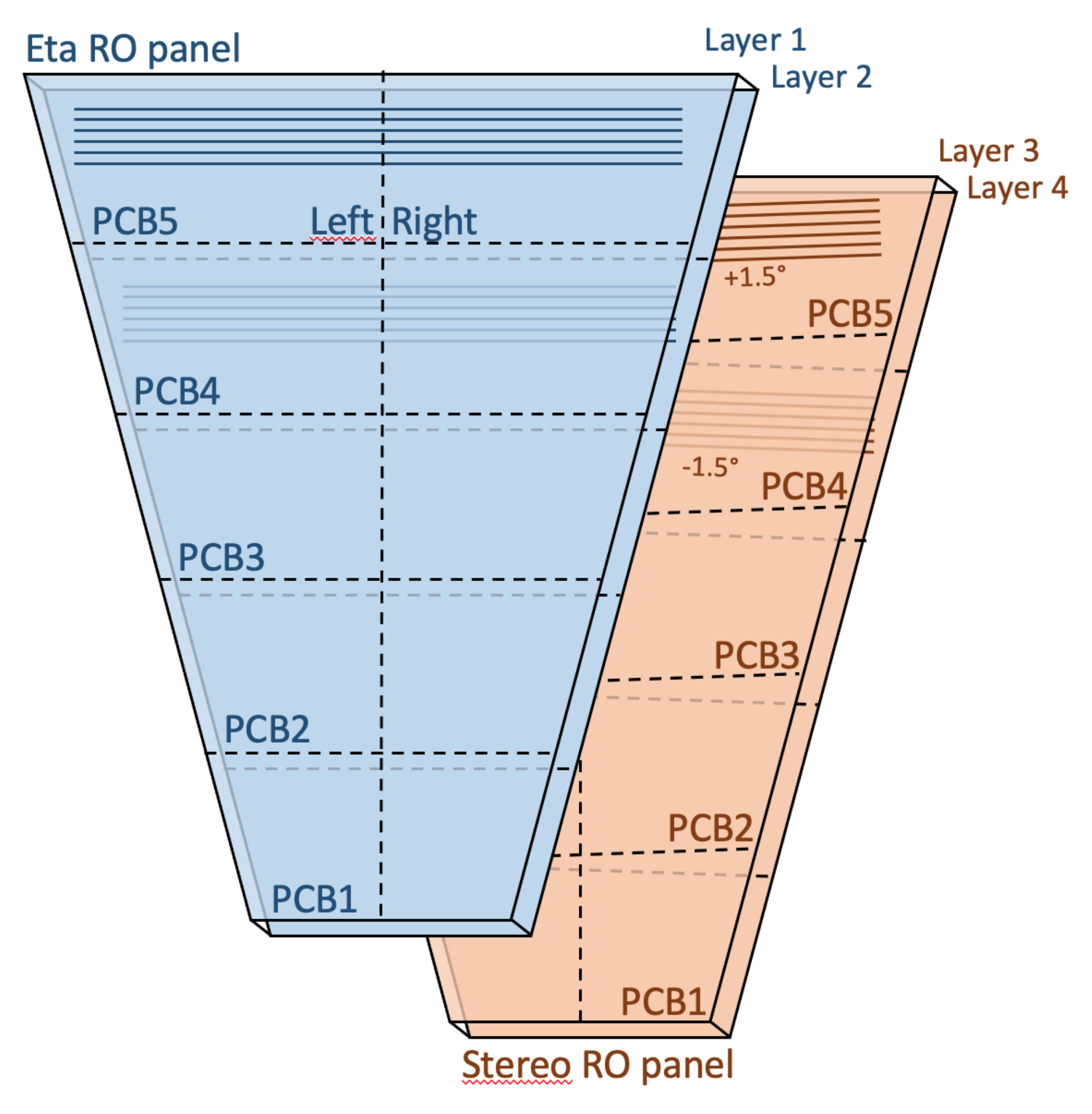}
	\includegraphics[width=\textwidth]{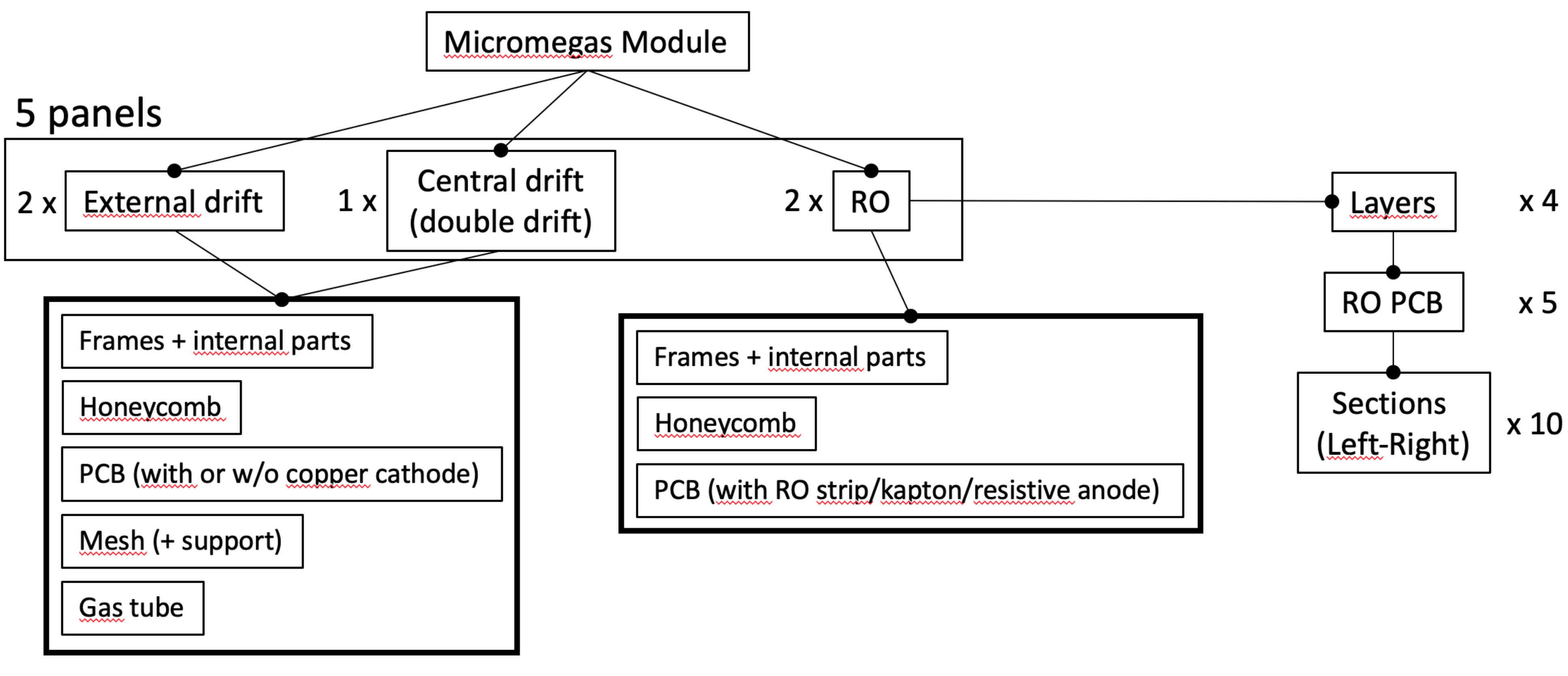}
	\caption{Top: Orientation of strips in different layers of LM1 modules. Each readout panel holds 2 layers (one on each side). Each layer is made of 5 boards, divided in 2 sections (left/right). Bottom: Breakdown structure of a LM1 MM module.}
	\label{fig:lm1_layout}
\end{figure}

Drift panels hold the drift electrode and the mesh, glued on a frame attached to the drift panel, defining a drift gap of 5 mm. The two external drift panels have an outer skin on one side and support a single cathode plane and mesh on the other side, whereas the central drift panel has cathode and mesh on both sides. \\  Readout strips and pillars constitute both sides of the two RO panels: the Eta panel, with strips in the precision direction ($\eta$) and the Stereo panel. The strips on the two sides of the Stereo panel are at opposite angles of $\pm$ 1.5$^\circ$ with respect to the strips of the Eta panel. Each layer is made of 5 PCBs aligned and glued together during panel construction and each PCB is divided in 2 sections, left and right as seen in Fig.~\ref{fig:lm1_layout}.\\

\begin{figure}[!h]
    \centering
	\includegraphics[width=7cm]{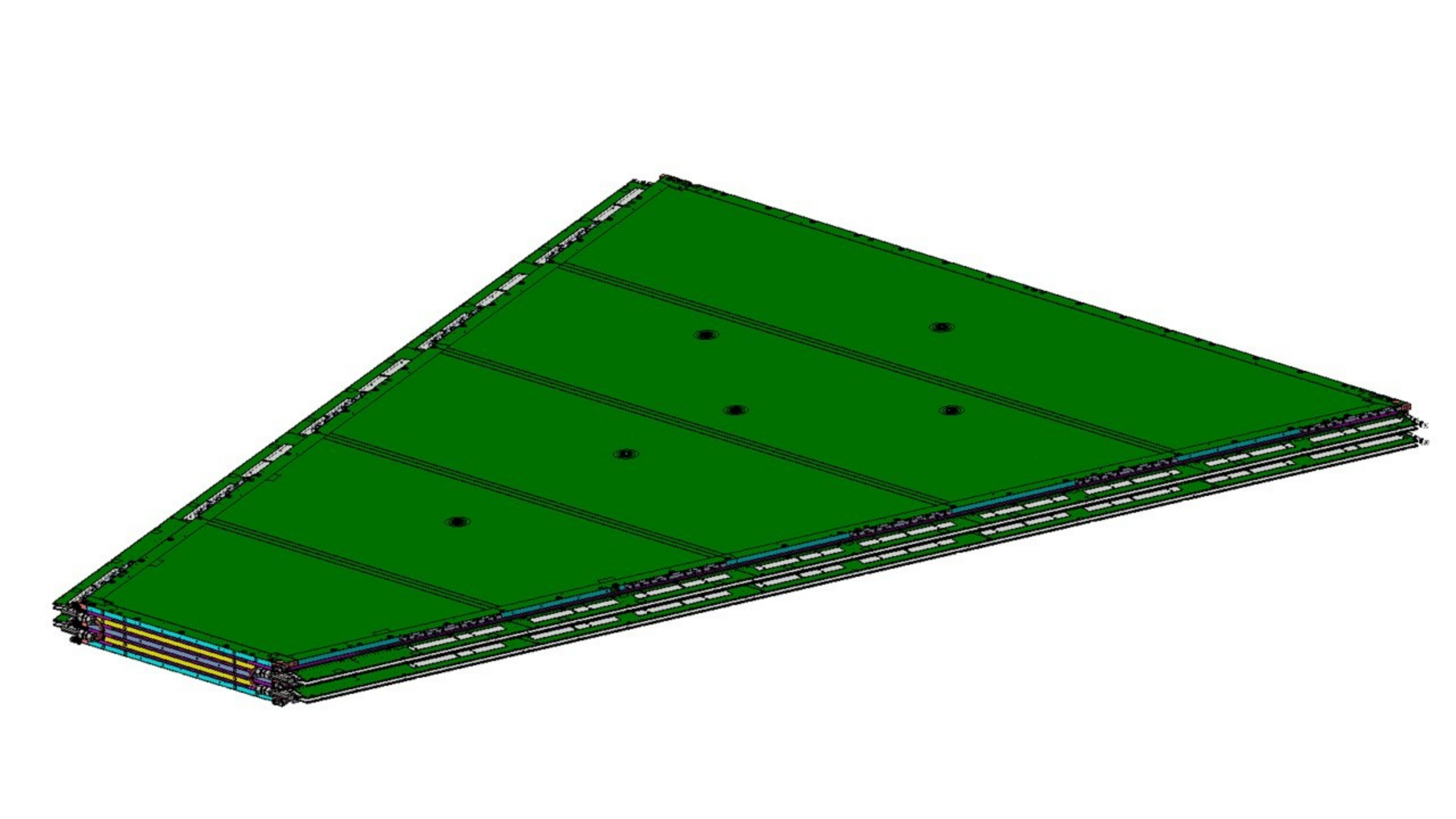}
	\includegraphics[width=7cm]{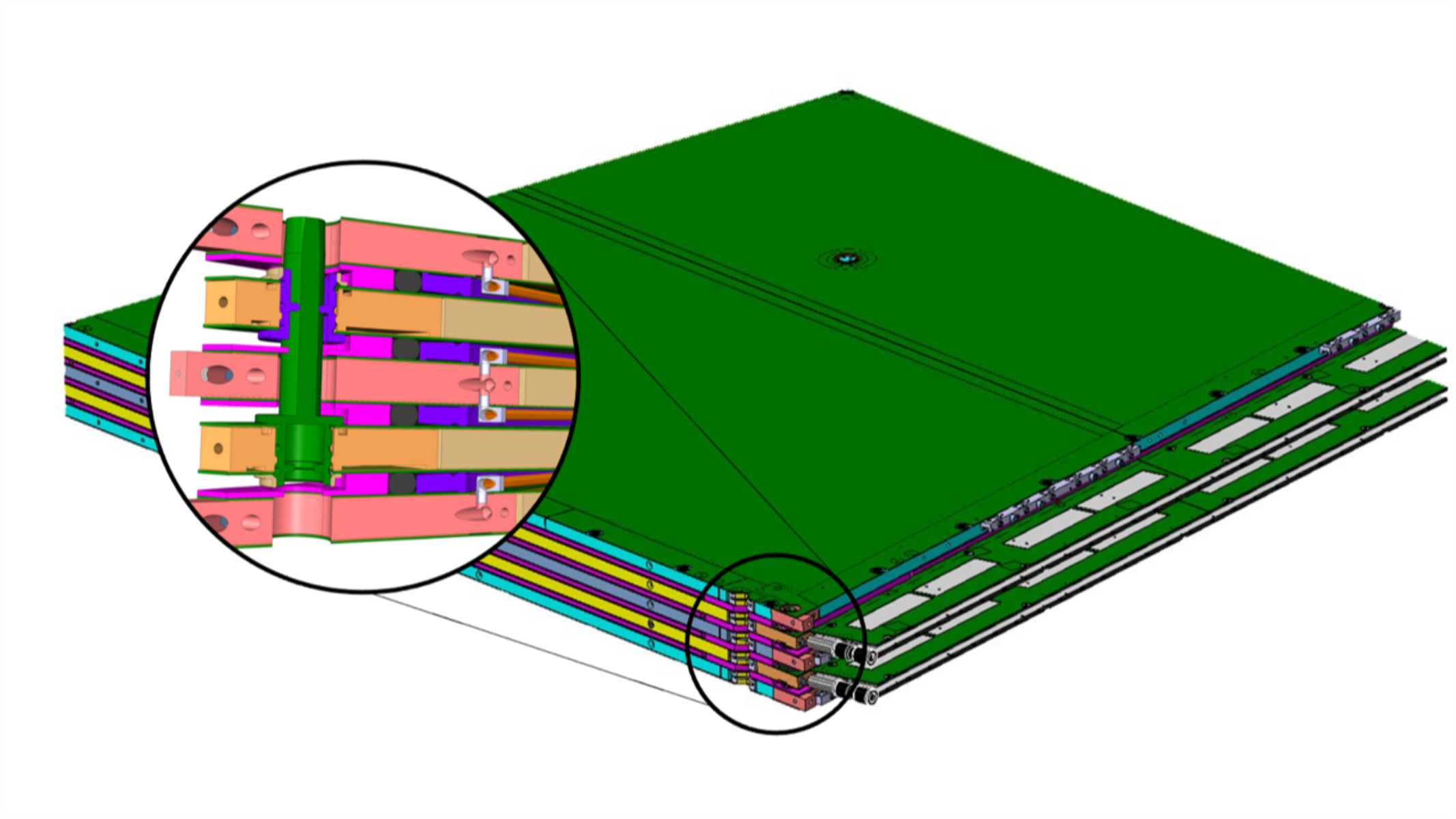}
	\caption{Left: Overview of an assembled quadruplet. The junctions between the 5 PCBS as well as the six interconnections that link the external sides of the outer drift panels are indicated by black circles. Right: Zoom on large base; details of gas and cooling inlets; transverse view of drift panels (cyan), readout panels (yellow) and gas-gap frames (pink). In the circle: cross-section of the module corner showing the alignment pin (green), the gas-gap frame (pink), the mesh frame (purple) and the gas distribution system (corner + gas insert + gas pipe). } 
	\label{fig:LM1-3D}
\end{figure}

LM1 modules are assembled by stacking panels using a dedicated assembly station (see Section~\ref{section:construction}).  A section of such a module is represented schematically in Fig.~\ref{fig:lm1_section} where panels and gas gaps are clearly identified. These modules are the largest Micromegas detectors ever built up to now, with a detection area of 3\,m$^2$ per layer, thus 12\,m$^2$ per module. In order to avoid deformations of the outer panels due to the gas pressure inside the quadruplet, panels are linked together at several points, depending on the size of the module (6 in the case of LM1). External sides of the outer drift panels are connected by a metallic rod. These interconnections allow a homogeneous deformation of all layers in the module. A general view of an assembled module is shown in Fig.~\ref{fig:LM1-3D}. The drift height is given by precision frames at the edge of each gas gap which are closed using a soft EPDM O-ring.


\subsection{Readout printed-circuit-boards (RO PCB)}
\label{section:pcb}
\begin{figure}[!h]
    \centering
	\includegraphics[width=\textwidth]{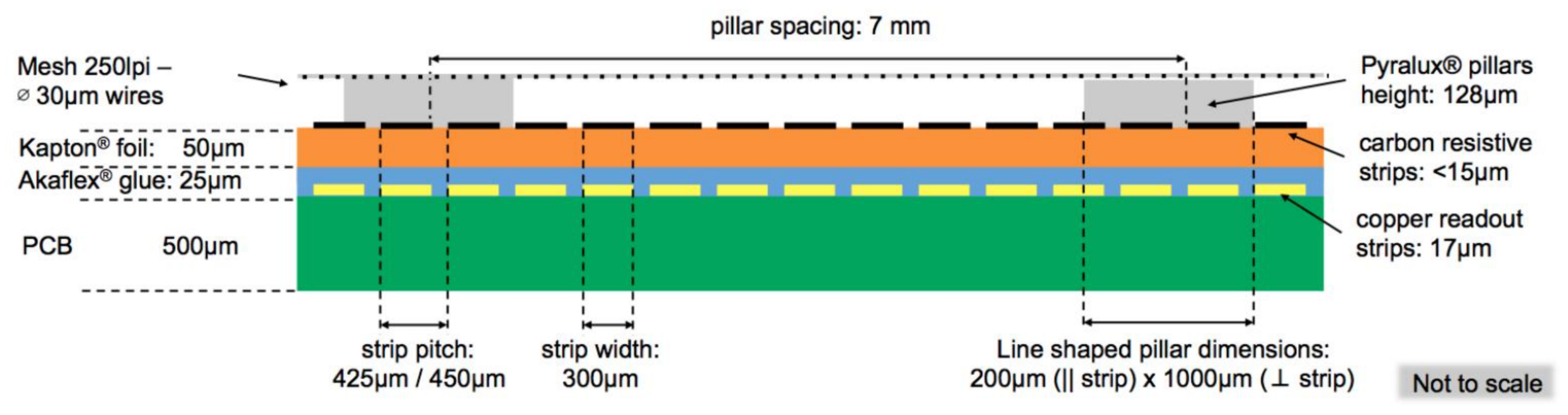}
	\caption{Structure of the RO boards. The mesh is also represented (250 Lines Per Inch).} 
	\label{fig:pcb_layout}
\end{figure}

\begin{figure}[!h]
    \centering
      \includegraphics[width=7cm]{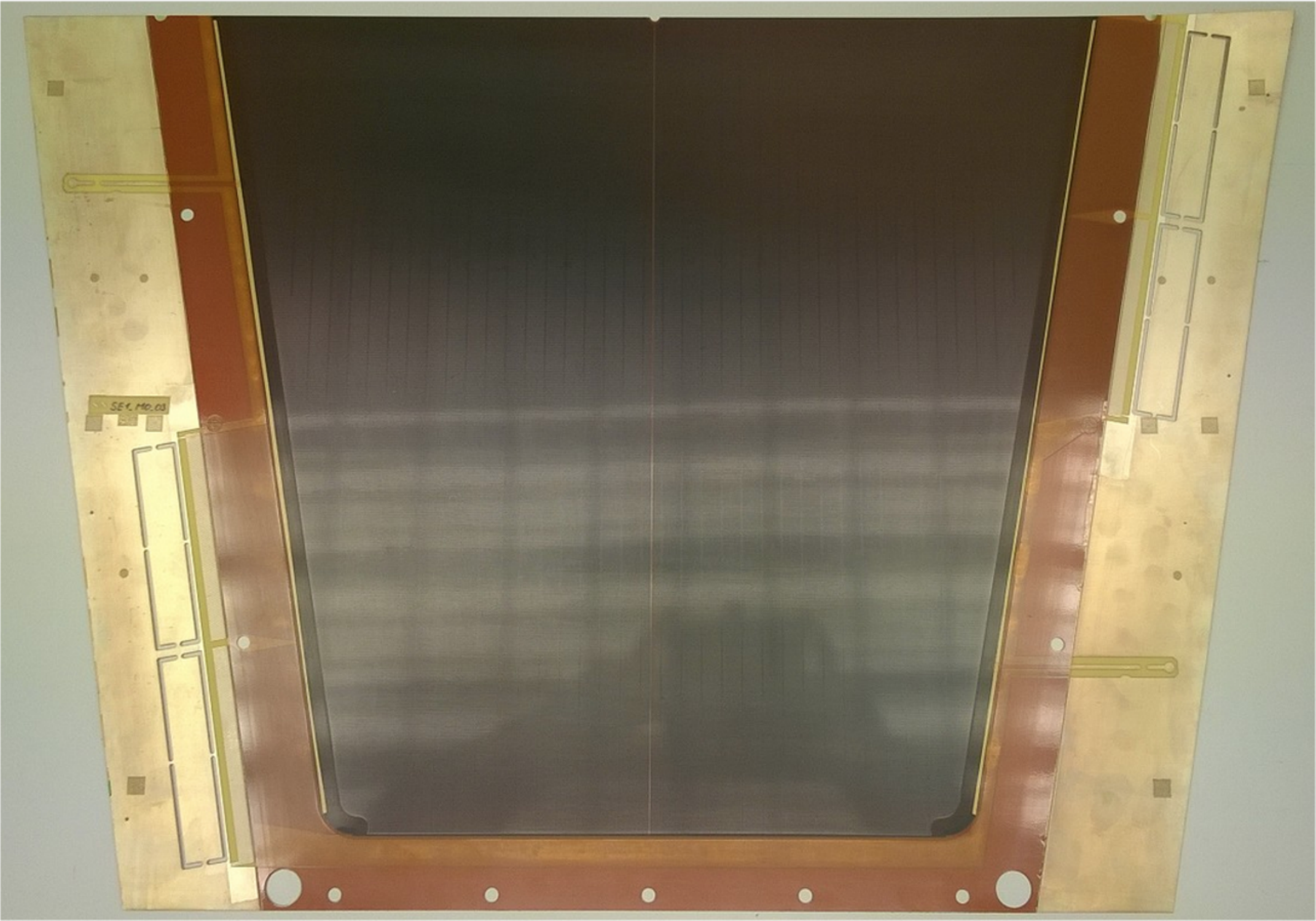}
      \includegraphics[width=7cm]{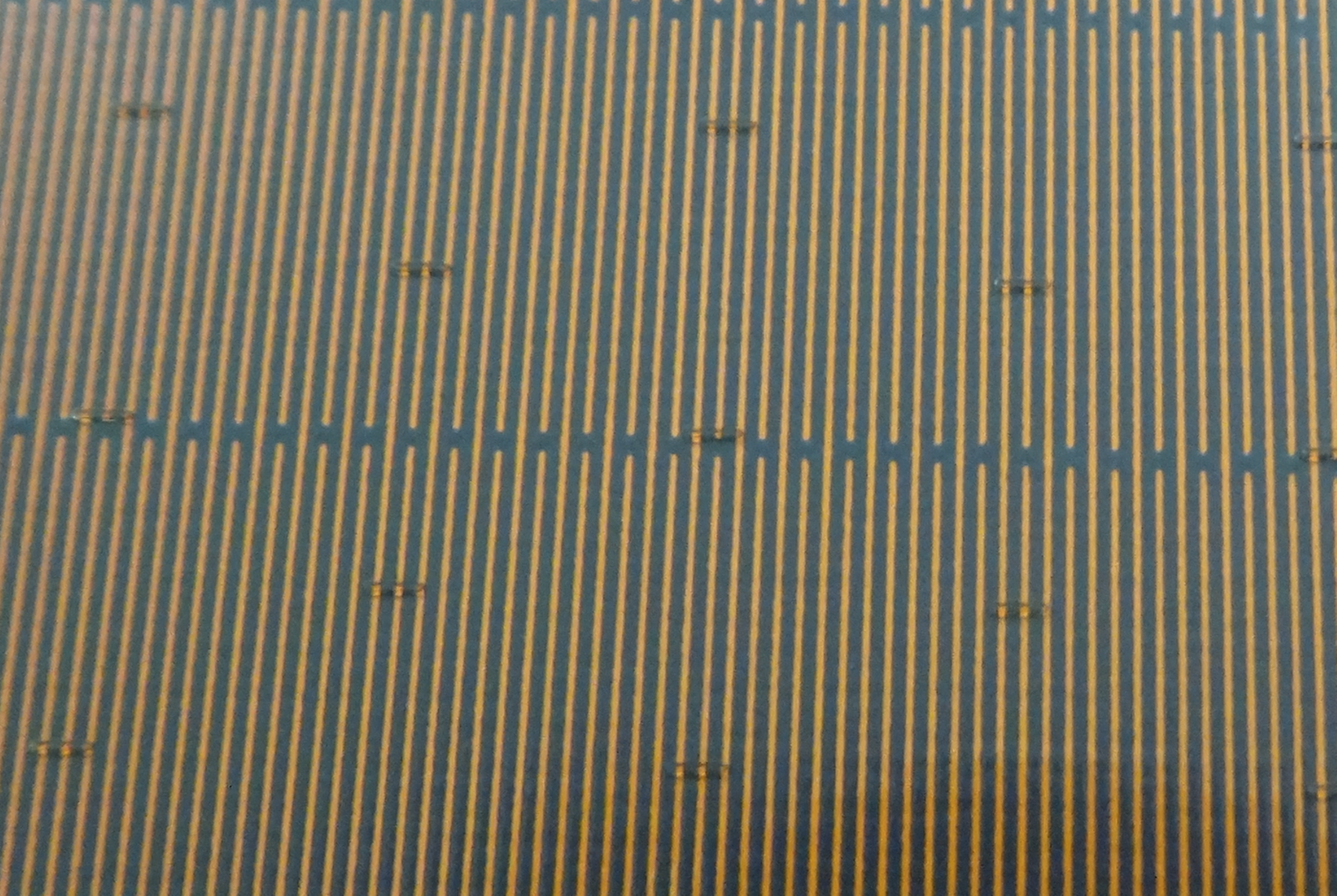}
	\caption{Left: Picture of a trapezoidal readout board (PCB1) already glued on a RO panel. Right: Detail of a Readout PCB where resistive strips and line shaped pillars are visible.} 
	\label{fig:pcb_detail}
\end{figure}

Together with the mesh, the RO boards define the amplification gap of the detector, thus the quality of materials and surfaces are critical for the detector to work in nominal conditions, in terms of gain, efficiency and position resolution. RO boards are manufactured in industry since specific tooling and large capacities are needed to produce them. RO boards are made by pressing at high pressure a stack of different layers (anode strips PCB, glue film and Kapton\textsuperscript{\textregistered} polyimide foil) and adding on top a network of insulating pillars by etching  a Pyralux\textsuperscript{\textregistered} foil (see Fig.~\ref{fig:pcb_detail}). Dimensions and values of interest are indicated in Fig.~\ref{fig:pcb_layout}.\\

Electronic amplification occurs between the mesh and the resistive strips plane. The signal on the copper strips is read by capacitive coupling through the Kapton foil. The resistive strips exactly cover the copper strips, except for a small gap in the centre of the PCB

The resistivity of the carbon resistive strips is about 800\,k$\Omega/$sq allowing the Micromegas detector to sustain the  high-rate particle environment. This structure was studied during
the MAMMA R\&D program~\cite{Jeanneau:2012pp,Alexopoulos:2010zz} where the adaptation of  Micromegas detectors to the HL-LHC environment was explored.\\

Each RO board is trapezoidal, with a height of 462\,mm and an opening angle of 33$^\circ$ . The short base of the smallest RO board measures 660\,mm, and the long base of the largest is 2008.5\,mm.

The overall dimensions of a RO panel are \SI{660}{mm} and \SI{2008.5}{mm} for the small and large bases respectively, for a height of \SI{2310}{mm} and a surface of \SI{3.1}{m^2}. A RO panel is made of five RO PCBs (see Section~\ref{fig:lm1_layout}) that are aligned precisely on the granite table during the panel gluing. The next Section will describe the infrastructure and tooling used for panel construction.

%

\section{Quadruplet construction and assembly}
\label{section:construction-assembly}
\subsection{Dedicated construction infrastructure}
\label{section:tooling}
\subsubsection{CICLAD clean room}
A new detector construction facility, CICLAD (Conception, Integration, and Characterization of Large Area Detectors), was set up with financial help from the French Ile de France region~\cite{sesame}. The core infrastructure of this facility is a 120\,m$^2$ clean room providing an ISO7 (Class 10000) cleanliness level in the main area, and an enhanced ISO5-level (Class 100)  in the protected areas where part of the construction and assembly take place. The clean room is equipped with two large-area (3.0 $\times$2.3\,m$^2$) instrumented granite tables, dedicated areas for tooling and washing, and special gateways for getting equipment in and out. 
\begin{figure}[!h]
    \centering
	\includegraphics[width=12cm]{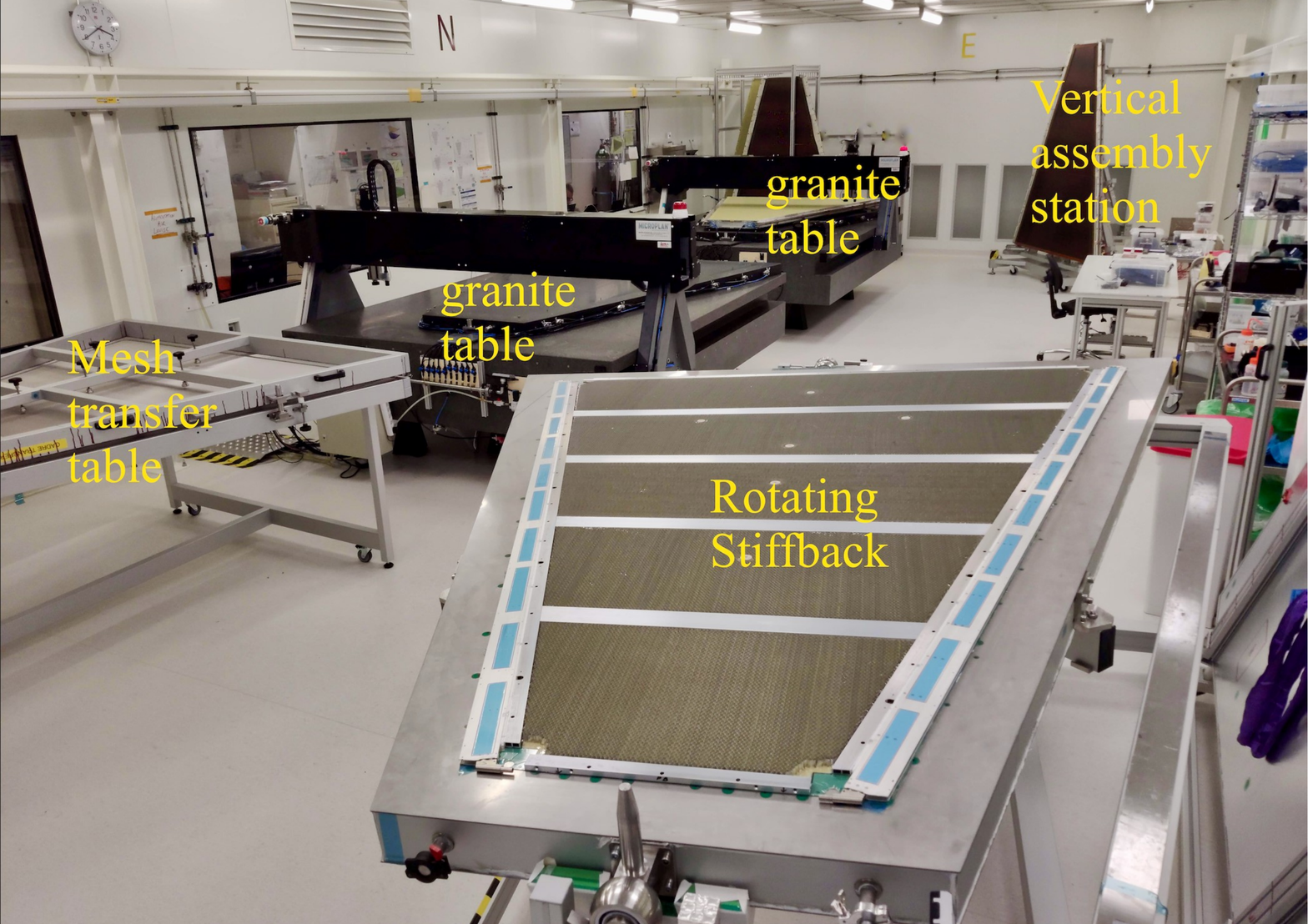}
	\caption{CICLAD clean room. Main tools for panel production and module assembly are shown.} 
	\label{fig:cleanroom}
\end{figure}
\subsubsection{Instrumented granite tables}
Two identical granite tables~\cite{Microplan} provide the high-quality planar references used to position precisely the various parts during construction of the panels. On top of the granite table upper surface, a secondary trapezoidal 8-cm thick granite sole carries high-precision metal pins to position the panel corners and each of the five PCB forming the panel outer skins. This sole is machined with a primary depressed air suction channel to force it to match the underlying granite top (30\,t attractive force), and five secondary suction channels to force the panel skins to match its top surface while gluing (adjustable downward pressure, typically 100\,mbar below $P_{atm}$).

\subsubsection{Measurement gantries}
\begin{figure}[!h]
    \centering
	\includegraphics[width=12cm]{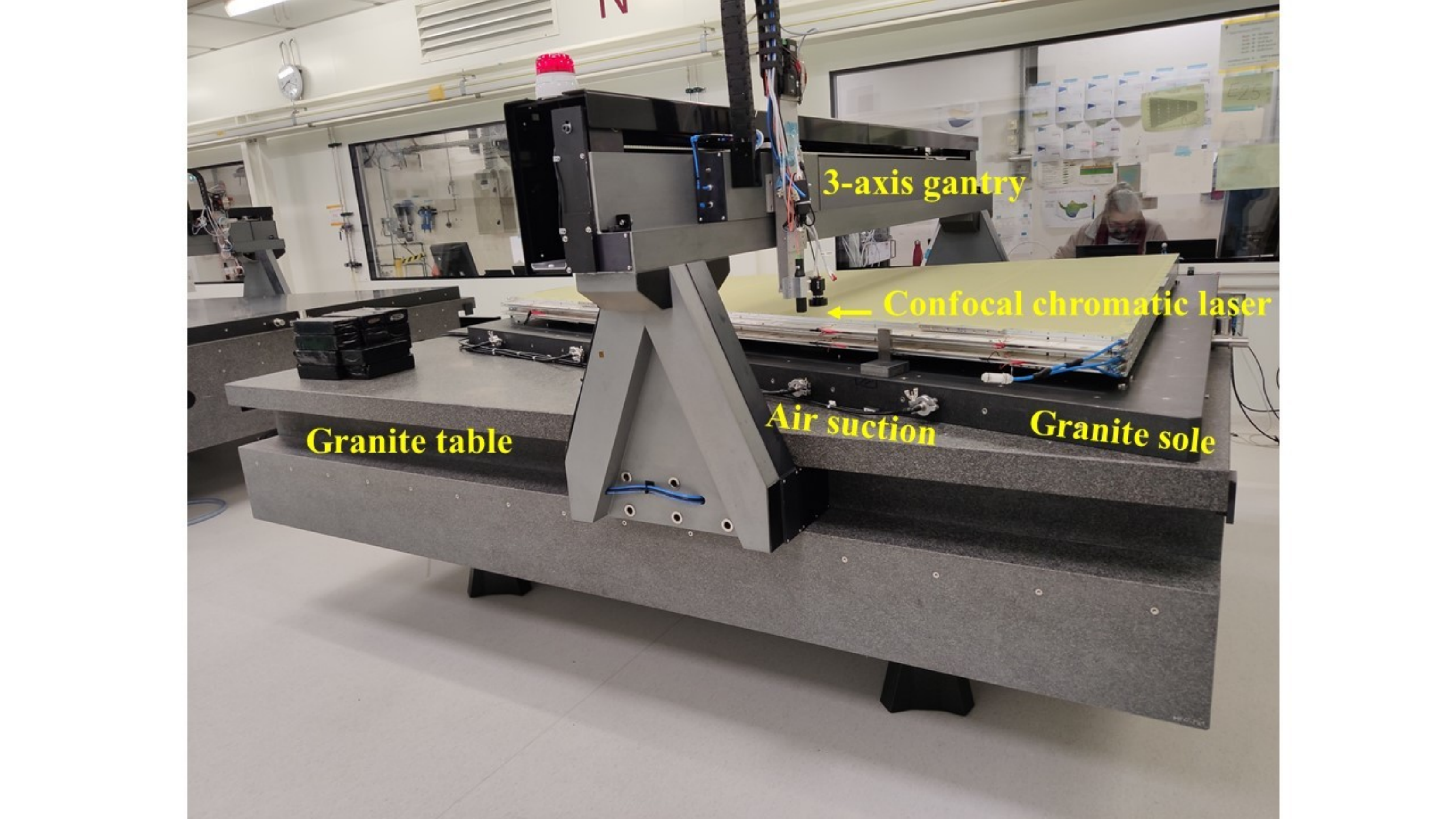}
	\caption{Planarity scan survey of a module.} 
	\label{fig:cleanroom2}
\end{figure}

Each granite table is equipped with a moveable 3-axis gantry (see Fig.~\ref{fig:cleanroom2}).
The gantry is made of high rigidity SiC beams, and provides fast air-cushion supported movements in the horizontal X and Y directions over a 2.4$\times$2.1 \,m$^2$ active area, using belt-driven displacements. Z movements use a  slower archimedean screw adjustment; this is acceptable since the measurement head position is mostly static in Z when doing XY planarity surveys.

\subsubsection{In-place measurements}
During production,  panel parts and elements are successively placed on the granite tables according to an incremental construction process (see~\ref{section:assembly}). Taking advantage of their accurate 3-D coordinate measurement capabilities, these tables are extensively used during this sequence to perform in-place planarity checks of panel components at crucial intermediate stages (skin placement before gluing, honeycomb and bar position check after gluing - see Section~\ref{section:drift}). Once panel construction is complete, they are again used for in-place measurements of the finished panel, both under suction and in relaxed mode. The final assembled quadruplet module is also measured before it leaves the clean room (Fig.~\ref{fig:Module3D}).
\subsubsection{Planarity Sensors}
Precise planarity measurements in the vertical Z direction rely on a contactless confocal chromatic laser device~\cite{Stil}. Depending on the precision required and on the element measured, a planarity survey lasts between 30 and 90 minutes, collecting several million points over the active panel surface.

\subsubsection{Software}
Dedicated in-house software was developed to control the gantry and automatically conduct the various measurement surveys and associated data logging (C++/Qt). Quick-look 3D basic analysis of the just-measured altitude point cloud at a particular production phase is made with open source tools~\cite{Cloudcompare}. In-house software tools (C++/Root) also provide advanced quick-look analysis used at intermediate step in the construction, and conduct the fine analysis leading to the final reports and database logs.

%
\subsection{Construction steps, integration and assembly}
\label{section:construction}
Quality checks ensure all items entering panel construction process are conformed to requirements. In particular, all mechanical parts (frames, honeycomb, corners) thickness is measured with accuracy better than 10\,\microns. Then
RO and drift panels are glued using the granite tables described in Section~\ref{section:tooling}. After additional processes, including wet cleaning, the panels are assembled by stacking on the assembly station. This section describes the different steps to complete an LM1 module.

\subsubsection{Drift and double-sided drift panels}
\label{section:drift}

Panels are sandwiches of honeycomb and PCB skins. They are glued in two steps in order to always have the side to be glued in contact with the reference table described in Section~\ref{section:tooling}:

\begin{itemize}
    \item the first side of the panel is made of five PCBs on which the copper cathode is printed. They are positioned and attached with vacuum to the granite table. Tape is used to close all openings and holes to apply the vacuum and avoid any leak of glue which will be applied on this first side;
    \item bars and honeycomb are put in place and this first side is cured in a vacuum bag overnight (see Fig.~\ref{fig:drift});
    \item the first side is removed from the granite table and attached with vacuum to a rotating aluminum-reference surface (stiffback - visible in  Fig.~\ref{fig:cleanroom});
    \item the PCBs of the second side (with or without copper cathode depending if it is a single or double drift panel) are prepared as for the first side (except honeycomb and bars) and then glue is applied;
    \item the stiffback is rotated with the first side on it and positioned over the second side, using precision shims to define the panel thickness;
    \item The glue is cured overnight and the panel can then be removed and
    equipped with the mesh frame, as described in  the next section. 
\end{itemize}

\begin{minipage}{\textwidth}
\vspace{0.5 cm}
  \begin{minipage}[c]{0.45\textwidth}
    \centering
	\includegraphics[width=6cm]{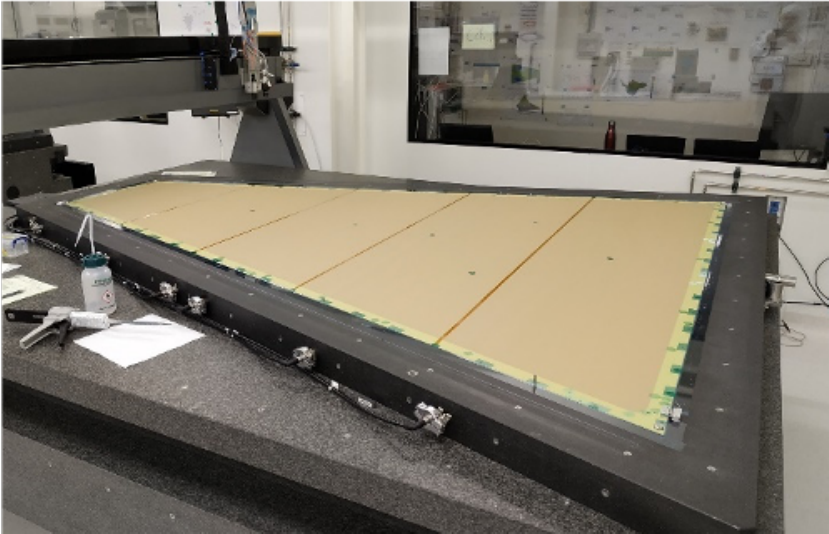}
	\includegraphics[width=6cm]{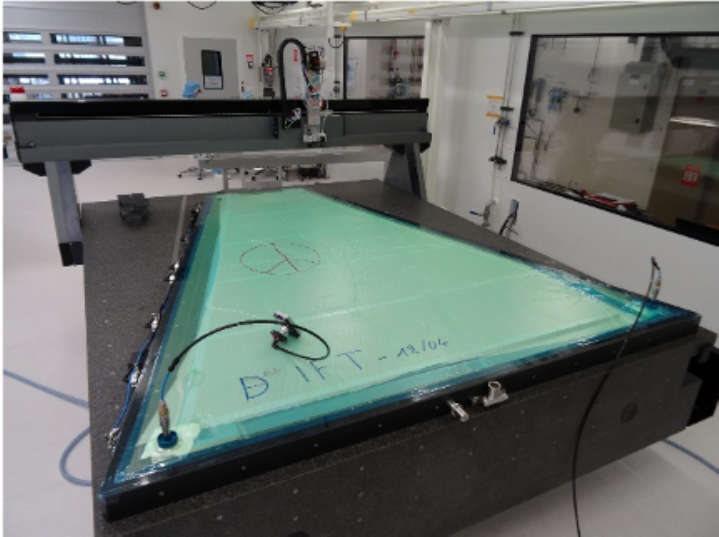}
    \captionof{figure}{Panel gluing: first side (top); vacuum bag (bottom).}
	\label{fig:drift}
  \end{minipage}
  \hfill
  \begin{minipage}[c]{0.45\textwidth}
    \centering
 	\includegraphics[width=6cm]{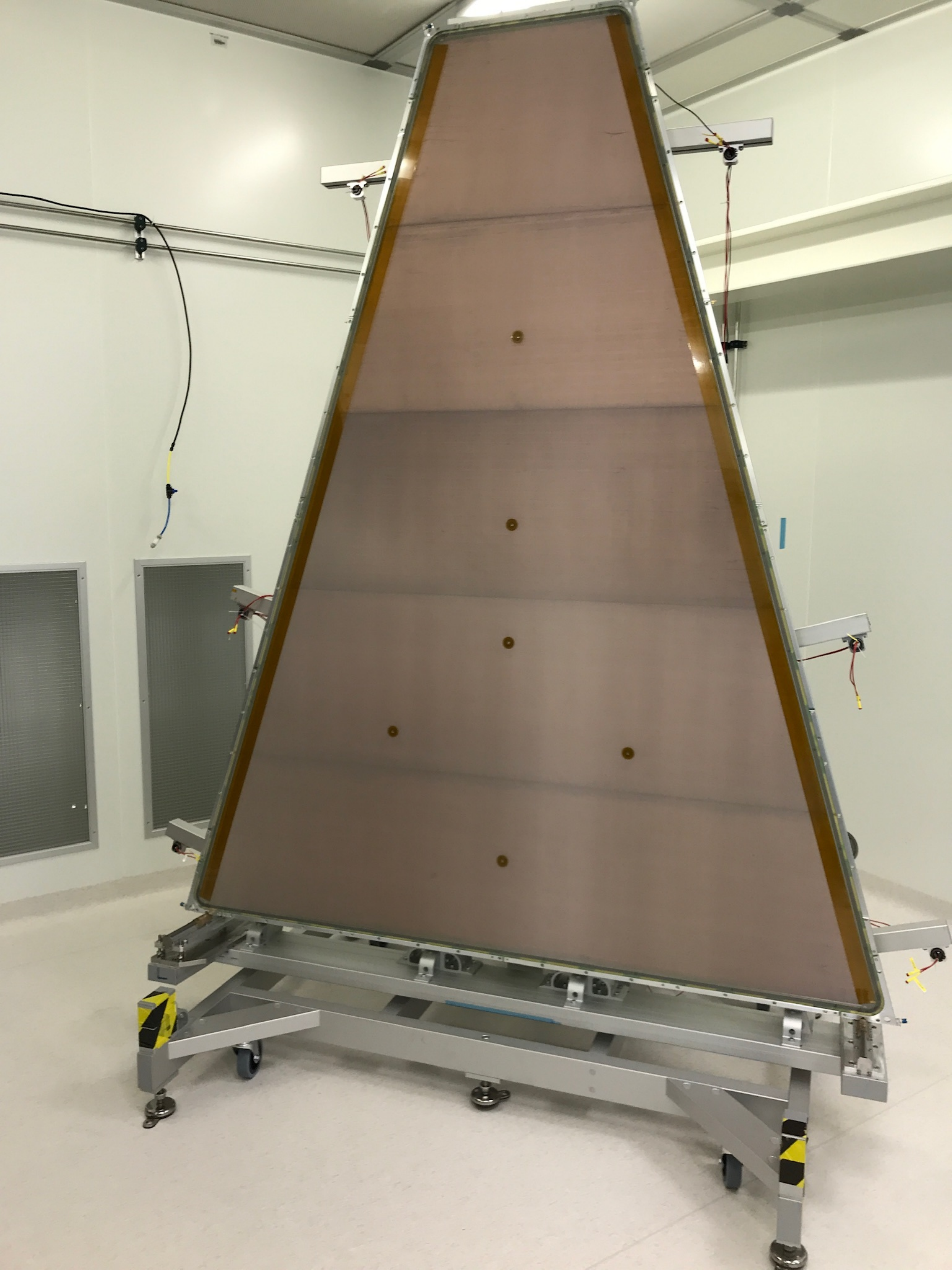}
    \captionof{figure}{A drift panel fully equipped. The five drift PCBs with their copper cathode are distinguishable through the greyish mesh stretched on it.}
	\label{fig:drift_panel}
  \end{minipage}
\vspace{0.5 cm}
 \end{minipage}

\subsubsection{Mesh stretching and assembly on drift panels}

The stainless steel mesh (250 Lines Per Inch - 71\,\microns aperture and 30\,\microns wire diameter – \si{2750}{mm} x \si{2240}{mm}) is stretched by means of a stretching table and glued on a transfer frame. During tension onto the transfer frame and again after transfer onto the drift panel, the mesh mechanical tension is checked in at least 36 points (in both the warp and weft directions) with a unidirectional tensiometer. The nominal tension is within the interval 7-10\,N/cm$^2$. Then the mesh is glued onto a frame previously fixed on the drift panel, in a way that the mesh tension is maintained and the gap between cathode and mesh is \si{5}{mm} (see Fig.\ref{fig:drift_panel} for a complete drift panel).


\subsubsection{Readout panels}
\label{section:readout}
RO panel gluing follows the same steps as for the drift panels except for a few specific points: positioning and mechanical reference, a precision pin for front-end electronics, resistivity measurement and adjustment.\\

\begin{figure}[!h]
    \centering
	\includegraphics[width=8cm]{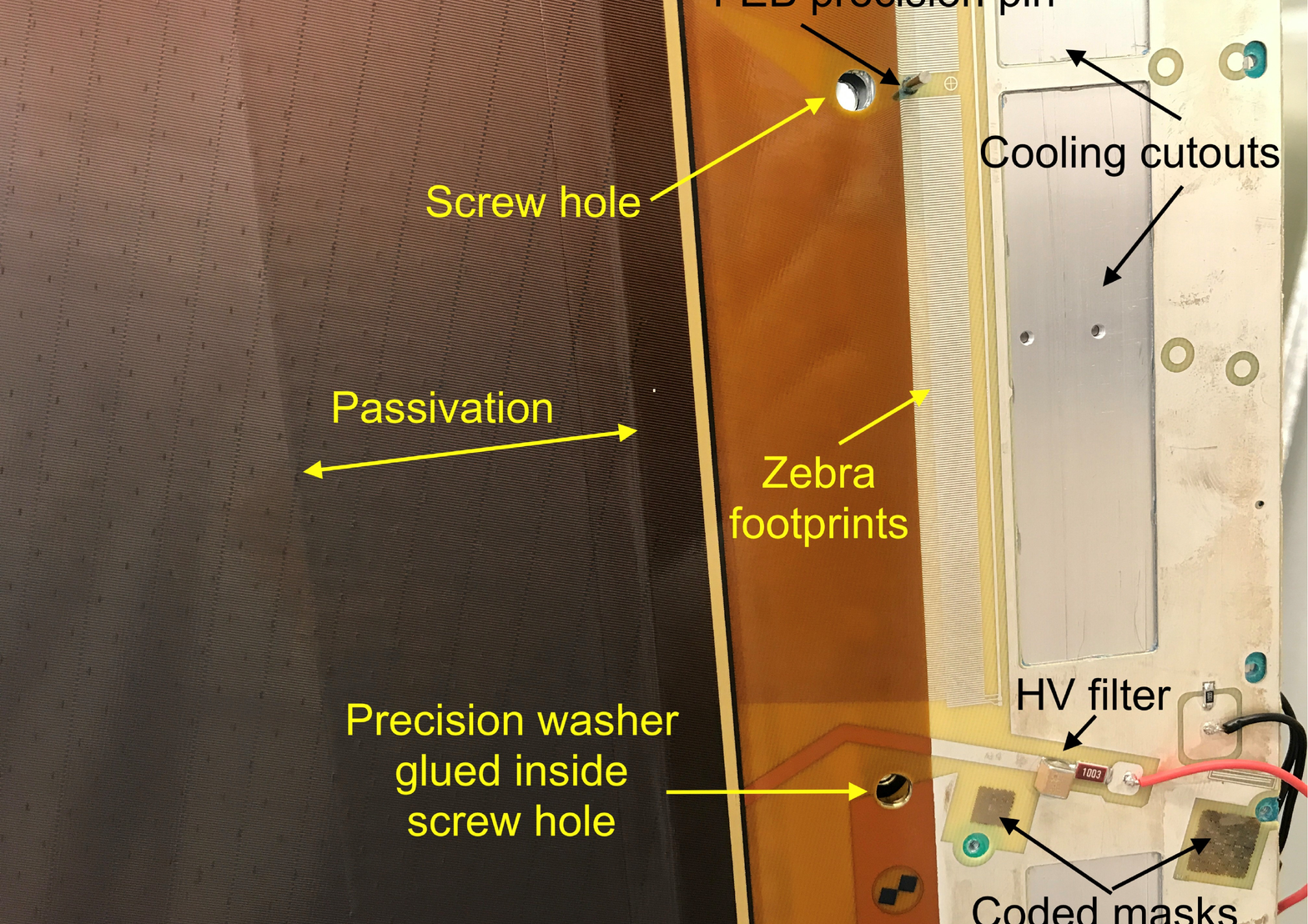}
	\includegraphics[width=10cm]{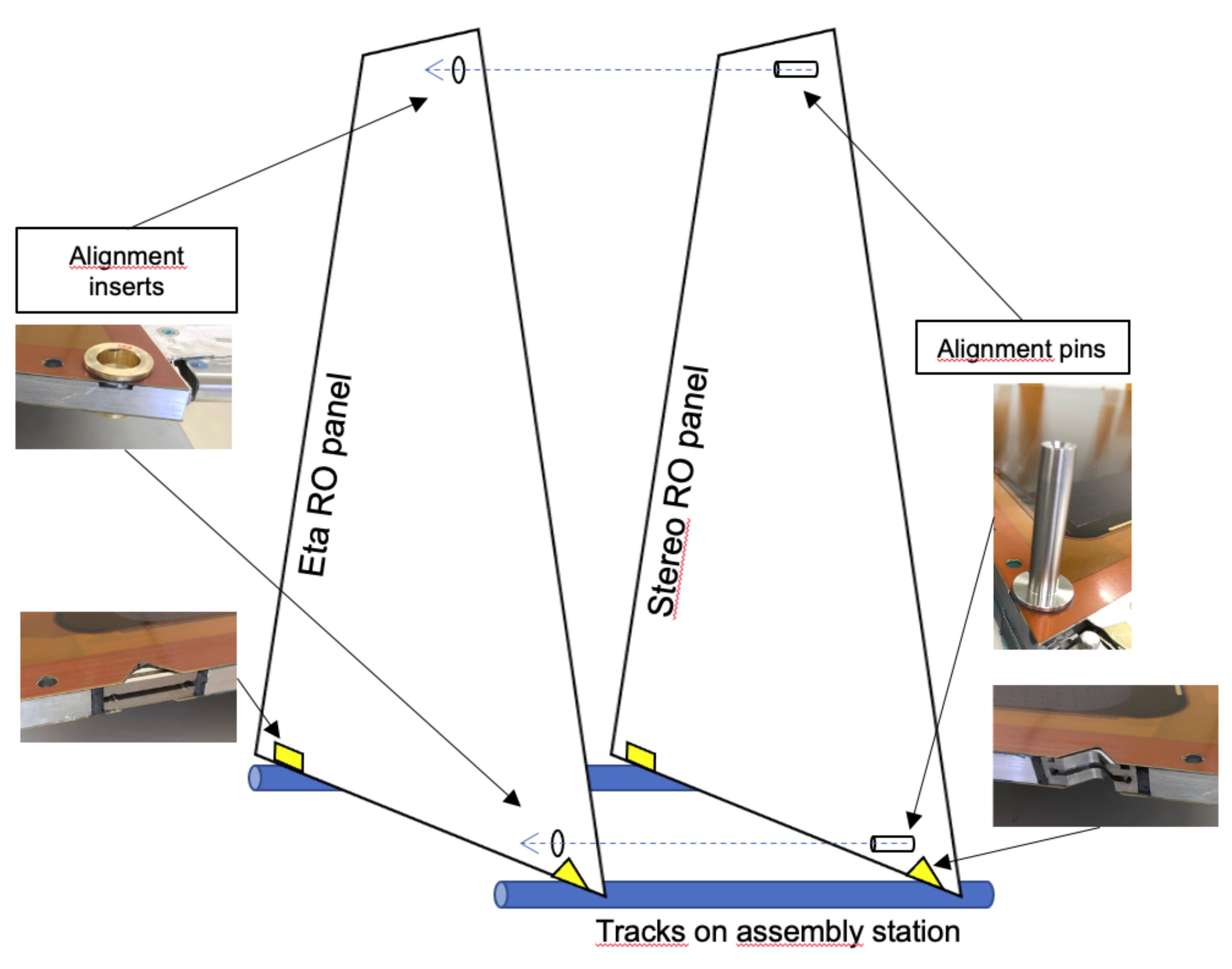}
	\caption{Top: Detail of a RO panel. Coded masks needed to glue the precision washer are indicated. The precision pin for electronics positioning and the RO strips footprints are also shown. Bottom: Positioning scheme of RO panels on the assembly station tracks (see Section \ref{section:assembly}). Alignment inserts and pins are represented.}
	\label{fig:ropanel_detail}
\end{figure}

\textbf{Specific gluing steps}\\
First, in order to position precisely the RO boards and thus the strips on one side, precision washers are glued on the PCB, using the coded masks that have been etched together with strips during PCB manufacturing (see Fig.~\ref{fig:ropanel_detail}, Section~\ref{section:pcb} and \ref{section:QC}). These washers will compensate the error of the absolute position of the precision pin on  the granite table.\\

Then precision inserts, called V (for V-shape) and L (for Line-shape), are glued inside the panel. They will be used to align Eta and Stereo RO panels during the stacking of panels on the assembly station.\\
Finally, two alignment pins are glued on the Stereo panel and corresponding inserts on the Eta panel.\\

\textbf{Finalization processes}\\

RO panels undergo then several steps before being ready for assembly:
\begin{itemize}
    \item excess of glue on the edges of the panel and cooling cutouts are removed (see Fig.~\ref{fig:pcb_detail});
    \item ten precision pins, necessary for the positioning of the Front-End Electronics on the edges of each PCB, is glued using a dedicated tool developed in Saclay and described in~\ref{section:QC};
    \item HV-filter components are soldered at dedicated positions on the HV connection line on the panel
    \item Wet cleaning and drying for three days at \SI{45}{\degreeCelsius}.
\end{itemize}

\begin{figure}[!h]
    \centering
	\includegraphics[width=7cm]{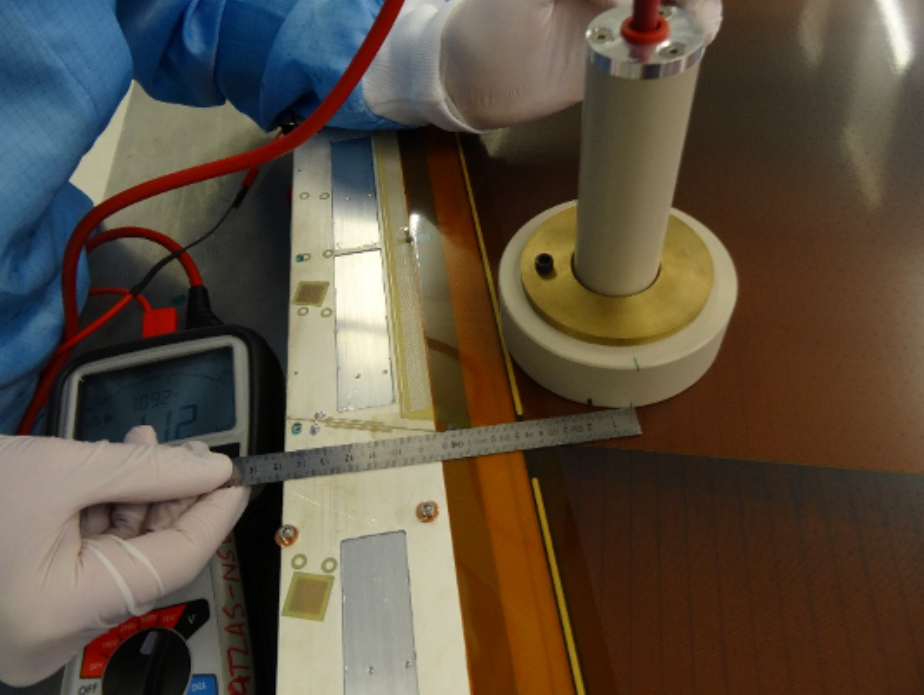}
	\includegraphics[width=7cm]{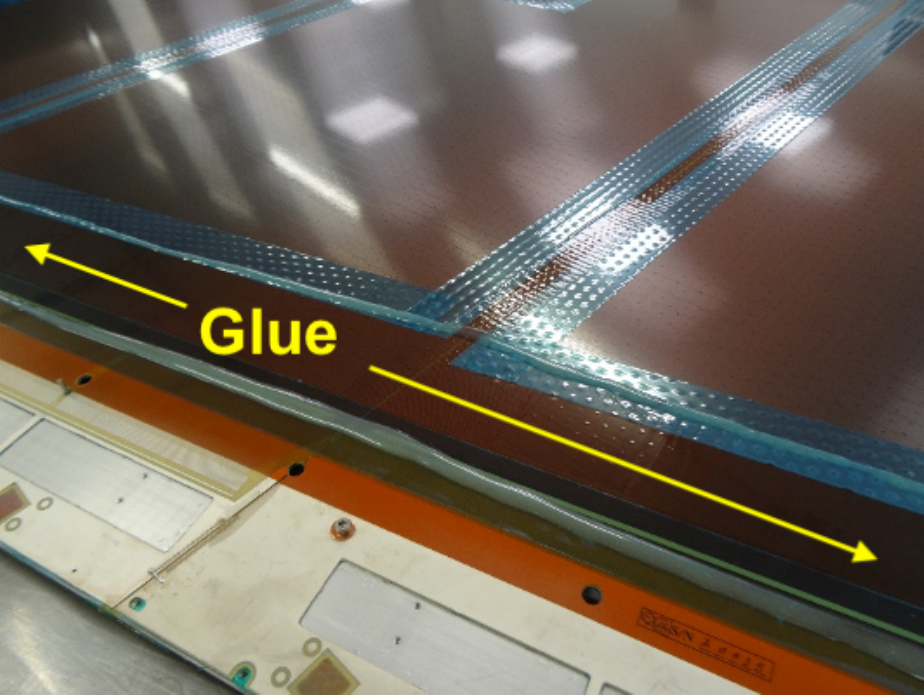}
	\caption{Left: Resistance measurement, with a 1\,cm$^2$ probe (hidden by the plastic disk), as a function of the distance from the edge. Right: Passivation with tape masking and glue application.}
	\label{fig:passivation}
\end{figure}

\textbf{Resistivity measurement and passivation}\\
The last step before final cleaning, is a check of the resistance on the edge of each PCB. A minimum value is needed to ensure a good high-voltage behaviour of the final detector, i.e. a minimal dark current and sparks rate at the nominal working voltage \SI{-570}{\volt} in a gas mixture of Ar/CO$_2$ (93/7\%). This checkpoint is mandatory before assembling the detector since this minimum resistance value is not always reached by the manufacturer.\\
 Resistance values are measured with a 1\,cm$^2$ probe between the HV connection and several points along the coverlay edge. If the resistance is less than  \SI{1}{\mega\ohm}, the coverlay wall is artificially increased by adding a layer of glue (Araldite\textsuperscript{\textregistered} 2011) until this threshold value is reached (see Fig.~\ref{fig:passivation}). This procedure is called {\it passivation} and is also applied in potential weak regions like around interconnections or on PCB junctions.\\
This additional step ensures  a very good HV stability of the final module but at the expense of a reduction of the active area (see Section~\ref{section:validation}).

\subsubsection{Washing and drying of panels}
All panels are washed (with water) in a dedicated part of the clean room that has been especially built for this purpose. Before being washed, the panels need to be prepared by masking all the assembly holes and the interconnections with tape, to avoid getting water inside the panel. After wetting the panels with warm water (\SI{40}{\degreeCelsius}), the panels are brushed with a micropolishing solution and thoroughly rinsed with warm tap water and scrubbed with a rotating brush.
This operation is repeated twice before using deionised water for the final rinsing. Afterwards the panel is  dried in an oven  for at least 72 hours at \SI{45}{\degreeCelsius}.

\begin{figure}[!h]
    \centering
	\includegraphics[height=5cm]{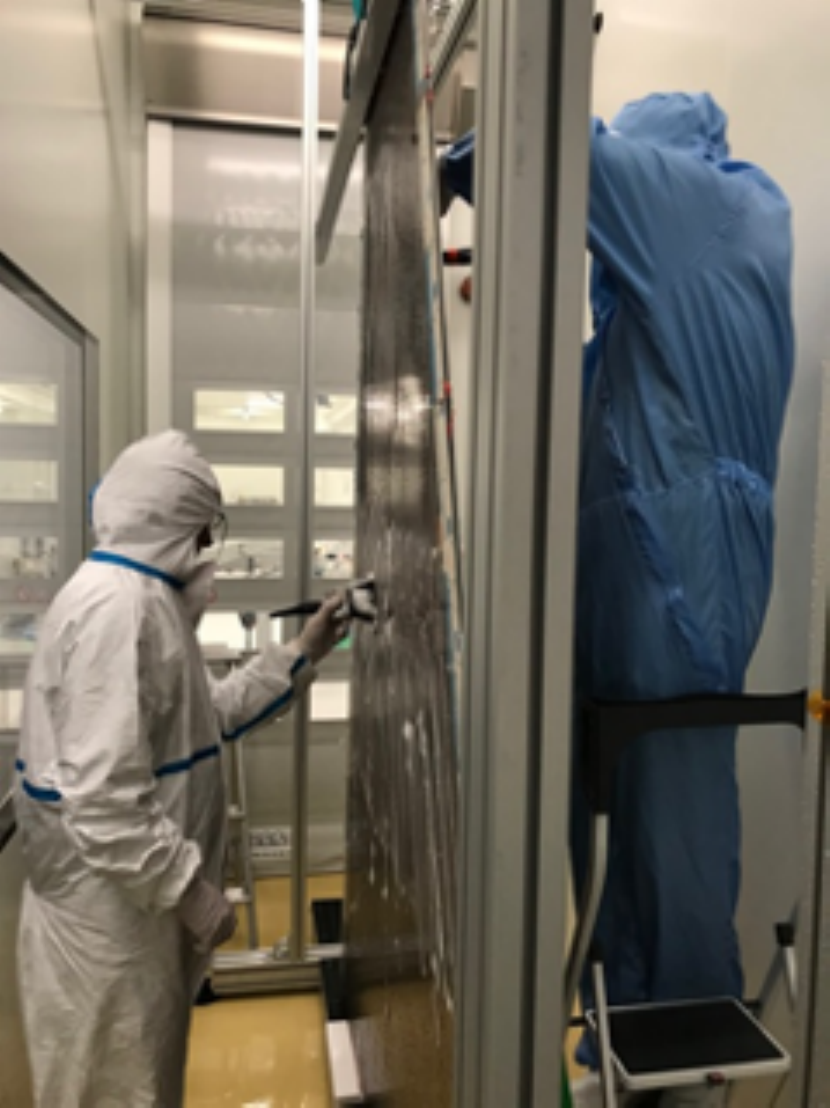}
	\includegraphics[height=5cm]{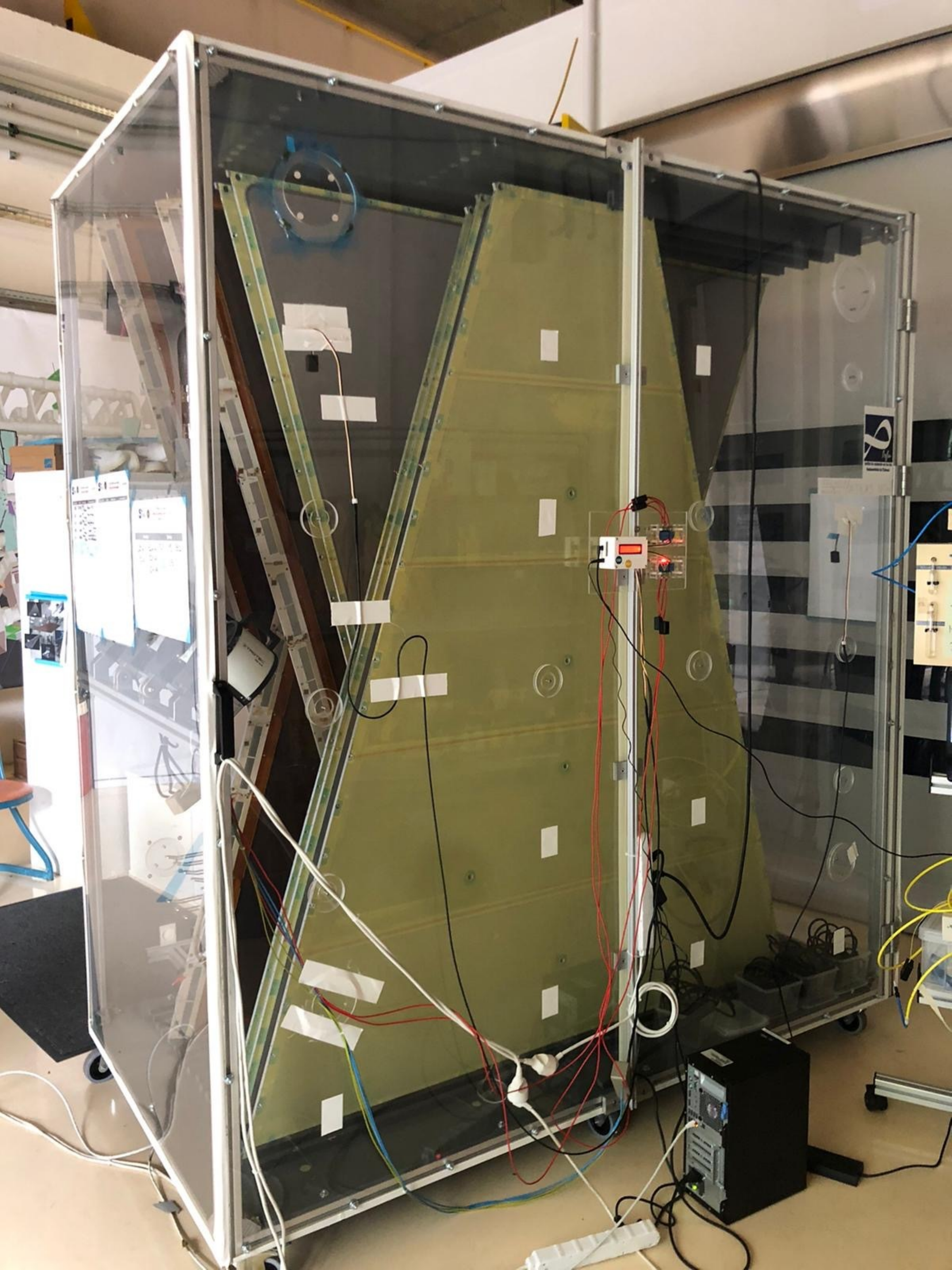}
	\caption{Washing cabin and drying oven.} 
	\label{fig:panel-shower}
\end{figure}

\subsubsection{Module assembly}
\label{section:assembly}

\begin{figure}[!h]
    \centering
	\includegraphics[width=5cm]{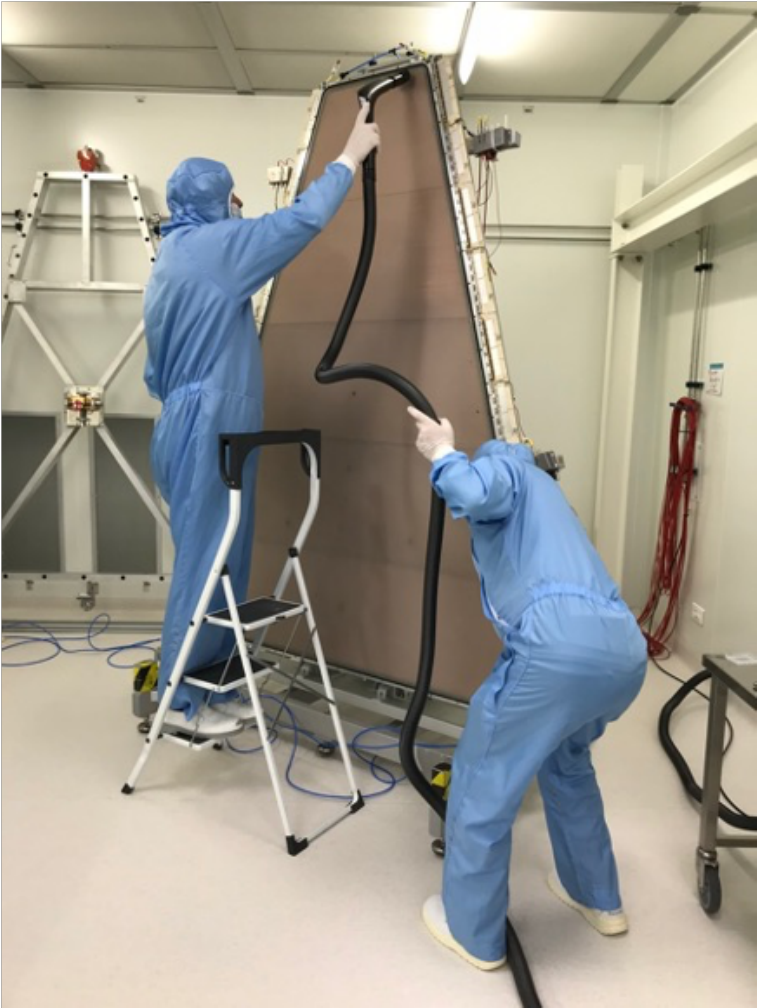}
	\caption{Vacuum cleaning of the drift panel before closing a gap.} 
	\label{fig:vacuum}
\end{figure}

Module assembly is the last step of construction in the clean room. The strategy chosen in Saclay is to assemble gap by gap and to test the HV stability at each step.
Before assembling one gap, the electrodes of the corresponding panels are dry-cleaned, using a vacuum cleaner and an antistatic roller to remove any remaining dust (Fig.~\ref{fig:vacuum}).\\

\begin{figure}[!h]
    \centering
	\includegraphics[width=6cm]{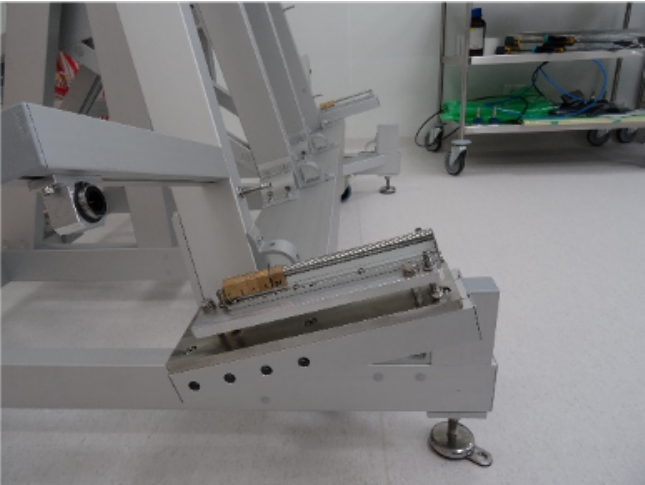}
	\includegraphics[width=6cm]{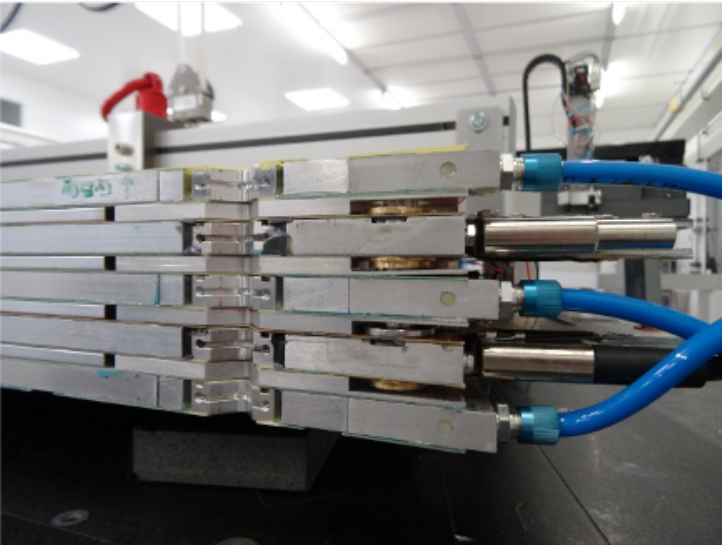}
	\includegraphics[width=6cm]{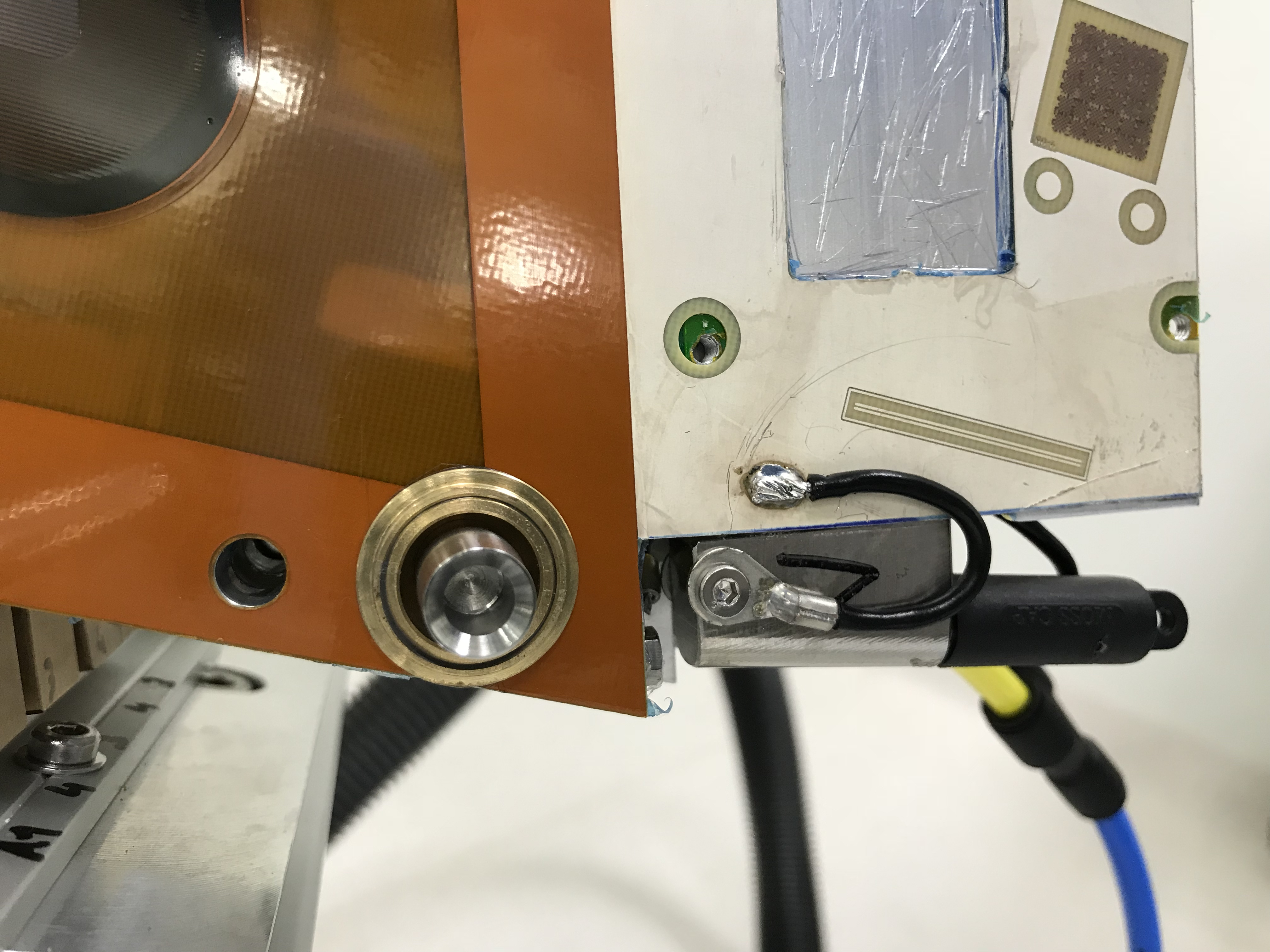}
	\caption{Left: Reference tracks on the assembly station on which lie the V-shape reference inserts of each panels during stacking. Right: Detail of the reference inserts in an assembled module. Bottom: View of the mechanical coupling between Stereo and Eta readout panels through precision pin and insert.} 
	\label{fig:assembly}
\end{figure}

Each RO PCB is divided in two HV sections (see Fig. \ref{fig:lm1_layout}) and tested individually. A section is considered as good if it can reach 850 V with less than 50\,nA of dark current, in the ambient air of the clean room (\SI{25}{\degreeCelsius} and 45\% of relative humidity). An automatic ramp-up program that is described in detail in Section~\ref{section:characterisation} is used for the conditioning of each gap (usually from 800 V to 850 V). If the dark current exceeds 50\,nA, the gap is dismounted and visually inspected, then dry-cleaned again before a new HV scan.
The mechanical alignment of RO panels, thus of strips, is achieved by sliding each panel on reference tracks using the reference insert glued during panel construction (see Section~\ref{section:readout} and Fig.~\ref{fig:pcb_detail}). Fig.~\ref{fig:assembly} shows details of the reference tracks on the assembly station and alignment inserts on the long base of the module. 

%

\section{Quadruplet Quality  Control}
%
\label{section:QC}
%
At each step of panel construction and module assembly, different tests and measurements are done to ensure the quality of constructed objects. 
The tests can be grouped in the following categories:
\begin{itemize}
\item insulation tests: between the cathode and the ground, the readout sectors and the ground;
\item planarity and thickness measurements of panels and quadruplets;
\item strip-position assessment by measuring the relative locations of readout PCB, using coded masks and a dedicated tool (see \ref{section:alignment});
\item Module gas-tightness.


\end{itemize}
In the next sections we will only describe planarity, alignment and module gas tightness measurements. Functional tests using a cosmic bench will be detailed in Section~\ref{section:validation}.


\subsection{Planarity}\label{section:planarity}

Planarity of Drift and RO panels is assessed using the gantry device described in Section~\ref{section:tooling}.
Its allows precise (5\,\microns resolution) and dense (more than 10 points per cm in both horizontal directions) height measurements over the panel surfaces.

Ideally the two panel sides are to be scanned both under suction and in relaxed mode.

However, very early in the production, the measurements with suction were possible only for one side, namely just after the panel gluing step described in Section~\ref{section:construction} with the panel still sucked on the sole.  
Indeed the other configuration with the flipped panel directly on the granite sole is prevented by the positioning pins embedded in the sole which do not fit anymore the positioning holes of the flipped panel which are asymmetric.

For measurement without suction, the panel is laid  on several precisely manufactured 25\,mm thick gauge blocks, placed onto the sole. The panel can be flipped over and both sides can be measured.



Calibration of thickness measurements relies on two precision gauge blocks located next to the panel.
Since the surfaces of the top panel and of these calibration gauge blocks are scanned together, the thickness of the panel can be measured in absolute terms.

 \subsubsection{Panel planarity}\label{section:panelplanarity}
 
\begin{figure}[hbt!]
  \centering
  \begin{subfigure}[b]{0.47\textwidth}
    \includegraphics[width=\textwidth]{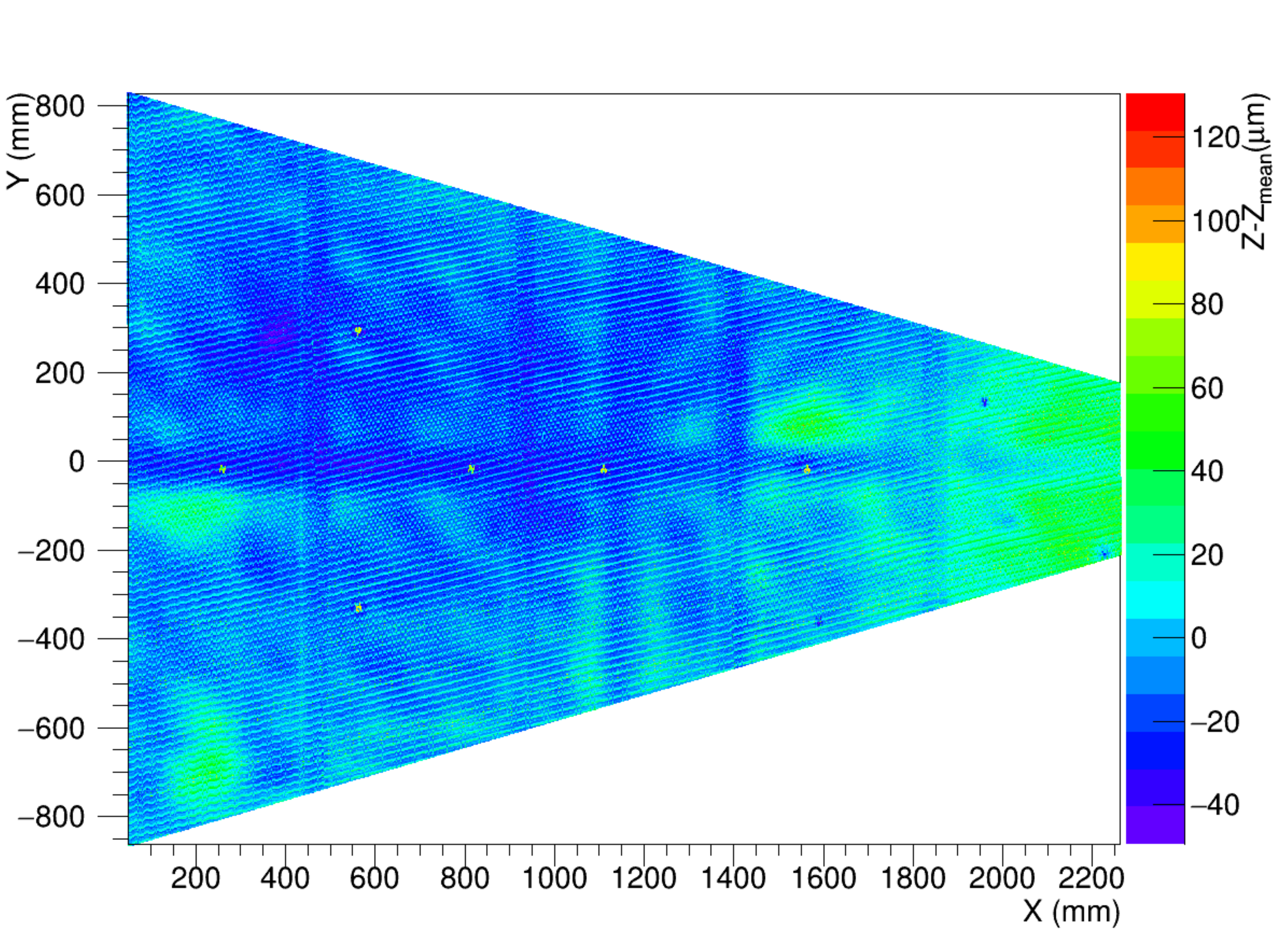}
    \caption{Map of all individual height measurements on a single side.}
    \label{fig:HMapIndiv}
  \end{subfigure}
  ~
  \begin{subfigure}[b]{0.47\textwidth}
    \includegraphics[width=\textwidth]{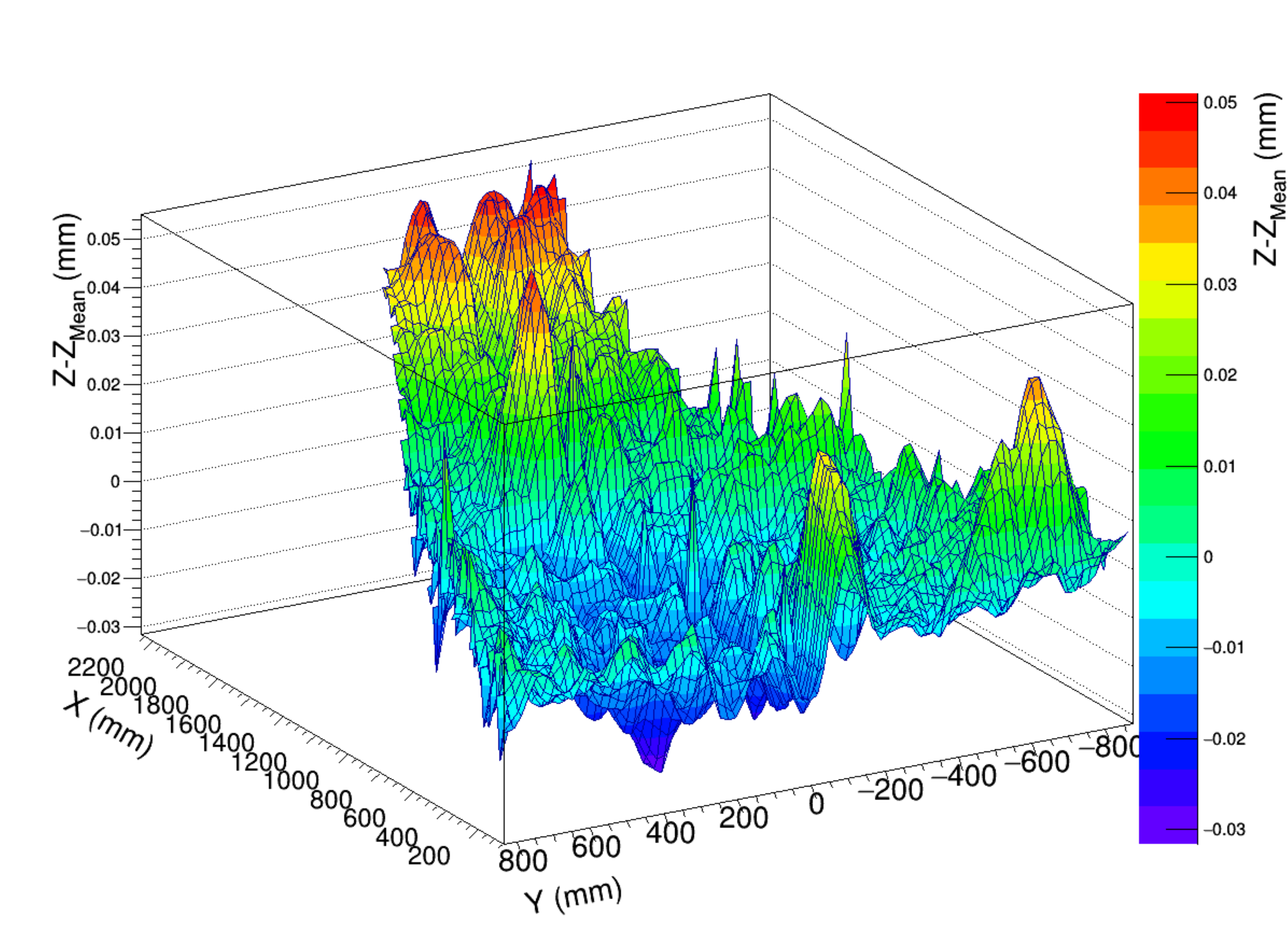}
    \caption{Averaged measurements over squares of 2\,cm by 2\,cm.}
    \label{fig:HMapAvg}
  \end{subfigure}
  \caption{
   Examples of panel height measurements.
  }
  \label{fig:Hmap}
\end{figure}

 Examples of panel-thickness measurements are shown in Fig.~\ref{fig:HMapIndiv} and \ref{fig:HMapAvg}. 
 The two figures correspond to the same scanned surface.
 Fig.~\ref{fig:HMapIndiv} shows the individual thickness measurements and therefore corresponds to the full resolution of the gantry (the bright spots at the interconnections locations are due to thin plugging stickers there).
 On  Fig.~\ref{fig:HMapAvg}, the measurements (few hundreds) have been averaged over 2\,cm by 2\,cm squares.
 These maps allow to spot and understand long range deformations or local defects which could impact the performance of the detectors.

\begin{figure}[hbt!]
    \centering
	\includegraphics[width=0.95\textwidth]{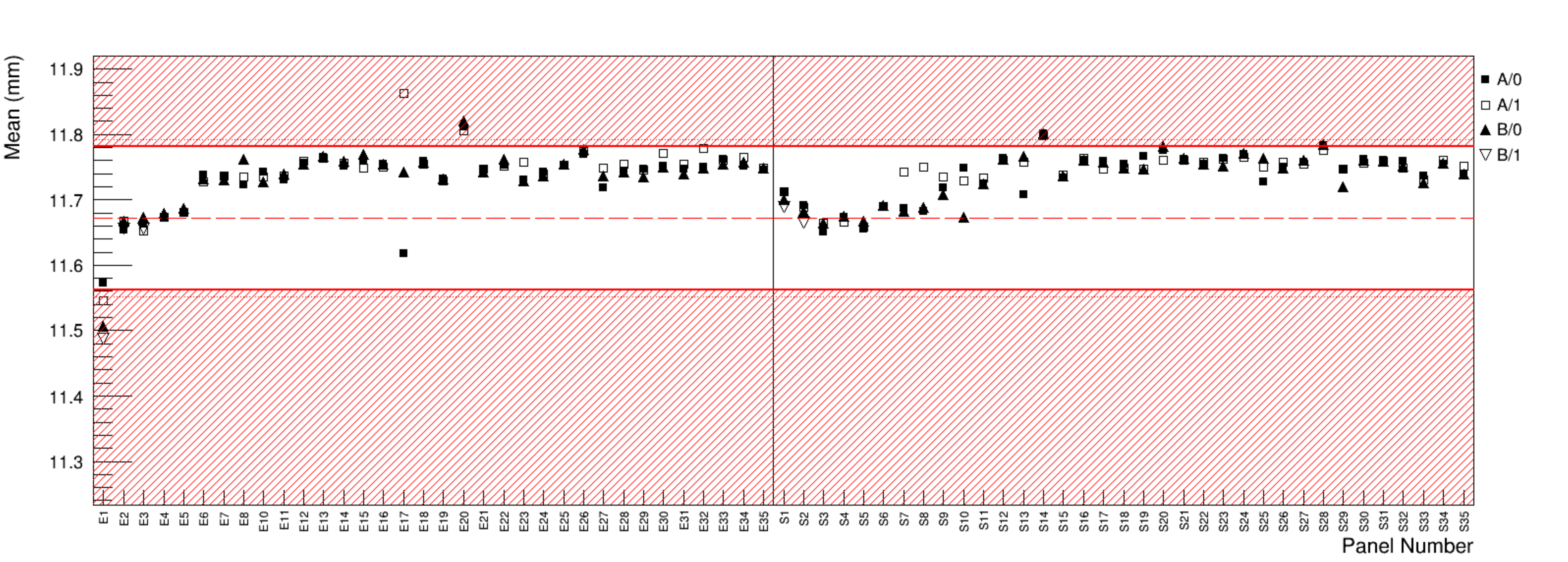}
	\caption{Mean thickness of panels as a function of the panel rank and for the two RO panels categories. Values are given for each of the four possible configurations (Side 1 or 2 [A/B], vacuum on or off [0/1]) of the measurements.} 
	\label{fig:PanelThicknessMean}
\end{figure}

\begin{figure}[hbt!]
    \centering
	\includegraphics[width=0.95\textwidth]{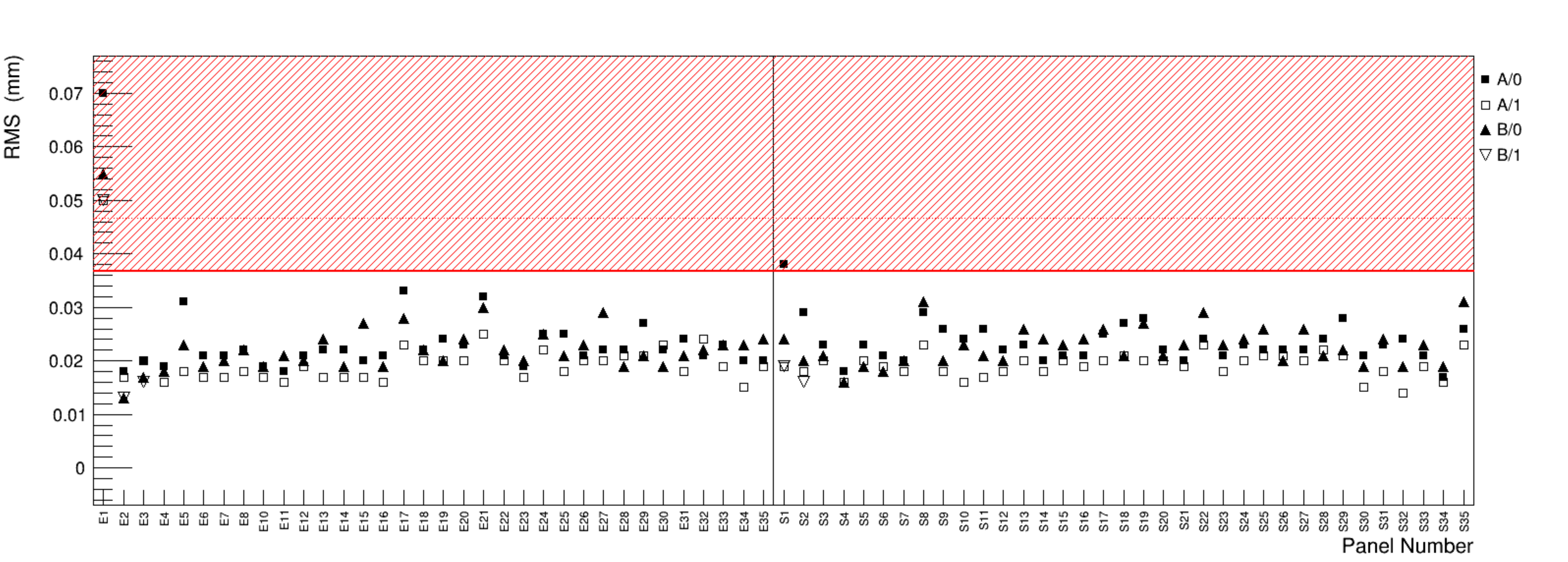}
	\caption{Panel thickness RMS as a function of the panel rank  (see caption of Fig.~\ref{fig:PanelThicknessMean} for the symbols explanations)} 
	\label{fig:PanelThicknessRMS}
\end{figure}

 \vspace{5mm}
 Objective quality criteria are based on the mean, the statistical dispersion and the maximal and minimal values of thickness distribution. 
 Fig.~\ref{fig:PanelThicknessMean} and~\ref{fig:PanelThicknessRMS} show for instance the average thickness and the RMS for the two categories (Eta and Stereo) of RO panels and the four possible measurements configurations (either panel side sucked or not onto the sole).  
 The expected thickness, defined as the distance from a resistive foil of one side to the top of pillars of the other side, is also shown on Fig.~\ref{fig:PanelThicknessMean}: \SI{11.672}{mm} .
Acceptance limits are indicated on Fig.~\ref{fig:PanelThicknessMean} and~\ref{fig:PanelThicknessRMS}: the average thickness should not differ from the nominal by more than 110\,\microns and the RMS should not exceed  37\,\microns (110/3 where 110\,\microns is considered as a 3\,$\sigma$ limit).

 As it can be seen on these plots, after a first period during which  processes are tuned and experience gained, in general constructed panels match the mechanical goal.
 
 
 Planarity data are very important to understand performance issues subsequently found in functional tests (cf Section~\ref{section:characterisation}) or in reconstruction performance.
 For instance it was found that local planarity defects such as $\sim$100\,\microns  deep depressions match low efficiency spots (see Section~\ref{section:characterisation}).

 The structural behaviour of panels has been modelled with a finite element model and shows good agreement with the measurements\cite{ROSSI201953}.




 \subsubsection{Module planarity}\label{section:modulesplanarity}
\begin{figure}[hbt!]
    \centering
	\includegraphics[width=0.6\textwidth]{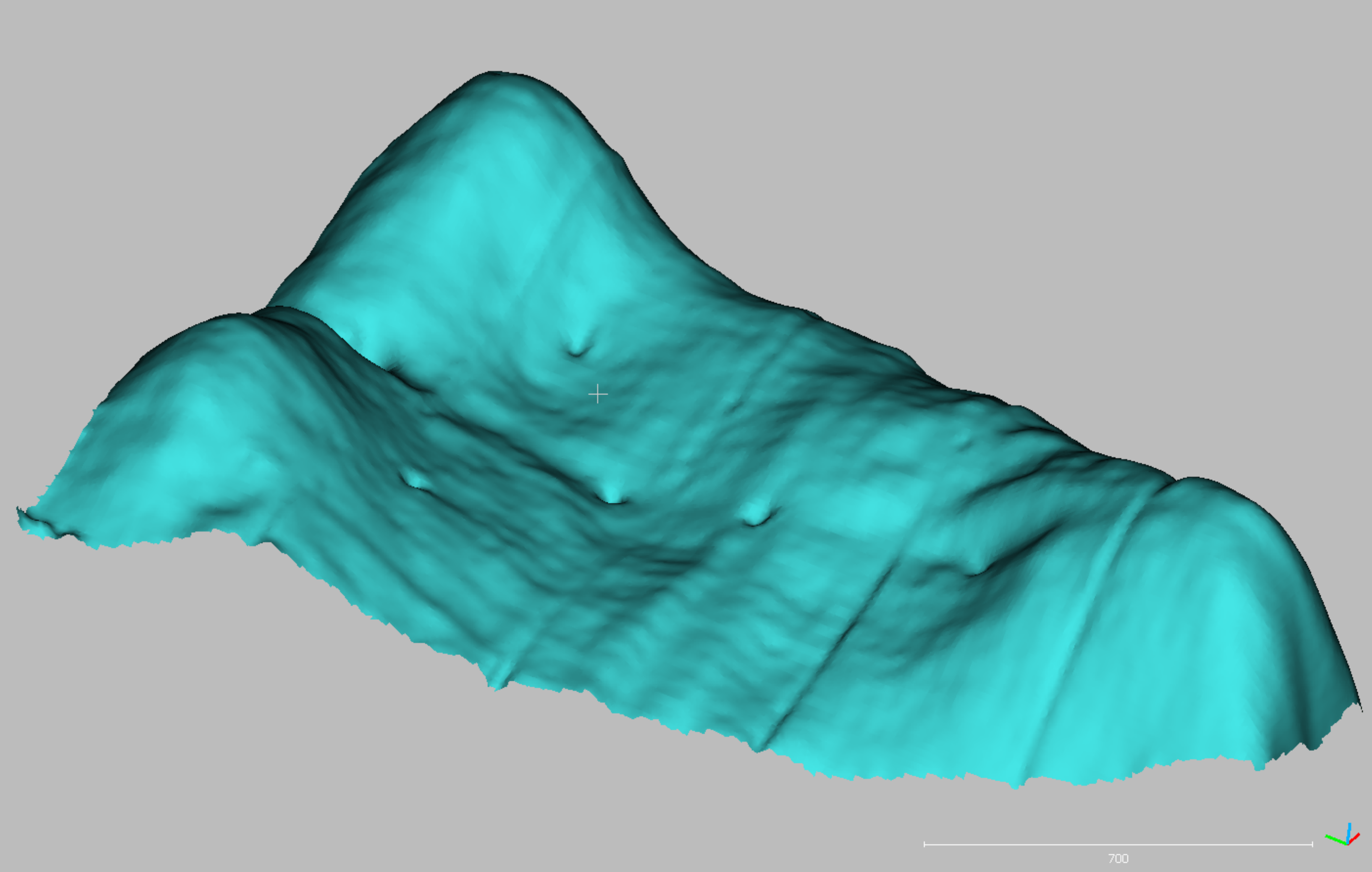}
	\caption{Typical reconstructed top surface of a module. Isometric projection. The scale bar applies to all 3 coordinates, but the Z coordinate is magnified by x1000 (X and Y in mm, Z in \microns). The observed peak-to-peak $\Delta$Z variation was 600\,\microns for this module (Module 21). } 
	\label{fig:Module3D}
\end{figure}

 The planarity of modules is assessed with the same means as for panel planarity. For this measurement, modules are placed on the granite table over 25\,mm thick pads and are flanked by calibration shims. Their external side (back panel external surface) is then scanned with the contactless optical sensor.
 Two areas were defined on this surface: a peripheral narrow band all around the module edges (“Edge” area) and the inner part (“Planarity” area). 
 Fig.~\ref{fig:Module3D} shows the typical shape of the resulting surface after reconstruction of the inner area. The junctions of the five PCBs can be seen as well as the six interconnections. The surface is shaped by the interconnections, creating a very flat surface in the centre of the module. 
 
\begin{figure}[hbt!]
    \centering
	\includegraphics[width=\textwidth]{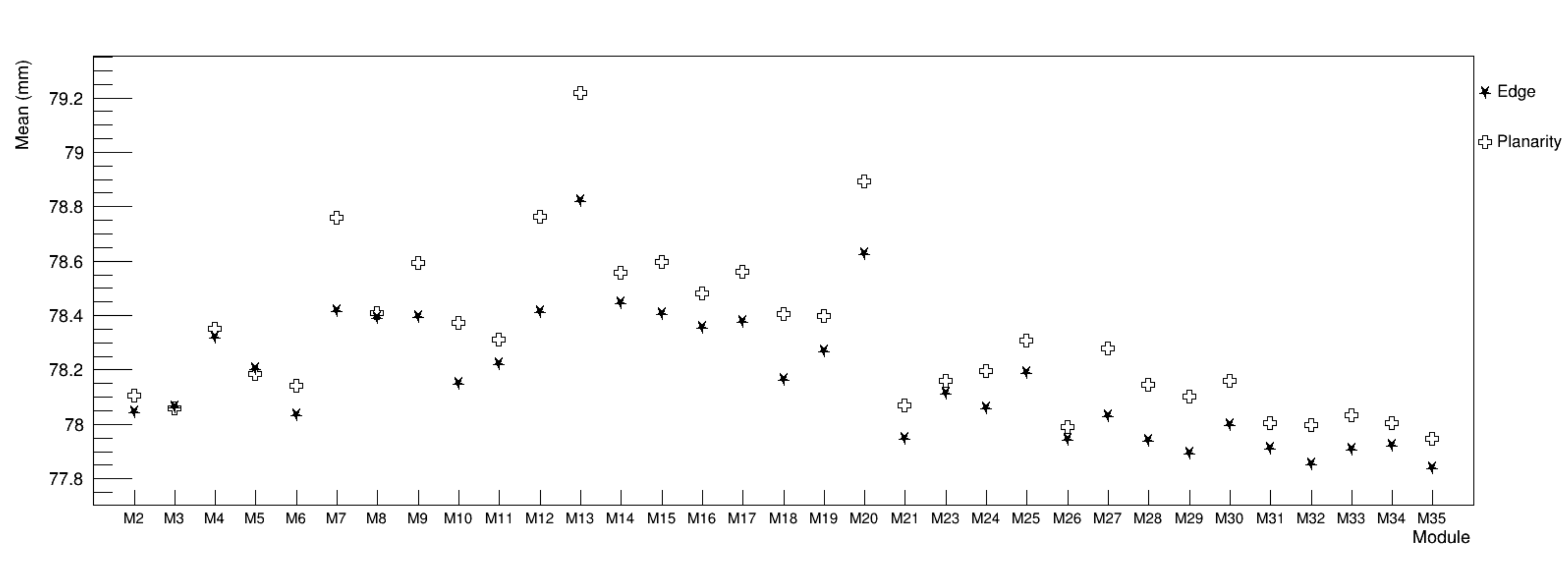}
	\caption{Mean thickness of Modules over the “Edge” and  “Planarity” areas} 
	\label{fig:ModulesMean}
\end{figure}

\begin{figure}[hbt!]
    \centering
	\includegraphics[width=\textwidth]{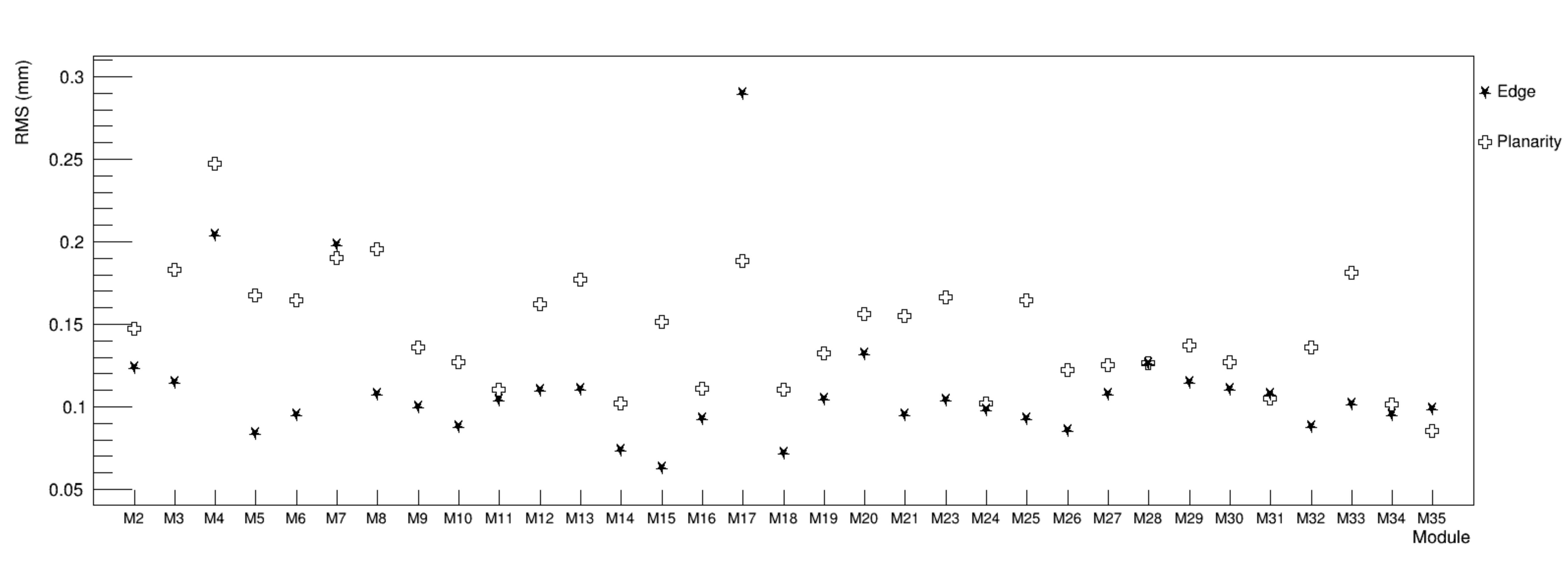}
	\caption{Module thickness RMS over the “Edge” and  “Planarity” areas} 
	\label{fig:ModulesRMS}
\end{figure}

 Fig.~\ref{fig:ModulesMean} and~\ref{fig:ModulesRMS} show the average and the RMS of the distributions of the measured  thicknesses for the “Edge” and  “Planarity” areas for all the modules.
 As it can be seen from these figures, except for a few accidents, modules produced were very consistent in thickness.

\clearpage
\subsection{Alignment}
\label{section:alignment}

\subsubsection{Rasnik masks: alignment markers on PCBs}

\begin{figure}[pth]
  \centering
  \includegraphics[width=0.3\textwidth]{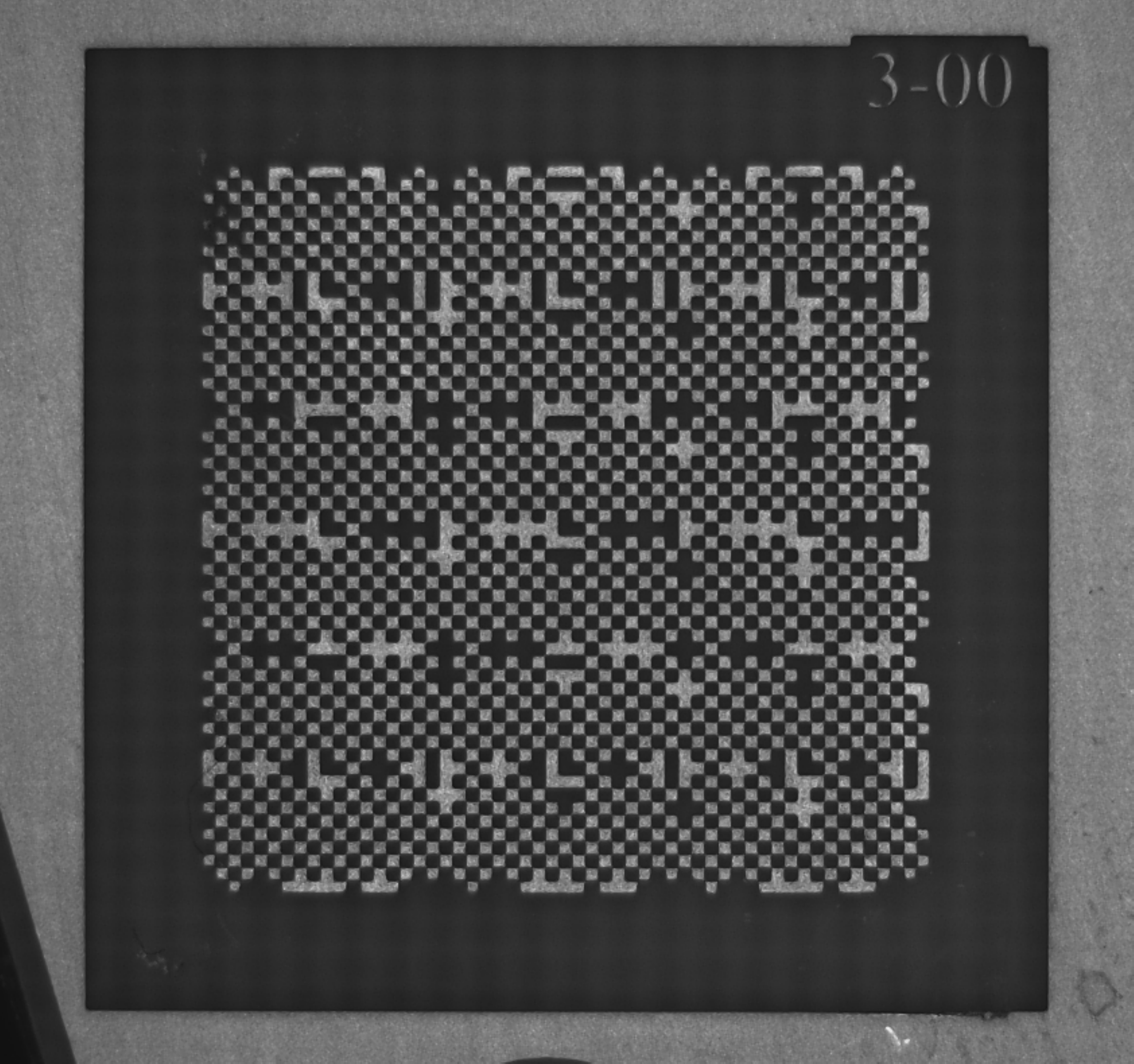}
  \caption{A Rasnik mask etched on PCB.}\label{figure:ali:rasmaskpicture}
\end{figure}

For the purpose of detector construction and measurement, precision markers are
etched on the PCBs, consisting of Rasnik masks \cite{Rasnik_Paper_Harry}. 
The rasnik masks are precision-etched chessboard patterns with squares of pitch \SI{220}{\micro\meter}, where squares of inverted colour encode absolute position references. They are etched in the same process as that of the read-out copper strips, ensuring the precision.  An example Rasnik mask is
shown in Fig.  \ref{figure:ali:rasmaskpicture}. Recording of the Rasnik masks with a calibrated camera enables determination of detector alignment parameters, using various set-ups, as will be seen in the next sections.

Tests of the precision of the Rasnik measurement on etched copper have been
performed, with controlled movements of a camera over a Rasnik mask: a high
precision of the Rasnik measurement is found ($<3~\si{\micro\meter}$). However,
over the larger size of the PCB, high accuracy residuals can be achieved only
if several deformation parameters are accounted for (elongation, sagging). PCB
deformations will be discussed in Section~\ref{section:ali:results}.

\begin{figure}[pth]
  \centering
  \includegraphics[width=0.8\textwidth]{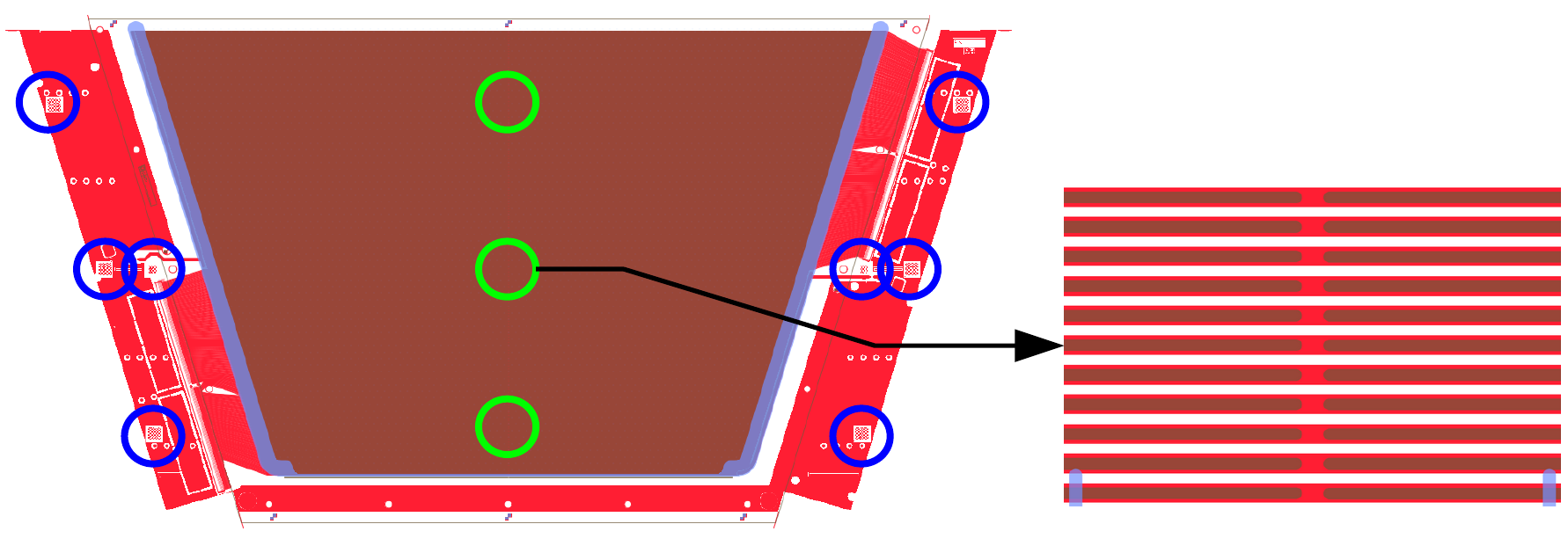}
  \caption{The layout of the Rasnik masks on a PCB (circled in blue), and
location of direct copper strip measurements in the gap of the resistive layer
(circled in green).} \label{figure:ali:masklayout}
\end{figure}

The layout of the Rasnik masks on a PCB is illustrated in Fig.
\ref{figure:ali:masklayout}. In total 8 Rasnik masks are etched on each PCB.
Each Rasnik mask has a code offset that is unique to a PCB type and to its
location in the layout, helpful for debugging purposes. In addition, through the
central gap of the resistive layer, the copper strips can be seen directly.
Three strip locations are defined along this central gap, as can be seen in Fig.
\ref{figure:ali:masklayout}.

\subsubsection{Precision washer positioning}

For the panel assembly procedure, a mechanical reference needs to be available
on the PCBs. For that purpose, precision machined washers are positioned and
glued with high accuracy on each PCB, using an opto-mechanical tool and the
Rasnik mask pairs which are etched along the central line of the PCBs.

\begin{figure}[pth]
  \centering
 \begin{subfigure}[c]{0.49\textwidth}
  \includegraphics[width=\textwidth]{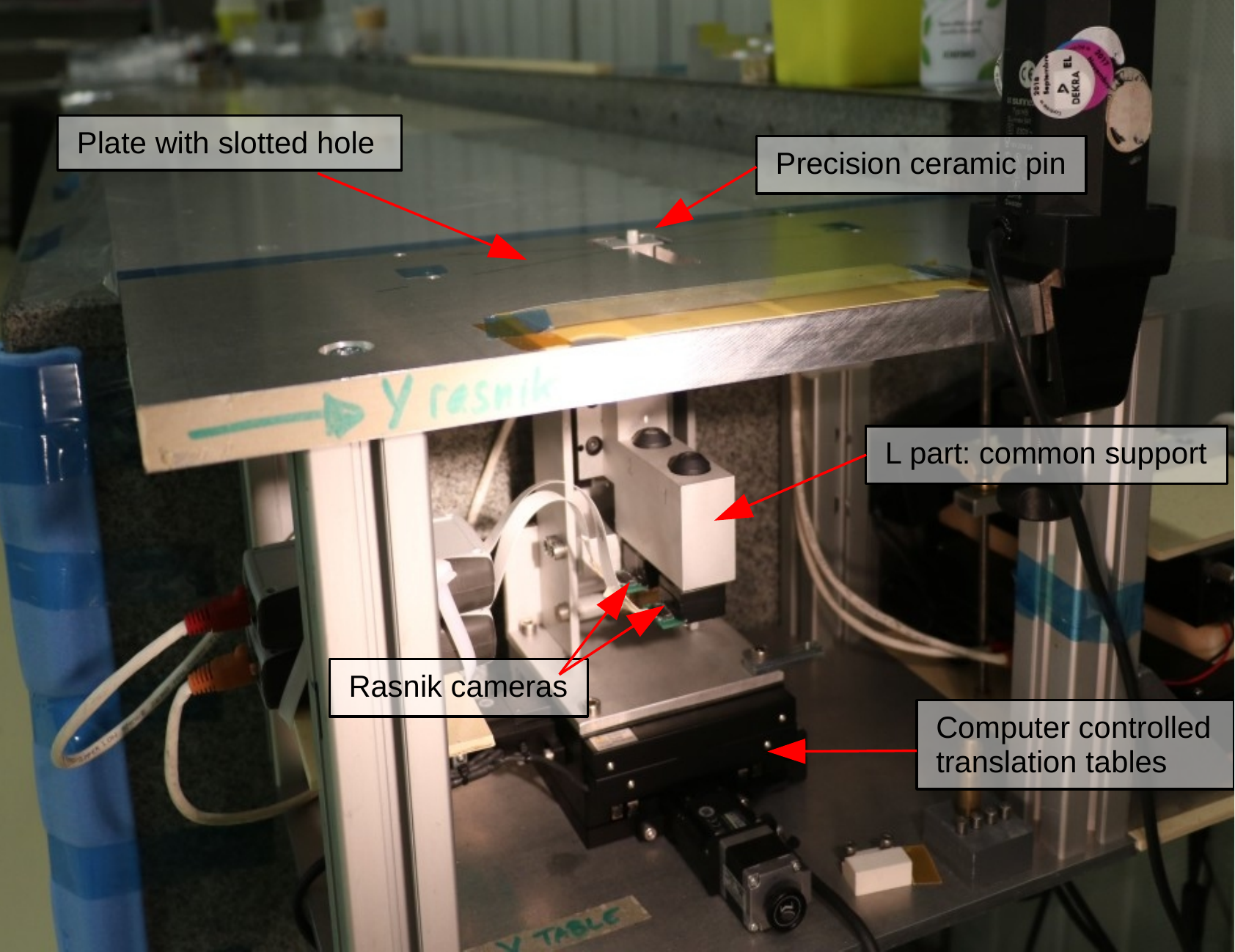}
   \caption{}
 \end{subfigure}
 \begin{subfigure}[c]{0.49\textwidth}
   \includegraphics[width=\textwidth]{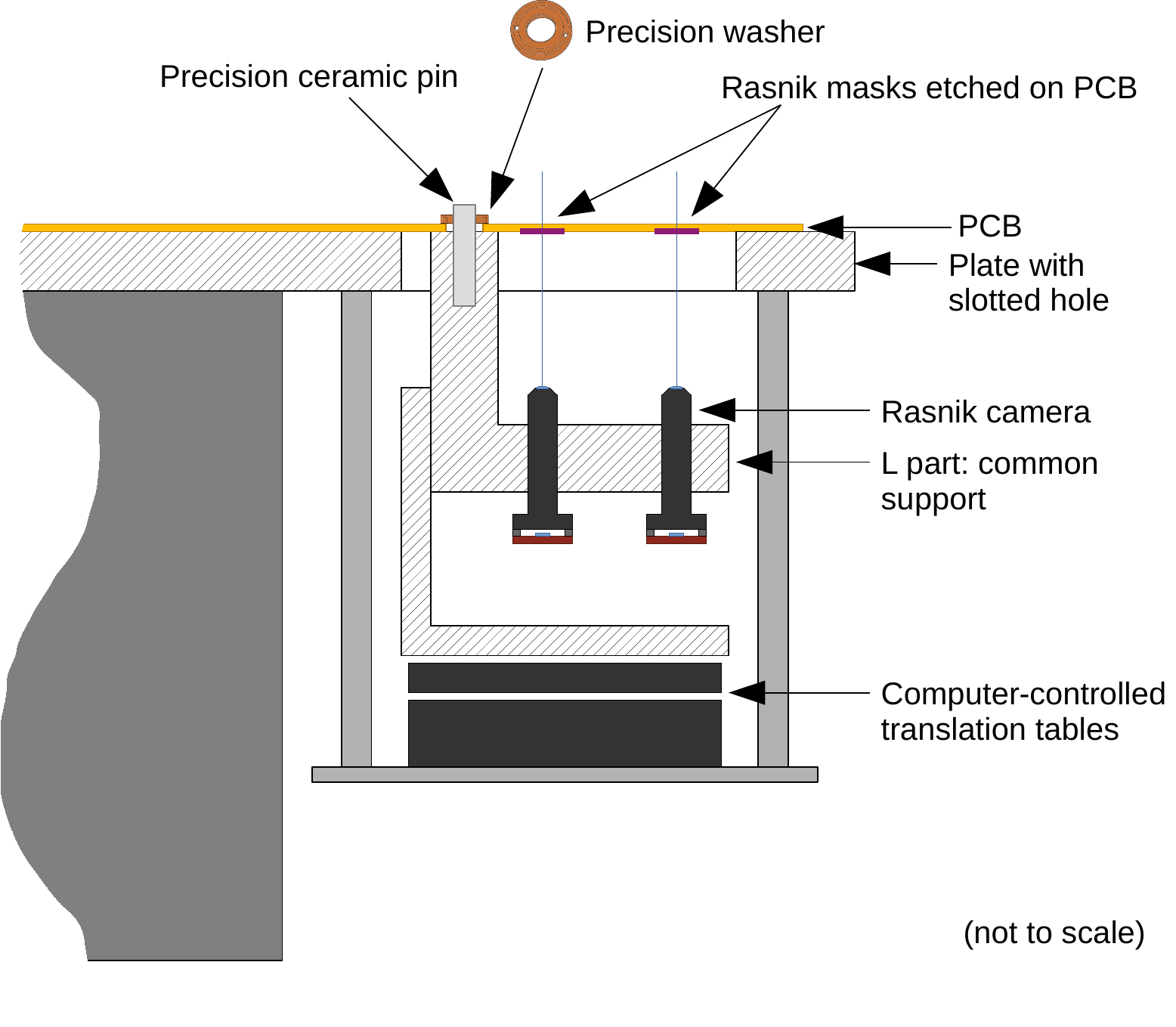}
   \caption{}
 \end{subfigure}
 \caption{The precision washer gluing tool (a) and principle of operation (b).}

\label{figure:ali:washertool}
\end{figure}

The washer gluing tool is presented in Fig. \ref{figure:ali:washertool}: it
consists of an L-shaped block holding together two Rasnik cameras and a ceramic
pin whose diameter is manufactured with 1\,\si{\micro\meter} precision. This block is
supported on computer-controlled translation tables. The ensemble is held under
a table with a cut-out such that the ceramic pin protrudes above. The washer
positioning procedure is the following: a PCB is placed on the table, a
washer is inserted on the pin, the masks are read-out by the two cameras,
and position corrections are computed and applied on the translation tables. At
that point, the washer is positioned with high accuracy with respect to the
Rasnik masks, and is checked with additional Rasnik readings. Glue is then
applied to fix the washer on the PCB.

To reach the required accuracy of the washer positioning, a calibration
procedure is carried out, which computes the positions and angles of the lines
of sight of the two rasnik cameras, with respect to the ceramic pin. The achieved accuracy is of a few \si{\micro\meter}. The full procedure is described in~\cite{Giraud}.


\subsubsection{Gantry CMM}

The measurement gantry was previously discussed in Section \ref{section:tooling}: it is
equipped with a camera sensor to which controlled displacements may be applied
with very good reproducibility.  As the stiffness and planarity of the granite
and SiC surfaces used for displacement are of high quality, this device is used
as a 2-dimensional optical CMM. Assembled read-out panels are placed in the
gantry, and survey is performed of each constituting PCB: the etched Rasnik
masks and the copper strips visible along the central gap of the resistive layer
are measured.

A calibration device is constructed to determine and correct displacement biases of the gantry, described in detail in~\cite{Giraud}.
%
%
%
%
The overall accuracy of the gantry measurement is modeled by the quadratic sum of the local (non-reproducible or non-linear) defects and an overall uncertainty on the scale. It is estimated to be
$5\,\si{\micro\meter} \oplus 12\,\si{ppm}$ ($2\sigma$ interval).

\begin{figure}[pth]
  \centering
  \includegraphics[width=0.6\textwidth]{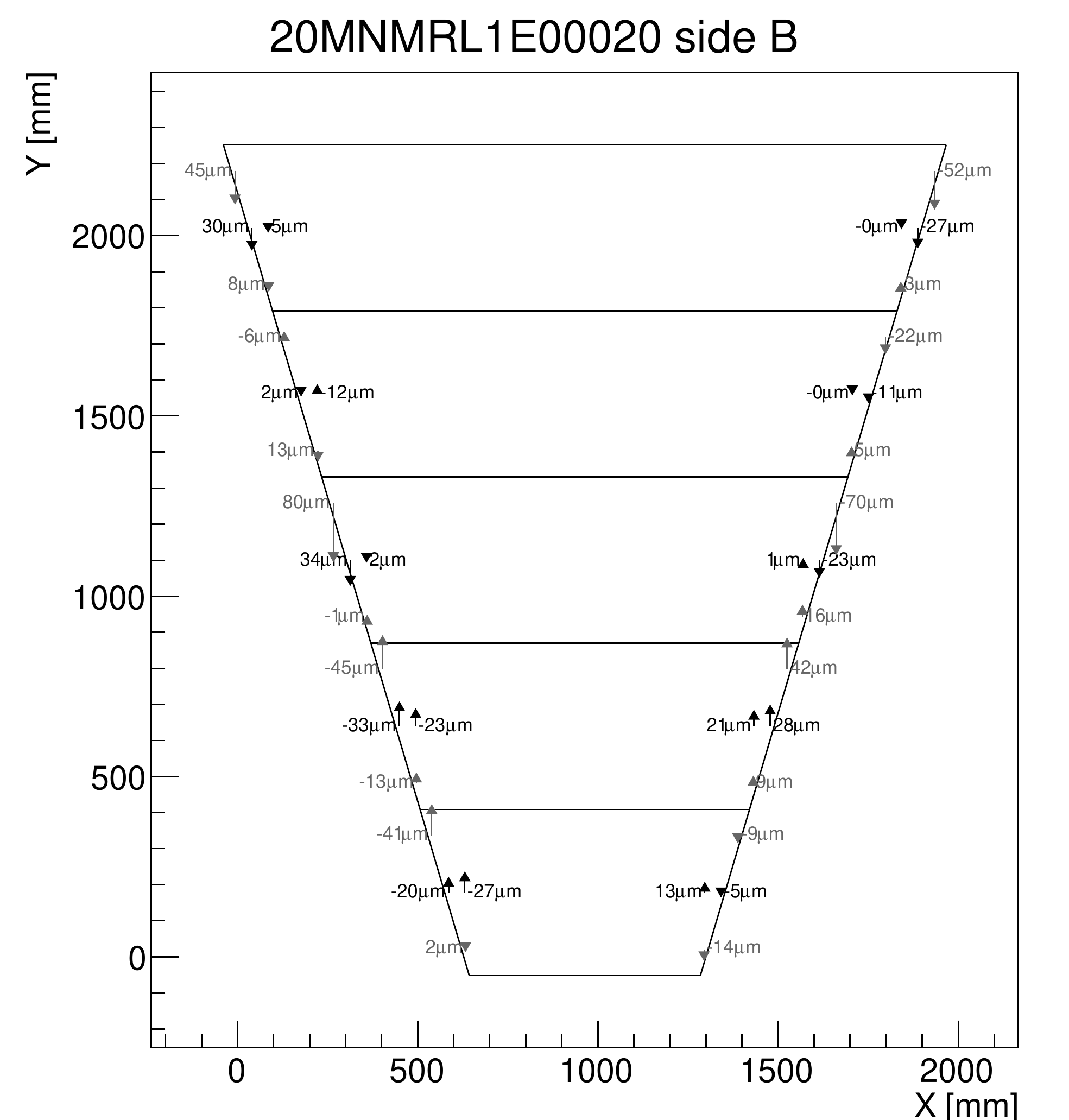}
  \caption{Example RO panel measurement with the gantry
  CMM.}\label{figure:ali:gantryexamplepanel}
\end{figure}

An example of a panel measurement is presented in
Fig.~\ref{figure:ali:gantryexamplepanel}: the position bias of the Rasnik masks
are given, in the precision coordinate of the detector. In this example, the
Rasnik masks along the central line of each PCB are measured within
\SI{35}{\micro\meter} to their nominal position, while the corner Rasnik masks
of each PCB may be displaced more, as an effect of PCB elongation bias.

\begin{figure}[pth]
  \centering
  \includegraphics[width=0.8\textwidth]{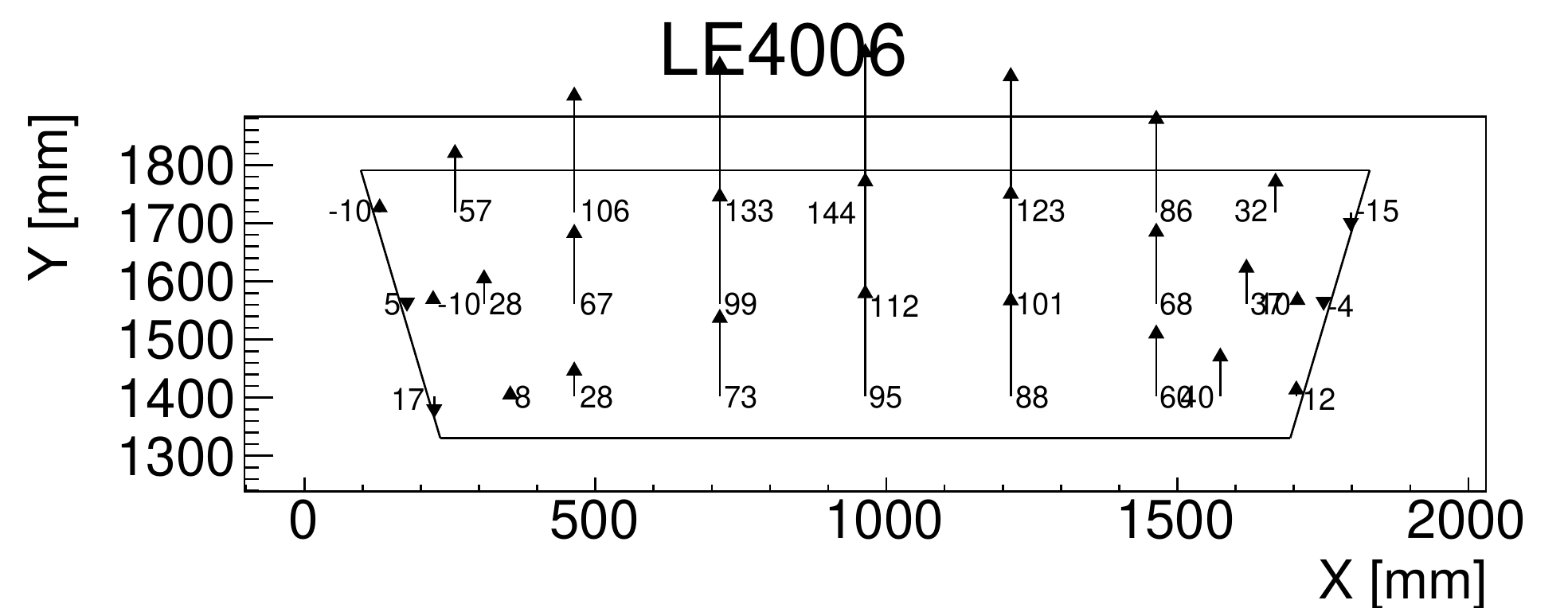}
  \caption{Gantry measurement of a (spare) RO PCB on which several cut-outs on the
  resistive layer have been performed, in order to measure the copper strips on multiple points across the PCB surface. The Right-most and left-most
measurements are Rasnik masks. The other measurements are copper
strips. The reported numbers are in \si{\micro\meter} and are deviations to the nominal along the precision (Y) direction.}\label{figure:ali:stripsagexample}
\end{figure}

An example of strip measurements is shown in
Fig.~\ref{figure:ali:stripsagexample}: on this particular board, several windows were cut out in the
resistive layer, in order to measure the copper strips in several points
of the PCB acceptance (only the central gap of the resistive layer is visible
otherwise). An overall sagging shape of the PCB is visible in this measurement,
of size $\sim 100\,\si{\micro\meter}$.

\subsubsection{Rasfork: position monitoring between the two layers of a panel}

\begin{figure}[pth]
 \centering
   \includegraphics[width=0.7\textwidth]{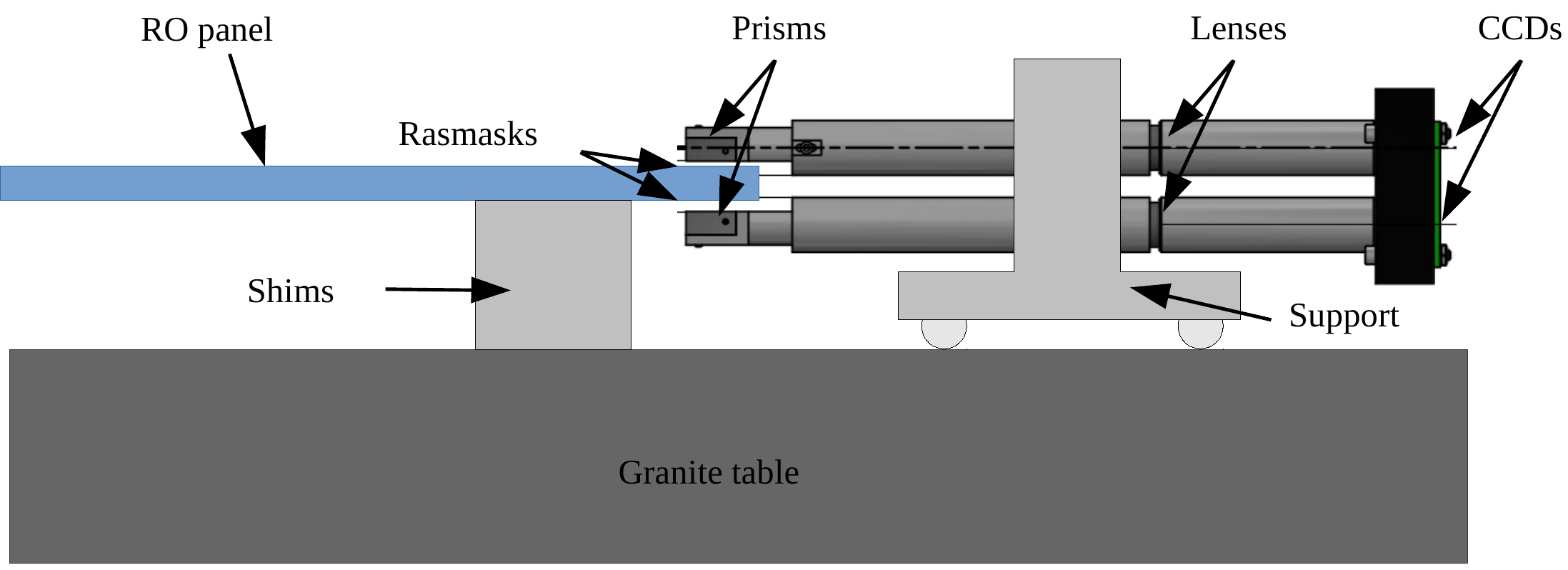}
 \caption{Rasfork principle.}\label{figure:ali:rasforkprinciple}
\end{figure}

Position monitoring between the two faces of a RO panel is performed using the
Rasnik masks etched on the PCBs with a dedicated tool called the Rasfork, see
Fig.~\ref{figure:ali:rasforkprinciple}.


A Rasfork is made of two tubes equipped with the necessary optics to record a
Rasnik mask, mechanically attached to a support block ensuring very good
perpendicularity reproducibility when sitting on granite. The optics in each tube
is composed of a prism, a diaphragm, a lens, and a CCD. The prism is working in
total internal reflection, deflecting the image of the Rasnik mask sideways.
The image is then focused by the lens on the CCD.

The RO panel is placed on shims on a granite table, and the Rasfork is inserted
on the side, where images of a pair of Rasnik masks are recorded. By
comparison of the two images recorded by the two tubes, the position bias of
the bottom Rasnik mask is reconstructed with respect to the top
Rasnik mask.

\begin{figure}[pth]
  \centering
  \includegraphics[width=0.6\textwidth]{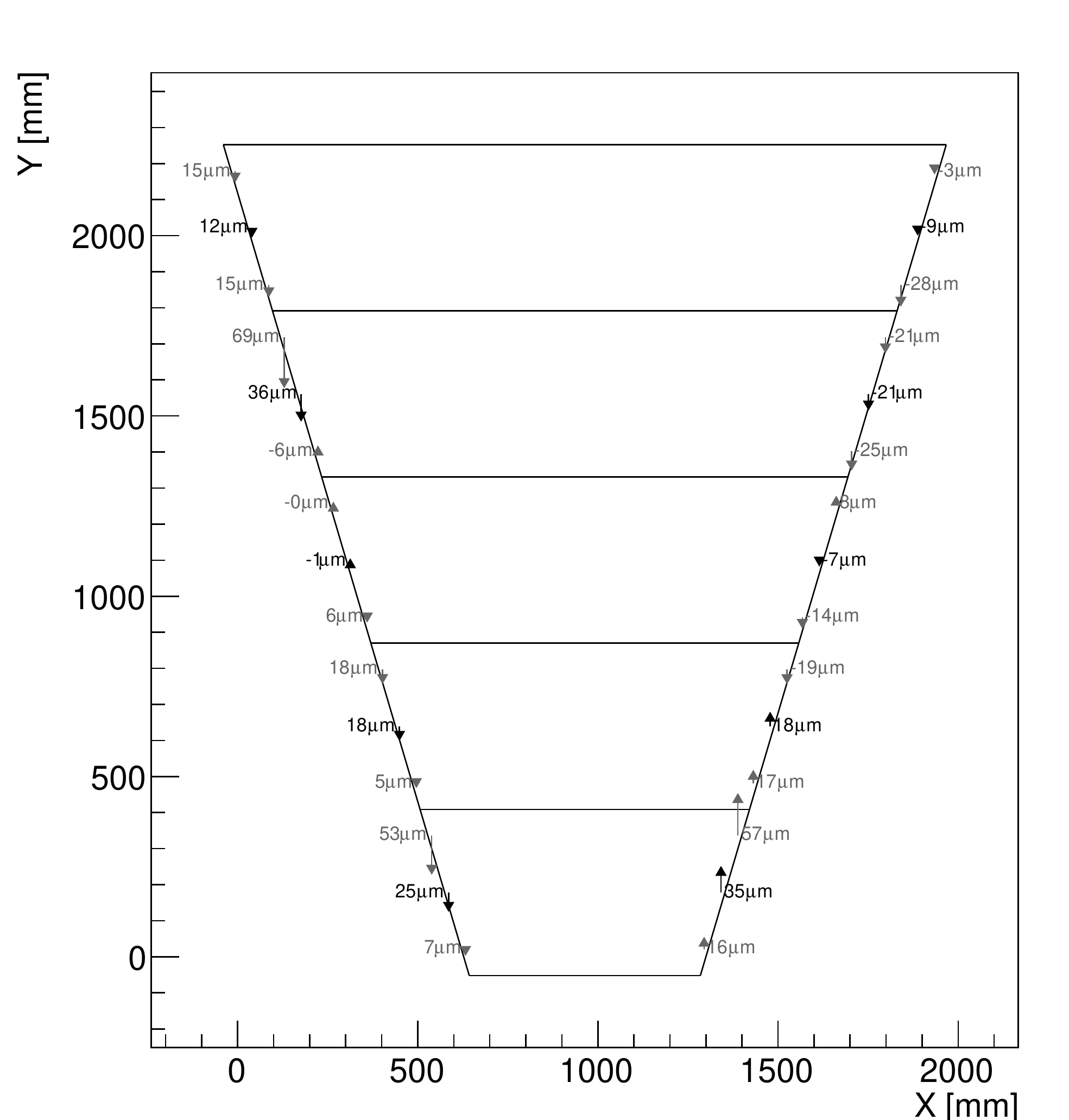}
  \caption{Example of a RO panel measured with the Rasfork. The reported measurements are the position differences in the precision coordinate Y between the Rasnik masks on the two faces of the panel. In black are the masks along the central line of the PCBs, in gray are the masks in the corner of the PCBs.}
  \label{figure:ali:rasforkpanelexample}
\end{figure}

An example Rasfork measurement of a read-out panel is presented in
Fig.~\ref{figure:ali:rasforkpanelexample}. For that particular panel, the largest measurement along the central line of the PCBs is of \SI{36}{\micro\meter}, informative of the quality of the panel construction. The systematic error of a Rasfork
measurement is estimated to be \SI{10}{\micro\meter} ($2\sigma$ interval),
dominated by illumination in-homogeneity of the recorded mask.

To monitor the positioning of the two panels of a quadruplet, a Rasfork device
with 4 tubes is also constructed, measuring the 4 layers of PCBs simultaneously.
The principle, layout, calibration procedure are similar to the 2-tube Rasfork.


\subsubsection{Results}
\label{section:ali:results}

A reconstruction of the 2D detector alignment parameters is performed in a global fit
combining all the available metrology measurements: gantry CMM, 2-tube Rasfork,
and 4-tube Rasfork.

\begin{figure}[pth]
  \centering
  \includegraphics[width=0.6\textwidth]{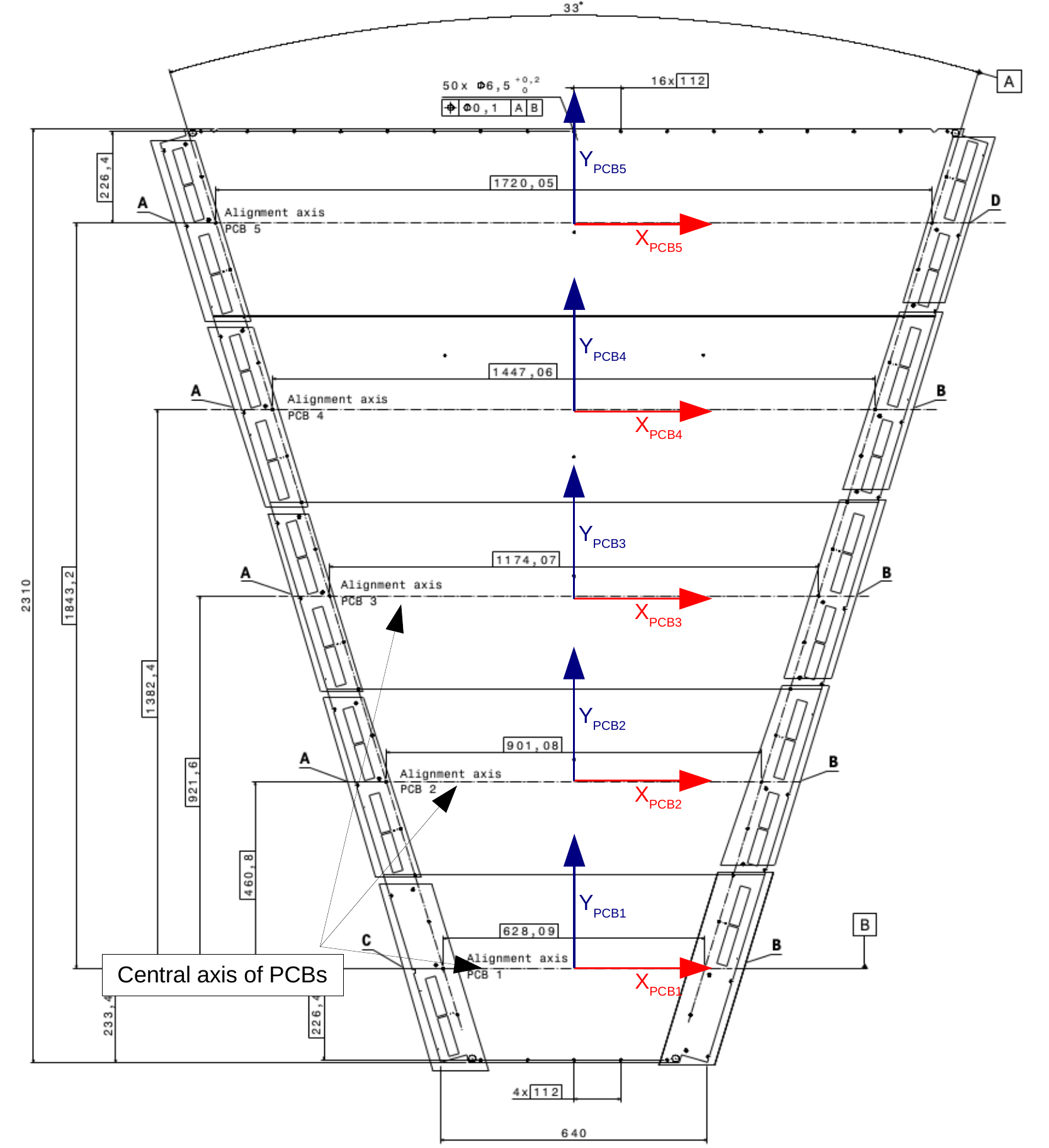}
  \caption{The PCB coordinate system used in the metrology reconstruction. The frames are aligned in the precision direction with the centres of the Rasnik masks along the central line of the PCBs, and with the nominal positions of the precision washers glued on each board.
    By definition, the coordinate systems of the layer coincides with that of the 5th PCB, and that of the panel coincides with that of the first layer. All coordinate systems are right-handed.}
  \label{figure:ali:reco_coordinatesystem}
\end{figure}

The reconstructed parameters are of several kinds: PCB deformation parameters
(elongation, sag), PCB position and rotation parameters in frame of layer, layer
position and rotation parameters in frame of panel, panel position and rotation
parameters in frame of module. The coordinate system used is shown in
Fig.~\ref{figure:ali:reco_coordinatesystem}. The RMS of the residuals between the Rasnik
measurements and the fitted model is of $7\,\si{\micro\meter}$
along the detector precision coordinate, illustrating the adequacy of the model,
and the compatibility of the different sources of measurements.

\begin{figure}[pth]
  \centering
  \includegraphics[width=\textwidth]{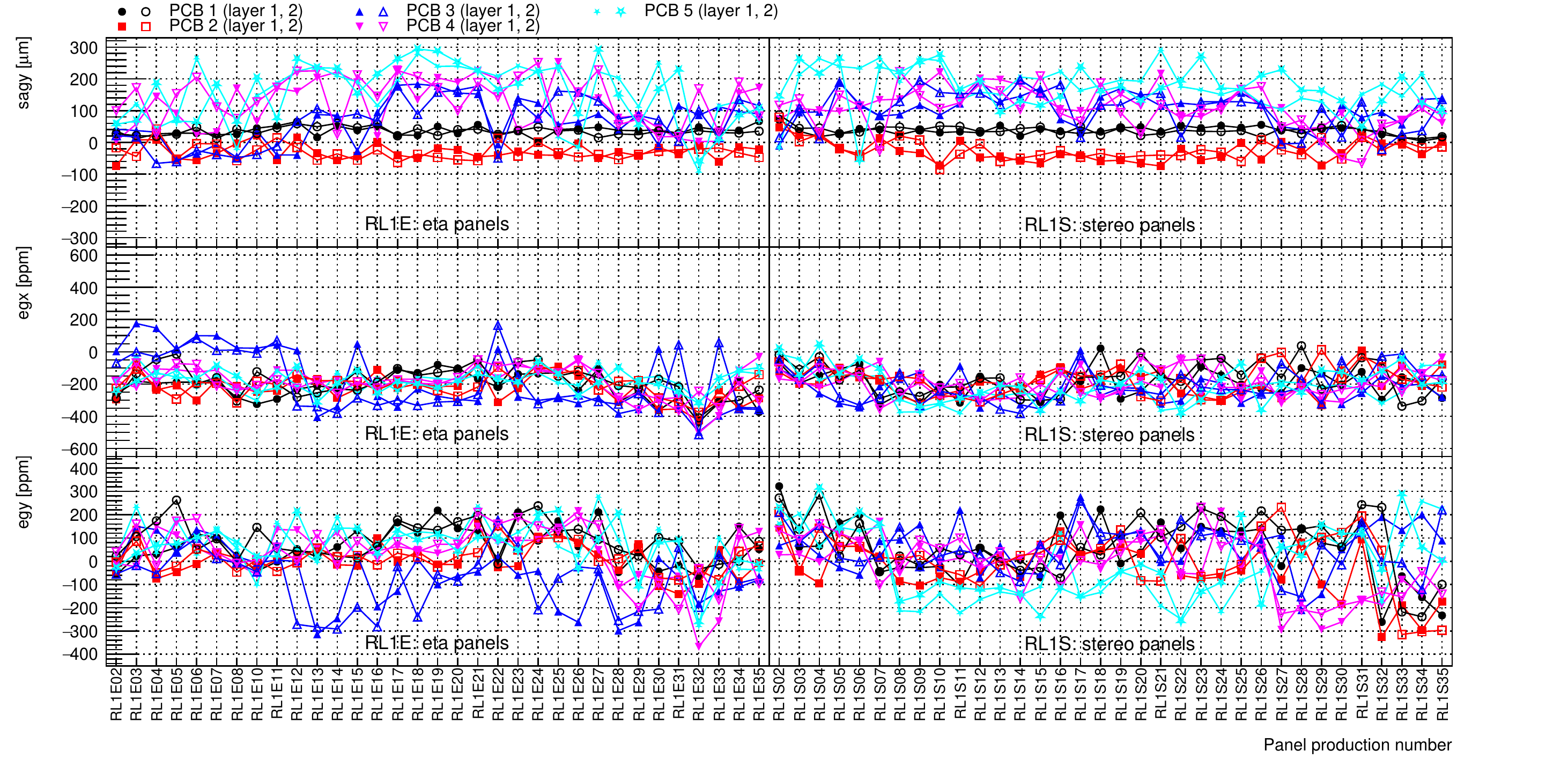}
  \caption{Metrology fit result: observed PCB deformation parameters, as a
  function of the panel production number. See text for the definition of these parameters. The different markers correspond to different PCB types and layers. The values depend typically on the PCB production batch, explaining the discretized behaviour for some of the boards. The parameter sagY is typically larger for the longer board types.}
  \label{figure:ali:metro_pcbdefo}
\end{figure}

The observed PCB deformation parameters are sizeable and are presented in
Fig.~\ref{figure:ali:metro_pcbdefo}. The plotted parameters are the PCB sag
along the precision coordinate (sagY), and the PCB elongation parameters along
the precision (egY) and second (egX) coordinates. PCB sag of up to
$300\,\si{\micro\meter}$ are observed, larger values found for the larger PCB
types. Elongation parameters lie typically within $400\,\si{ppm}$, and vary with
the PCB production batch. Additionally are also fitted parameters describing the half-difference of elongation between the two sides of the PCB, along the precision (degY) and second (degX) coordinates. The parameter degX models effectively the non-parallelism of the strips, and its RMS over all the boards is of 50 ppm. The RMS of the strip measurement residuals is of \SI{9}{\micro\meter}, hinting to the fact that possible higher-order non-parallelism effects are small.

\begin{figure}[pth]
  \centering
  \includegraphics[width=\textwidth]{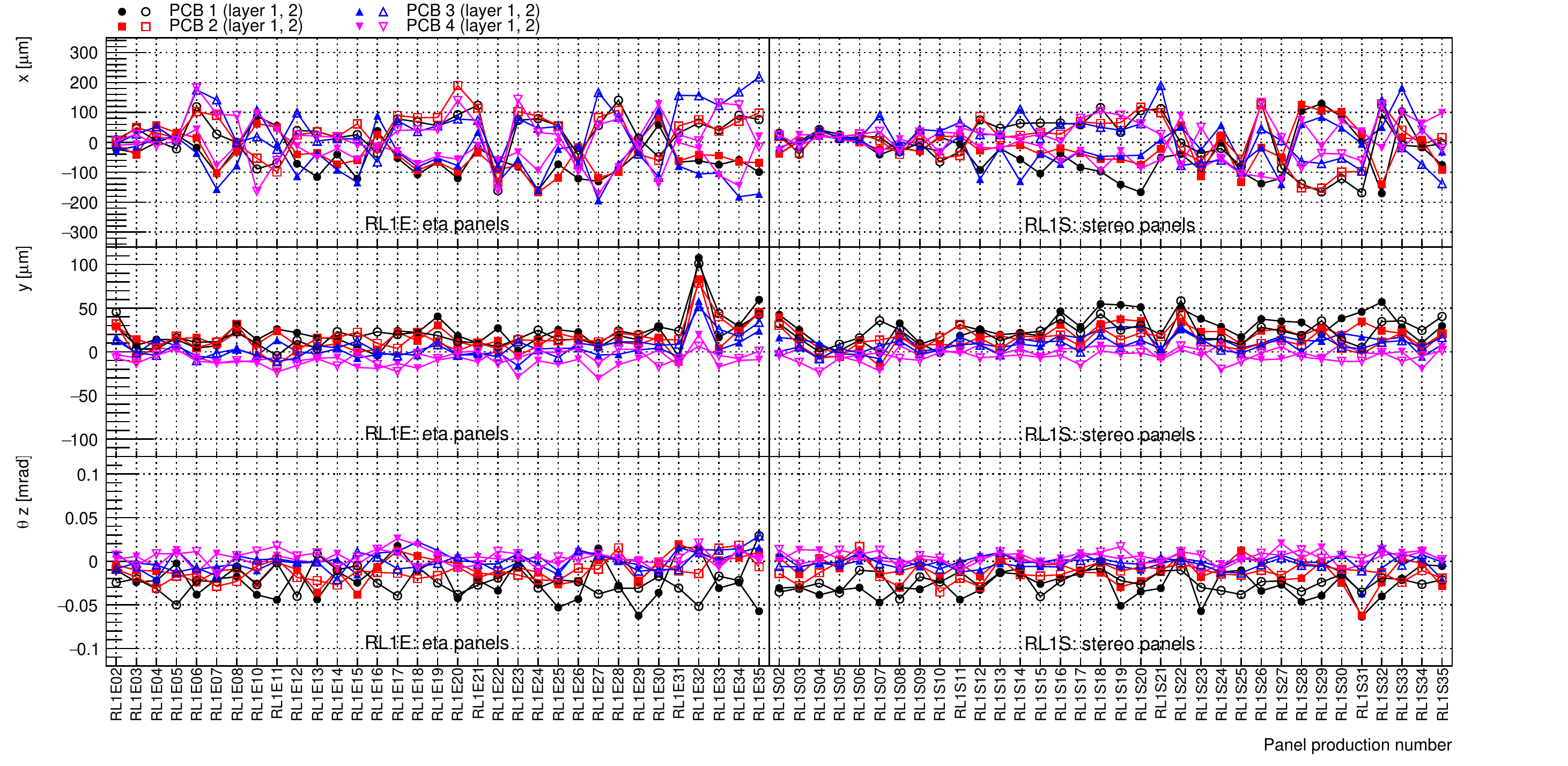}
  \caption{Metrology fit result: PCB position ($x$ and $y$) and rotation (about the $z$ axis, $\theta_z$) parameters, as a function of the panel production number. The different markers correspond to different PCB types and layers. The reported values are deviations of the different PCBs in their nominal frame (see Fig.~\ref{figure:ali:reco_coordinatesystem}). The layer frame is fixed on the 5th PCB, whose parameters are thus zero by definition, and are not reported in this plot.}
  \label{figure:ali:metro_pcb}
\end{figure}

PCB position and rotation parameters in the frame of read-out panel layer are
presented in Fig.~\ref{figure:ali:metro_pcb}. The RMS of the position bias along the
precision coordinate ($y$) and of the rotation bias ($\theta_z$) are \SI{17}{\micro\meter} and \SI{16}{\micro\radian}, respectively, representative of the high quality of the panel construction. A few outliers are however present.

\begin{figure}[pth]
  \centering
  \includegraphics[width=\textwidth]{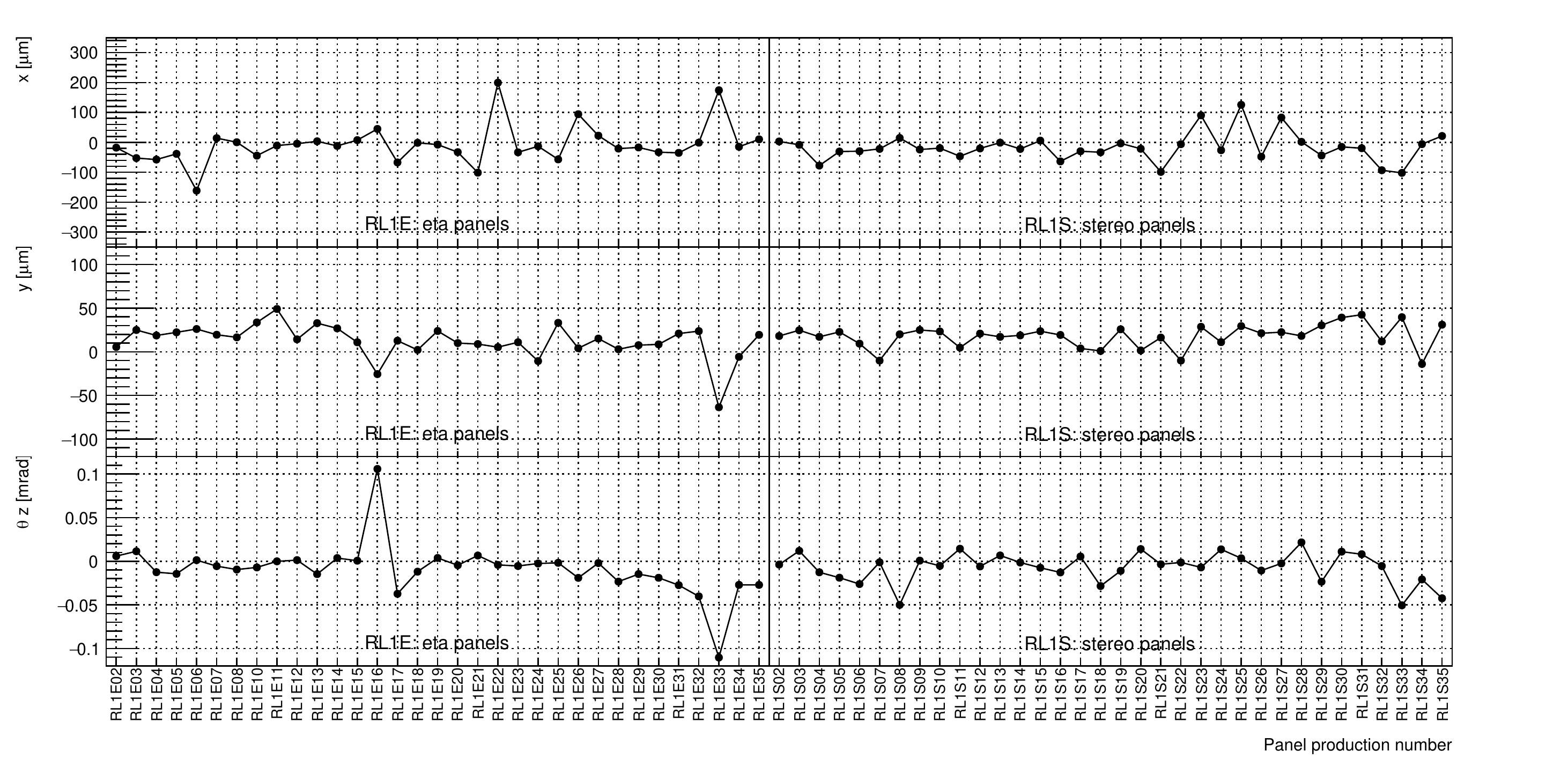}
  \caption{Metrology fit result: second layer position($x$ and $y$) and rotation (about the $z$ axis, $\theta_z$) parameters with respect to the first layer, as a function of panel production number. The reported values are deviations with respect to the nominal layer frame, which coincides with that of its 5th PCB (see Fig.~\ref{figure:ali:reco_coordinatesystem}). The panel frame is fixed on the first layer, whose parameters are thus zero by definition, and are not reported in this plot.}
  \label{figure:ali:metro_layer}
\end{figure}

Layer position and rotations biases in the frame of the panel are presented in Fig.~\ref{figure:ali:metro_layer}. Along the precision coordinate, the translation ($y$) and rotation ($\theta_z$) parameters have an RMS of \SI{17}{\micro\meter} and \SI{24}{\micro\radian}, respectively, with the presence of several outliers in the production. The overall alignment quality of the panel construction is very good.

\begin{figure}[pth]
  \centering
  \includegraphics[width=\textwidth]{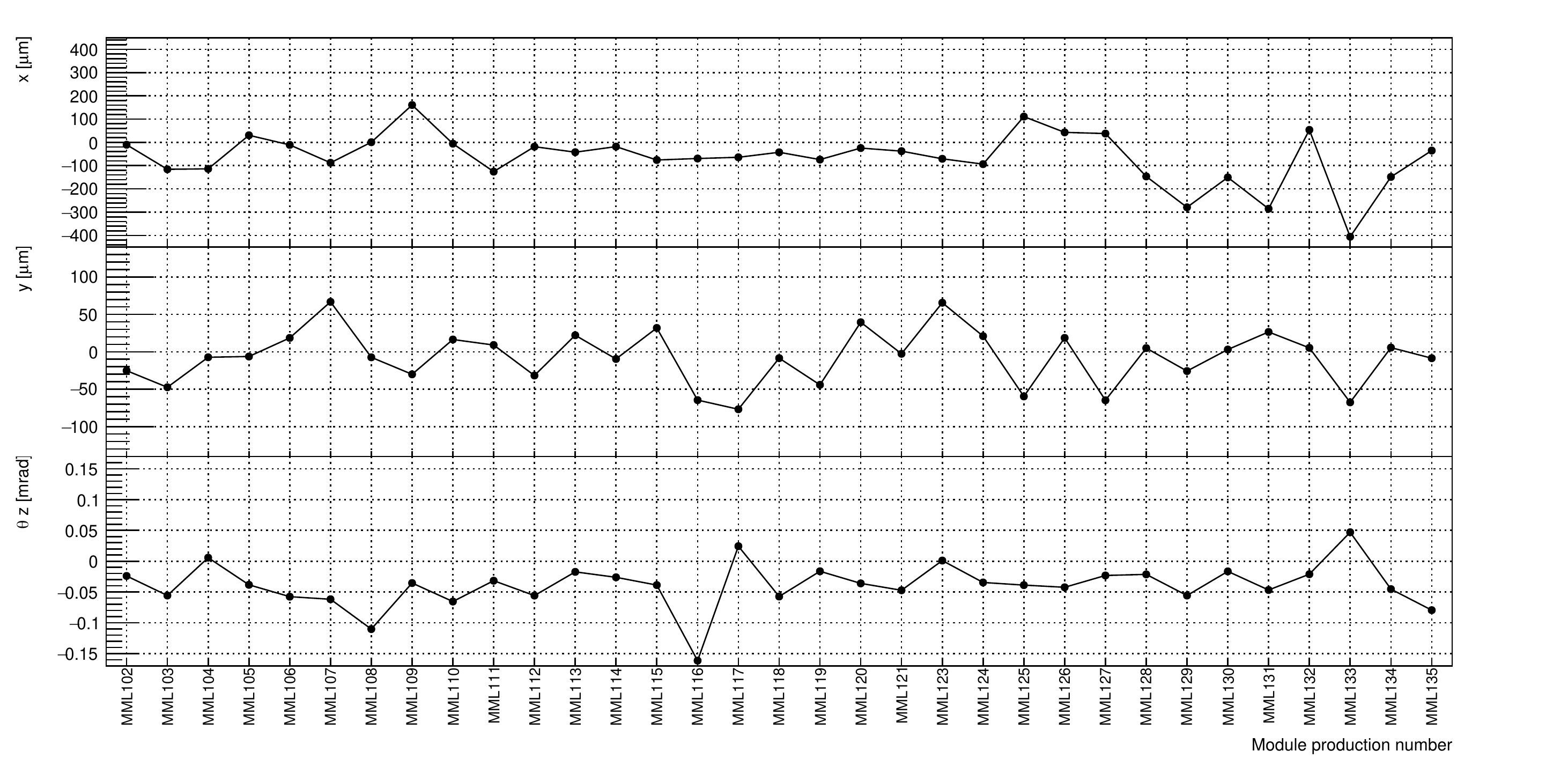}
  \caption{Metrology fit result: stereo panel position ($x$ and $y$) and rotation (about the $z$ axis, $\theta_z$) parameters, with respect to the eta panel, as a function of module production number. The parameters are expressed in the nominal frame of the stereo panel (which coincides with the nominal frame of the 5th PCB of the first layer, see Fig.~\ref{figure:ali:reco_coordinatesystem}).}
\label{figure:ali:metro_panel}
\end{figure}

In Fig.~\ref{figure:ali:metro_panel} are presented panel position and rotation biases in the frame of the assembled module. The translation ($y$) and rotation ($\theta_z$) have an RMS of \SI{37}{\micro\meter} and \SI{36}{\micro\radian}, respectively. The alignment quality of the module assembly is of lower quality than what is achieved with the panels, reflective of the difficulty of the assembly procedure.

The PCB material (FR4) is sensitive to humidity. The result of the metrology fit
presented here is thus valid at the end of panel construction. An additional
gantry CMM measurement is performed after the drying process, resulting in a
correction to the overall elongation of order $100\,\si{ppm}$. As the internal structure of the panel is in-homogeneous (most notably because of the presence of the cooling bars), this elongation is not uniform across the detector. This results in an additional PCB sag that is of the order of $50\,\si{\micro\meter}$. 


\subsection{Module gas tightness}
\begin{figure}[hbt!]
    \centering
    \includegraphics[width=0.7\textwidth]{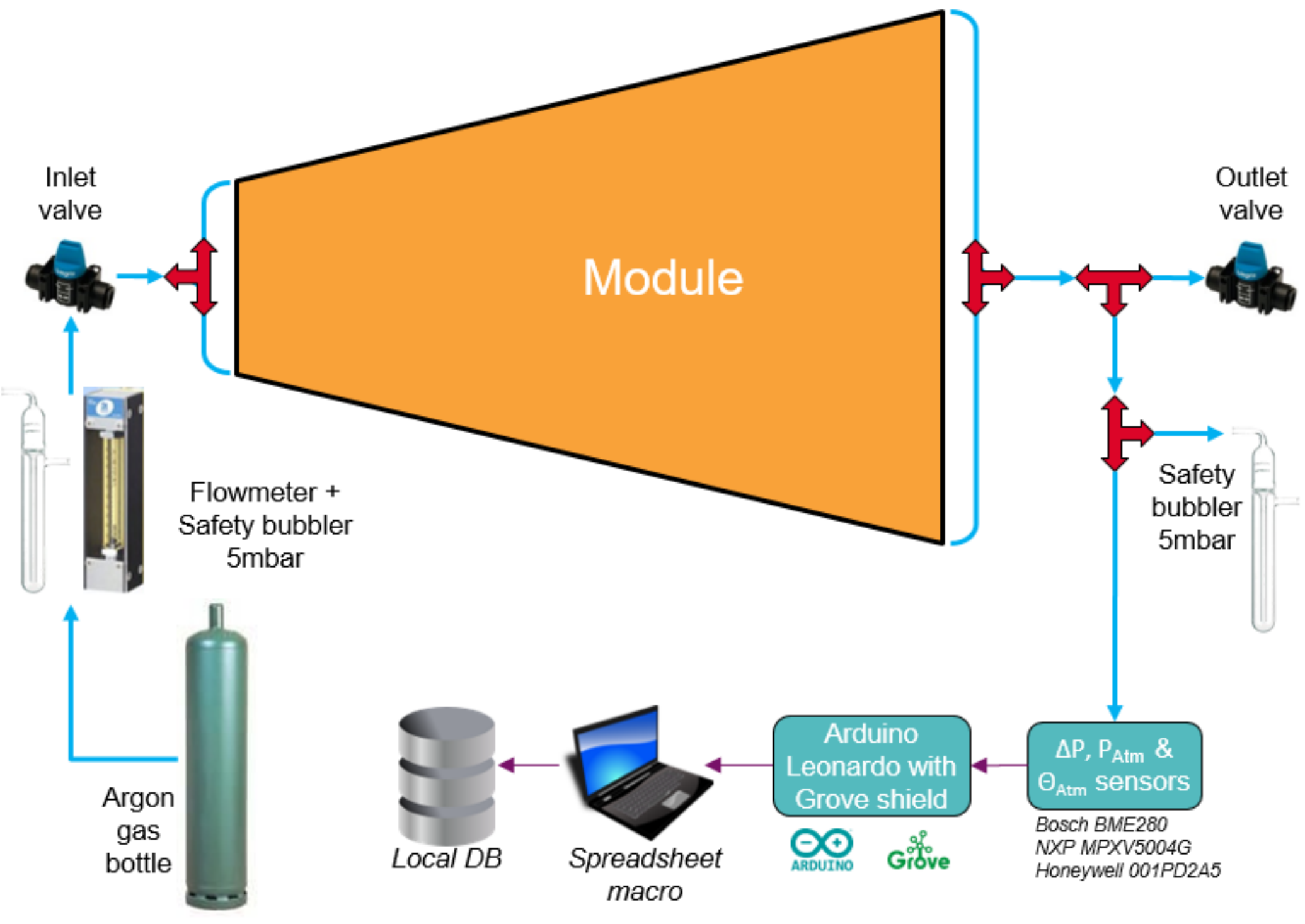}
    \caption{Module gas tightness setup: gas bottle, inlet safety panel, inlet valve, outlet safety panel with outlet valve, an Arduino device associated with atmospheric sensors (absolute pressure, temperature, relative humidity) and pressure differential sensor, a laptop on which a dedicated spreadsheet document is installed. Data are then transferred to a local database.}
    \label{fig:ModulesGT_Setup}
\end{figure}

\begin{figure}[hbt!]
    \centering
    \includegraphics[width=0.8\textwidth]{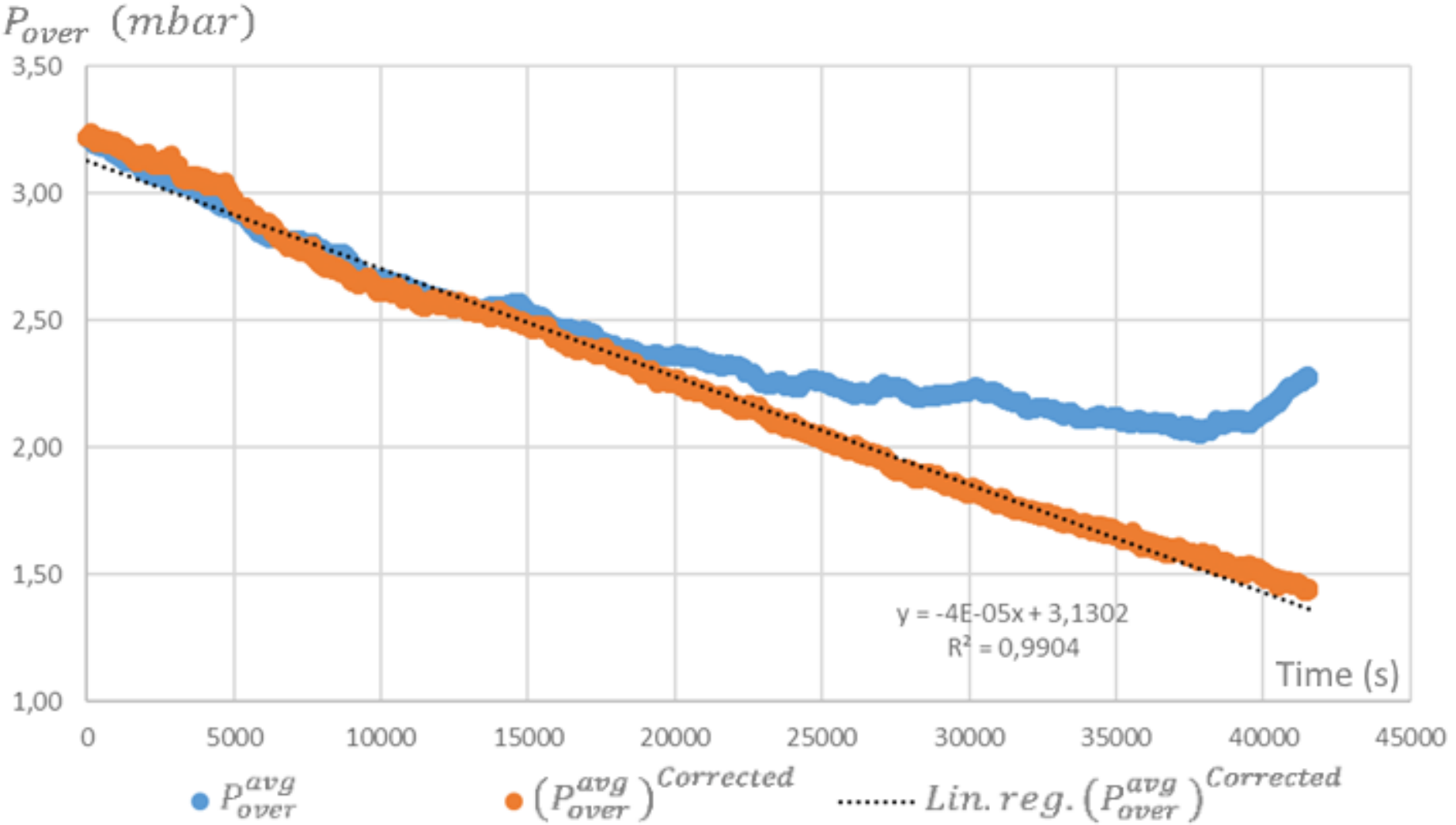}
    \caption{Typical corrected curves - Module 15.}
    \label{fig:ModulesT_CorrectedCurve}
\end{figure}

Measurement of module gas-tightness is based on the pressure drop method. The main idea is to increase the inner pressure by flushing the module with Argon and then close the inner gas volume. The decrease of this overpressure, corrected by the variations of the atmospheric pressure, is then recorded as a function of time. The set-up is  shown in Fig.~\ref{fig:ModulesGT_Setup} and a typical recording is shown in Fig.~\ref{fig:ModulesT_CorrectedCurve}.

The module gas tightness acceptance threshold has been defined as \(10^{-5}\,\min^{-1}\) ~\cite{qaqc}, to be understood as the maximum acceptable volume of leak gas relatively to the  detector volume per time unit.

To determine the gas tightness value, the overpressure decreasing curve is computed over time, with a linear approximation. This method is conservative, considering leak rate is decreasing as the overpressure does.
To ensure repeatability and comparison between construction sites, recording is done between 3.2 and 2.7\,mbar of overpressure. A safety bubbler device is used to limit at 5\,mbar the maximum overpressure seen by the module (see Fig.~\ref{fig:ModulesGT_Setup})



\graphicspath{...}

The correction depends on a few parameters: time, overpressure values, atmospheric pressure, temperature and module ``gas stiffness".
This last parameter reflects the module capacity to “convert” overpressure variation into gas volume variations (i.e. coupling strength).
This stiffness is determined through a dedicated test: a calibrated volume of gas is inserted by steps into the module and the overpressure variation is followed. Knowing the initial volume of gas and the atmospheric pressure, it is then possible to find the stiffness law. For overpressure below 5\,mbar, this law is linear and the corresponding slope is used in the correction formulas.

The leak rate $F_{Leak}$ is extracted from the overpressure $P_{over}$  time variations by correcting first for the time variation of the atmospheric pressure $P_{atm}$ and of the temperature $T$ 
\begin{equation}
  \frac{dP^{Corrected}_{ over } }{dt} 
\simeq
  -    \frac{\beta F_{leak}  }{ P^{r}_{ over }/P_{atm}^0 }  P_{ over } 
  \label{equ:LEAKformula1}
\end{equation}
  with  
\begin{equation}
P^{Corrected}_{ over } (t) =
 {P_{ over }} (t)
 + 
 \beta
 { P_{atm}  }  (t)
 -
  \beta
 \frac{P_{atm}^0}{T^0} 
 { T  }  (t) 
   \label{equ:LEAKformula2}
  \end{equation}
where $\beta$ is the module stiffness, $P_{atm}^0$, $T^0$ are initial values of the atmospheric pressure and of the temperature and $P^{r}_{ over }$ is the reference overpressure. An example of a corrected overpressure evolution curve (orange curve) is given in Fig.~\ref{fig:ModulesT_CorrectedCurve}.

\begin{figure}[hbt!]
    \centering
	\includegraphics[width=1\textwidth]{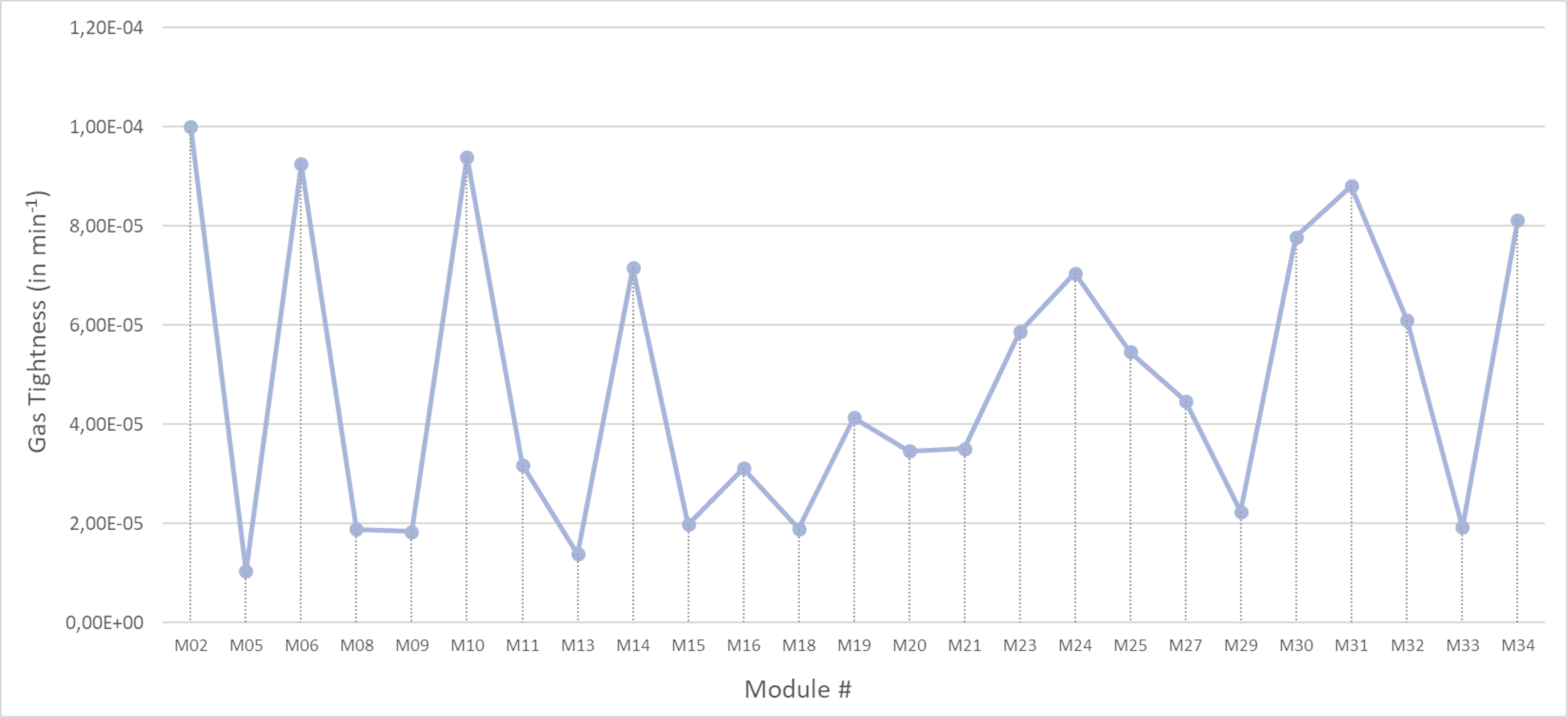}
	\caption{Module Gas Tightness measurements. The gas tightness has been defined as the volume of leak gas normalised by the  detector volume per time unit.} 
	\label{fig:ModulesGT_Stat}
\end{figure}

An overview of the  gas tightness measurements for the first 25 modules is given in Fig.~\ref{fig:ModulesGT_Stat}.

\section{Validation with cosmic rays}
%
\label{section:validation}
\subsection{Description of the cosmic ray test bench}
\begin{figure}[!h]
    \centering
	\includegraphics[width=0.6\textwidth]{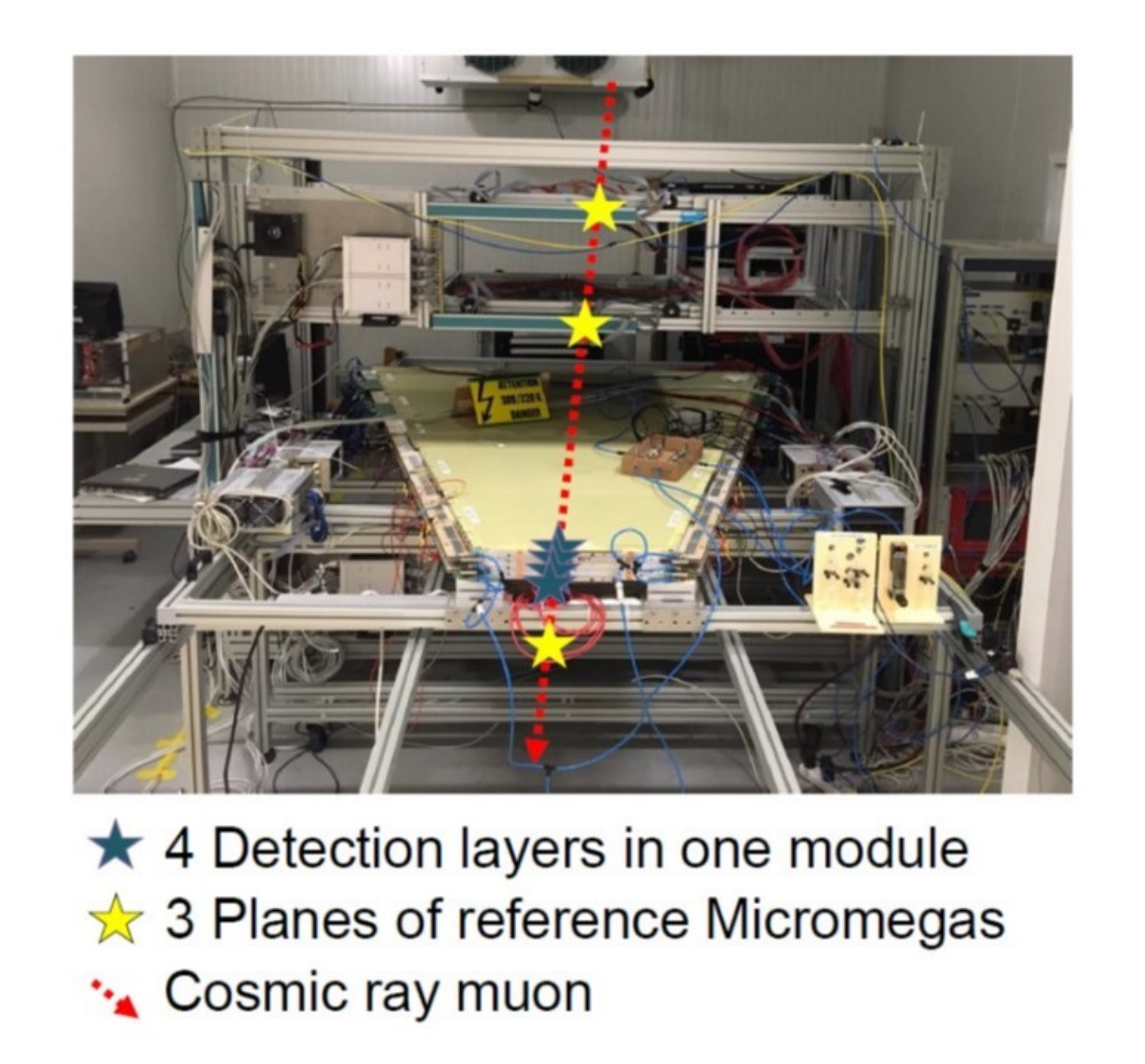}
	\caption{Photo of the M$^3$ structure with three layers of Micromegas detectors and an LM1 quadruplet in test. The LM1 quadruplet sits on a tray that allows to slide it the acceptance of the M$^3$ external trackers. On top of the LM1 module two external Micromegas trackers are visible, the third tracker is below the module, not visible on the picture but indicated with a star.} 
	\label{fig:cbphoto}
\end{figure}

An existing telescope~\cite{Bouteille:2016wdv}, previously developed for muon tomography, was adapted in order to test and validate LM1 modules with cosmic rays. The M$^3$ telescope, used as an external tracker, consists of an aluminium structure supporting three layers of Micromegas detectors. To this initial structure was added a tray that allows to slide the LM1 quadruplet to scan the surface of the modules in three steps, covering 1\,m$^2$ of the quadruplet in each step. A photo of the modified structure with an LM1 quadruplet in test can be seen in Fig.~\ref{fig:cbphoto}. 
Each Micromegas external-tracker layer has an active area of 1\,m$^2$ and is composed of four bulk~\cite{Giomataris:2004aa}  detectors of 0.5 $\times$ 0.5\,m$^2$ with 1\,cm drift gap. The detector readout is 2-dimensional, made of 1024 strips with a pitch of 488\,$\mu$m connected to only 61 channels thanks to the factor 16 multiplexing~\cite{Procureur:2013yea}. A Kapton foil with resistive strips of \SI{1}{\mega\ohm}$/square$ is glued on top of the 2-dimensional readout~\cite{Bouteille:2016wdv} allowing higher gains and larger clusters sizes. These M$^3$
detectors are flushed with an Argon-5\% Isobutane mixture.

\subsection{Readout and acquisition electronics}
The final NSW electronics, so called  VMM~\cite{Alexopoulos:2020wea}, were not available when the LM1 series modules had to be validated and characterised in 2017.

An in-house available solution was looked into leading to the choice of the {Dead-timeless Readout Electronics ASIC for Micromegas }(DREAM) developed at CEA for the CLAS12~\cite{Burkert:2020akg} experiment at Jefferson Laboratory.

DREAM~\cite{7097517} was designed to cope with the high strip capacitance (up to 200\,pF) of the CLAS12 Micromegas detectors achieving at the same time  comfortable signal to noise ratio well above 10. This implied that the equivalent noise charge of the detection chain should be $\sim2500$\,e$^-$ for the $140-200$\,pF range of the total input capacitance. By the time of development, none of the existing HEP ASICs could deliver the required performance.

These electronics were suitable for the LM1 cosmic ray test bench (CTB) as they had been designed and had been tested with success in very different conditions, in particular in CLAS12 and in muon tommography set-ups~\cite{Bouteille:2017jty}. Moreover the required number of channels could be produced in sufficient number and in due time for the characterisation of the LM1 production.\\


Each LM1 module contains $\approx$\,20k channels. In order to lower the cost and the complexity of the set-up and profiting from the fact that the cosmic ray multiplicity is rather low (3-4 strips per event), a multiplexing strategy, similarly to the one used in the \mcube detectors,  was adopted to reduce the required number of electronics channels. Two different multiplexing factors (2 and 4) were considered but tests of the signal to noise ratio led to choose a multiplexing factor of 2. Interface multiplexed cards were produced to connect DREAM electronics to the LM1 Zebra \textsuperscript{\textregistered}~\cite{Zebra} connectors.\\

\begin{figure}[!h]
    \centering
	\includegraphics[width=\textwidth]{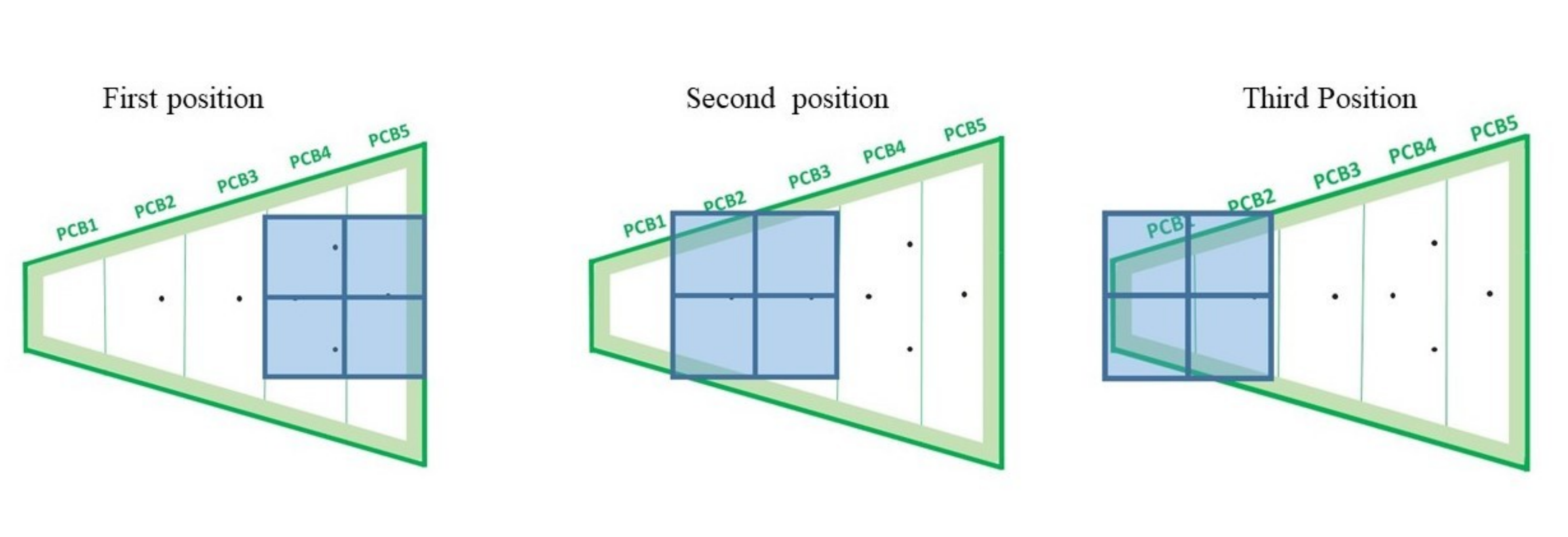}
	\caption{Sketch of the procedure to scan the LM1 quadruplet with the \mcube tracker. The cosmic muon scan starts with the two largest PCBS (4 and 5), the second position is devoted to the two intermediate PCBS (3 and 2) and the last position is devoted to PCB1. The interconnections are indicated by black dots in the sketch. The green band around the border of the module represents the average passivated area.} 
	\label{fig:scan-CTB}
\end{figure}

%
\label{section:characterisation}
\subsection{Tests protocol}
Once the quadruplet is assembled and the quality checks have been completed, the quadruplet is installed on  the CTB to undergo a series of validation procedures to ensure that it can be sent to CERN and mounted on the NSW.

\subsubsection{Humidity curating}
\begin{figure}[!h]
    \centering
	\includegraphics[width=0.7\textwidth]{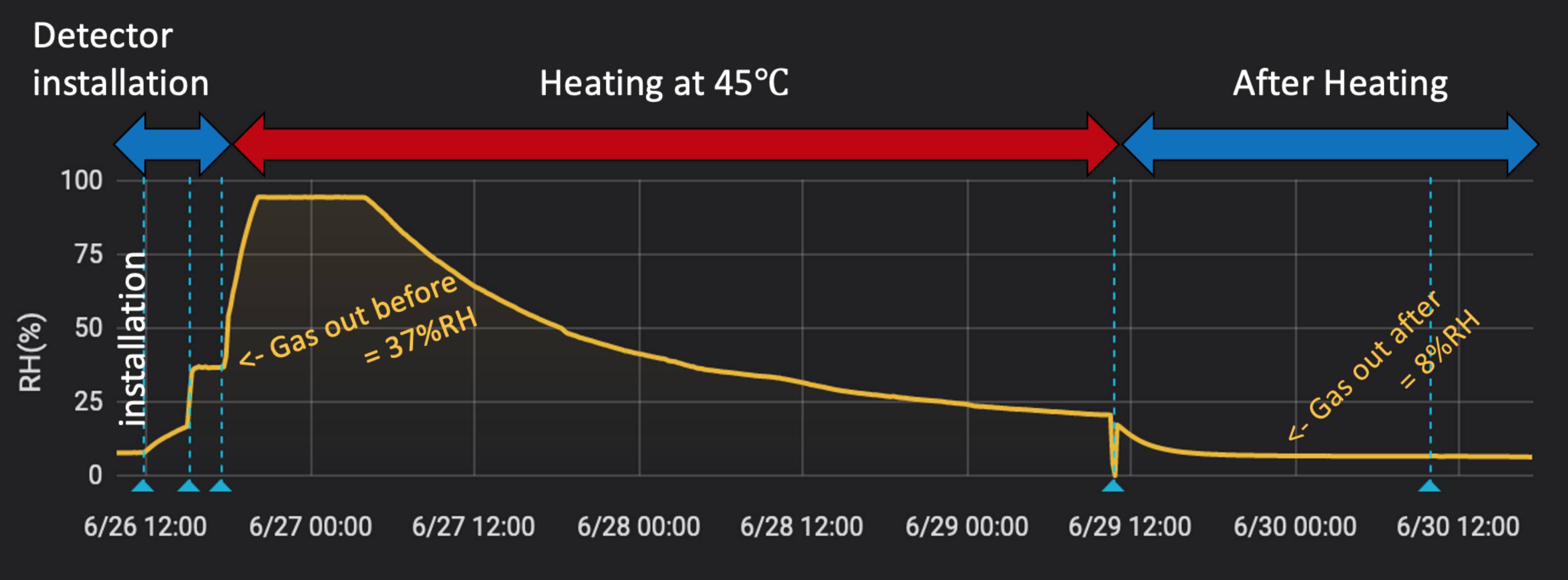}
	\caption{Gas relative humidity in \% before and after the heating. Just after the connection of the module on the CTB gas system, the module output relative humidity is 37\%. Then it is heated at \SI{45}{\degreeCelsius} to help water out-gassing during three days. Once a stable mode is achieved (the humidity has decreased), the heating is turned off. The detector output reaches then 8\% RH.} 
	\label{fig:humidity}
\end{figure}

Since above a certain level of humidity ($>10\%$) out-gassed by the detector, high voltages instabilities have been observed, high-voltages are not ramped-up immediately after the quadruplet is installed on the CTB. Instead, the module is heated and dried until it reaches  a threshold of 10\% relative humidity (RH) at \SI{20}{\degreeCelsius}, for an Ar/CO2 gasflow of \SI{15}{\liter\per\hour}.
Fig.~\ref{fig:humidity} shows the evolution of humidity during treatment of one LM1 quadruplet (M20) by heating it for three days.

\subsubsection{High-Voltage Ramp-up}
\begin{figure}[!h]
    \centering
	\includegraphics[width=\textwidth]{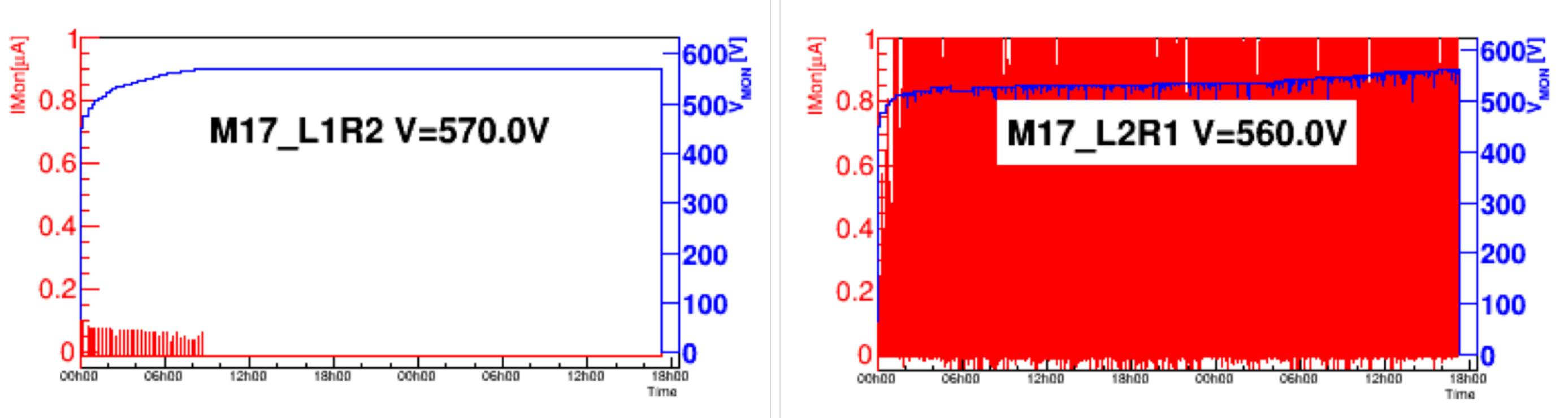}
	\caption{Ramping up the high-voltage using automatic program on LM1-M7 over 42 hours. The first HV steps are quite large up to 500\,V and only after the ramping-up is done in small  steps.
	Most of the sectors looks like (left) where the only current observed comes from the increasing of the voltage during the first 8 hours. This is a good sector, reaching the nominal HV at \SI{570}{\volt}. The right plot shows a problematic sector where the program takes a lot longer to raise the HV. The current in red covers the graph because of the long time range covered here, it is in fact only \SI{9.1}{spikes\per\minute}.}
	\label{fig:rampingup}
\end{figure}

Once the detector is dry enough, the next step is  high voltage ramp-up on the resistive strips in order to reach the nominal gain. Unlike smaller detectors, it was found that only a slow and careful rise of the high voltage while monitoring the current consumption of each of the 40 HV sectors, could reach the best results. To operate  all the HV channels of the quadruplets and test bench in the best conditions, the CAEN power supply SY5527LC equipped with 4 positive HV cards (A1561HDP and A1821H) and a negative HV card (A1821H) for the drift electrodes were used. This system has a current resolution of the order of nA and allows,via a C++ library~\cite{CAENwrapper}, a fine control of the system. Using this library, we have developed a program to automatise the high voltage rise by a series of pre-programmed steps. Depending on the current behaviour, we can either follow a pre-programmed sequence or act otherwise. The standard procedure chosen to ramp up the CTB is:
\begin{itemize}
    \item if the current is above \SI{0.6}{\micro\ampere} for \SI{20}{\second} of a \SI{30}{\second} sliding interval, the voltage is lowered by \SI{10}{\volt} to avoid tripping;
    \item if the current is above \SI{50}{\nano\ampere} for \SI{30}{\second} over a duration of \SI{300}{\second}, the program does not increase the HV;
    \item otherwise the high voltage is increased to the value of the next step.

\end{itemize}
An example of ramping up is shown in Fig.~\ref{fig:rampingup}.

The voltage steps and current thresholds are fully configurable by the user, and each HV channel has an independent offset correction to avoid dependence on the power supply intrinsic channel calibration.

During data taking, the program is used in {\it protection mode} where it lowers the voltage in case of large current to avoid tripping. It is much safer to lower the voltage in steps than to let it trip and fall directly to \SI{0}{\volt}, because the large coupling between sectors through the readout strips and the mesh makes it very dangerous to have the HV on one sector fall suddenly to \SI{0}{\volt}.

The code uses the CAEN library~\cite{CAEN} and is available here~\cite{gitlab-maxence}.

\subsubsection{Connection and test of the readout electronics}
\begin{figure}[!h]
    \centering
	\includegraphics[width=0.6\textwidth]{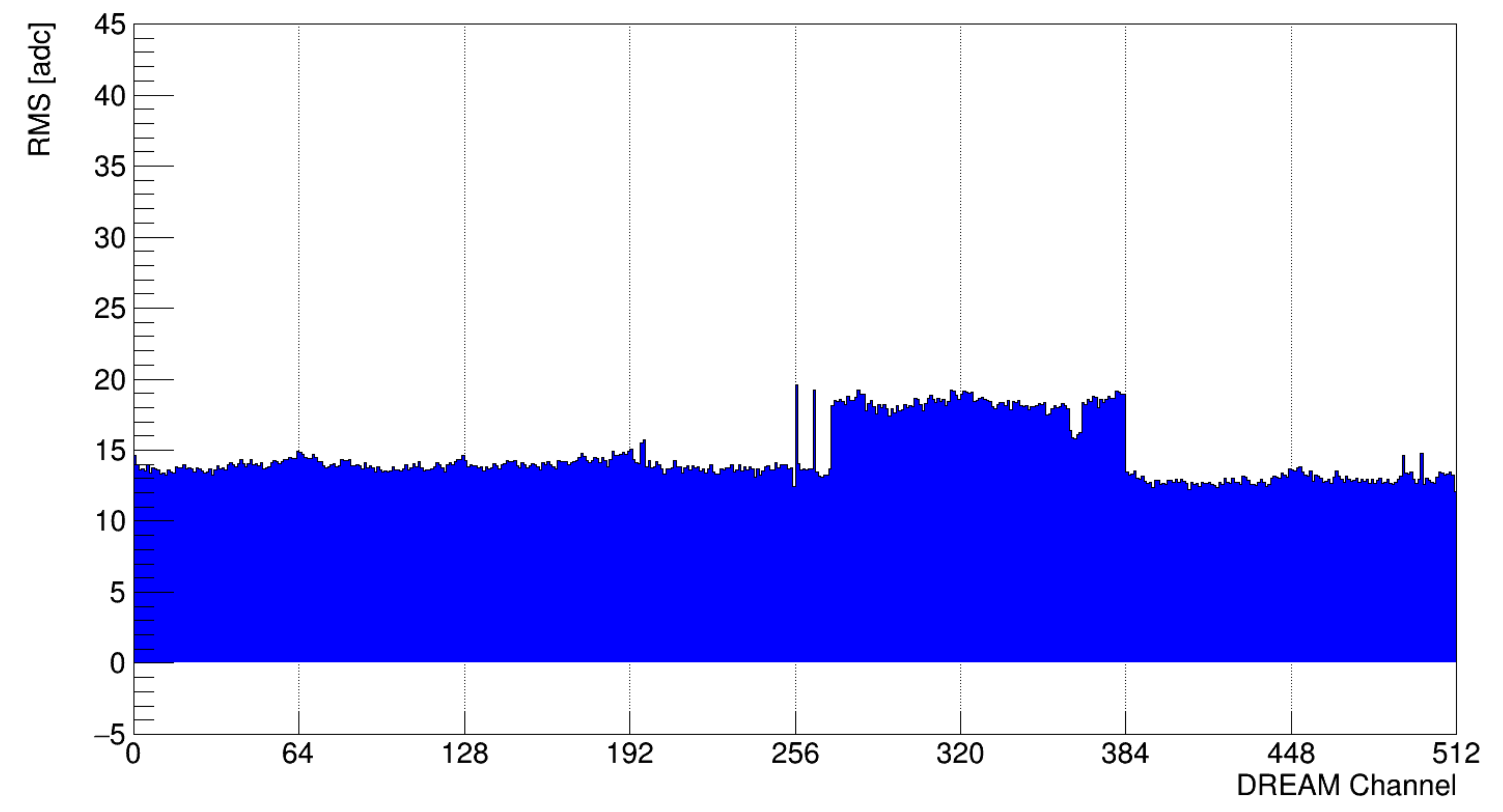}
	\caption{RMS of the pedestals for one Front End Unit card (512 channels) with a mean value of 13  ADC counts over 4096 representing roughly \SI{2}{\femto\coulomb} RMS. The small increase between channels 256 to 354 indicates a bad connection, confirmed by Fig.\ref{fig:hitmaps}.}
	\label{fig:pedestals}
\end{figure}

\begin{figure}[!h]
    \centering
	\includegraphics[width=0.8\textwidth]{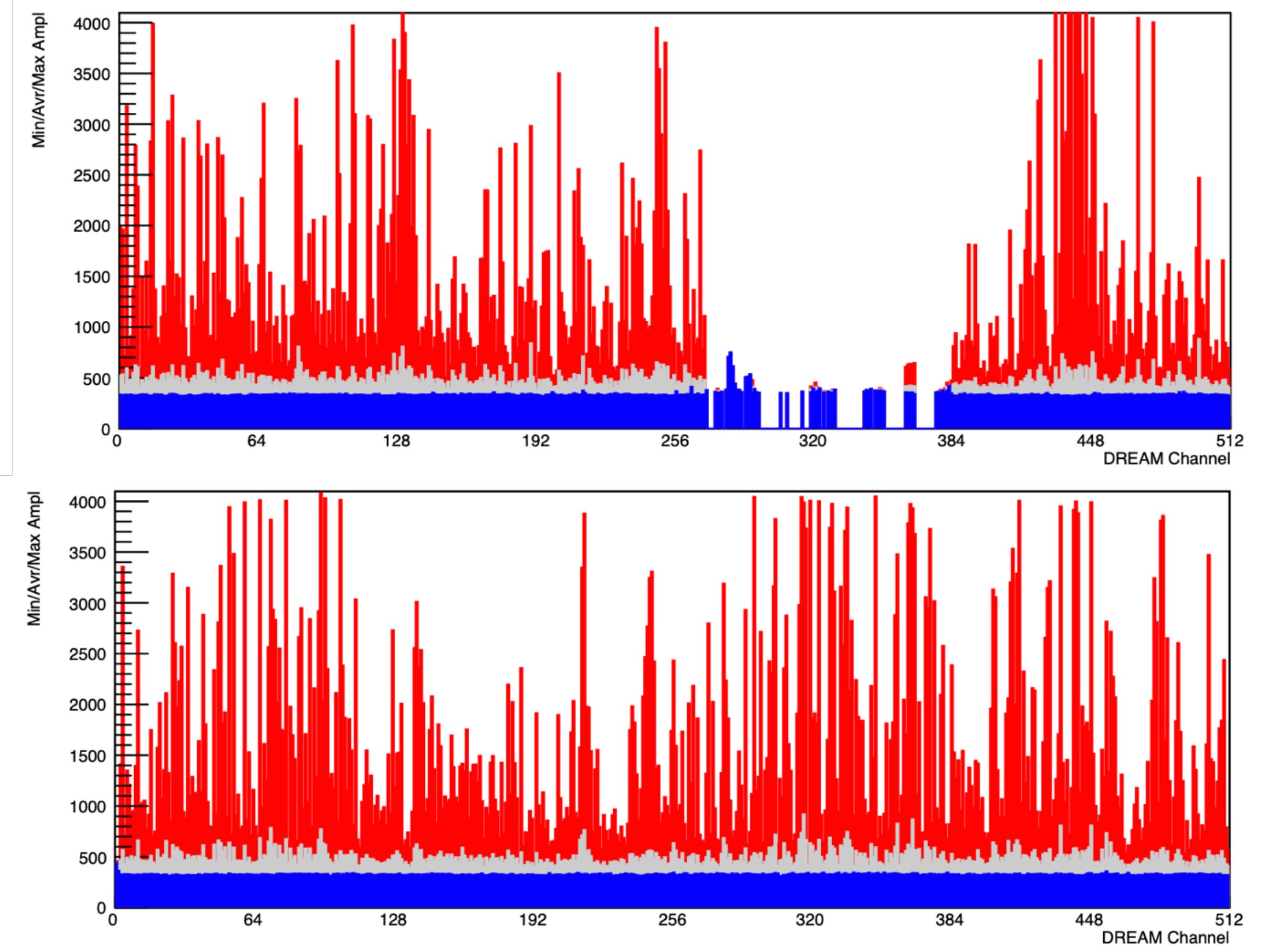}
	\caption{Hit maps for the 512 channels of one Front End Unit (same as Fig.\ref{fig:pedestals}). The red, grey, and blue represent respectively the maximum, average, and lowest, values recorded over a one-hour run. The upper graph shows almost no signals on channels 256 to 354 due to a connection problem. The lower graph shows the same hitmap  after re-fixing the connector to fix the issue.}
	\label{fig:hitmaps}
\end{figure}

During the gas flushing of the detector and before testing its performance, the first step is to connect the electronics boards to the module. The interface cards are connected after a thorough cleaning to remove any dust from the footprints on the readout PCBs and from the Zebra connectors on the interface cards. Since faulty connections on the Zebra connectors are common, two independent checks are performed before taking data. First by looking at the pedestal RMS, it is possible to see unconnected areas, since the noise of the electronics channels is proportional to the connected capacitance. Lower or higher noise indicates a possible mis-connection, as seen in Fig.~\ref{fig:pedestals}.
Once bad connections have been improved by repositioning the interface cards, the quadruplet is slid into the \mcube tracker for one hour of data taking. From this short run  a rough hit and amplitude map is extracted, as shown in Fig.~\ref{fig:hitmaps}, in order to check the connections again before starting long runs for the complete data taking.

Since the \mcube reference trackers cover only  \SI{1}{\meter\squared}, this operation needs to be repeated three times to scan the full length of the detector. For the same reason, the external parts of the three largest PCBs  do not fit in the acceptance of the \mcube tracker. They are not scanned, so as to complete a full scan of the module within five days and fit in the tight production schedule.

\subsection{Data analysis}
\begin{figure}[!h]
    \centering
	\includegraphics[width=0.7\textwidth]{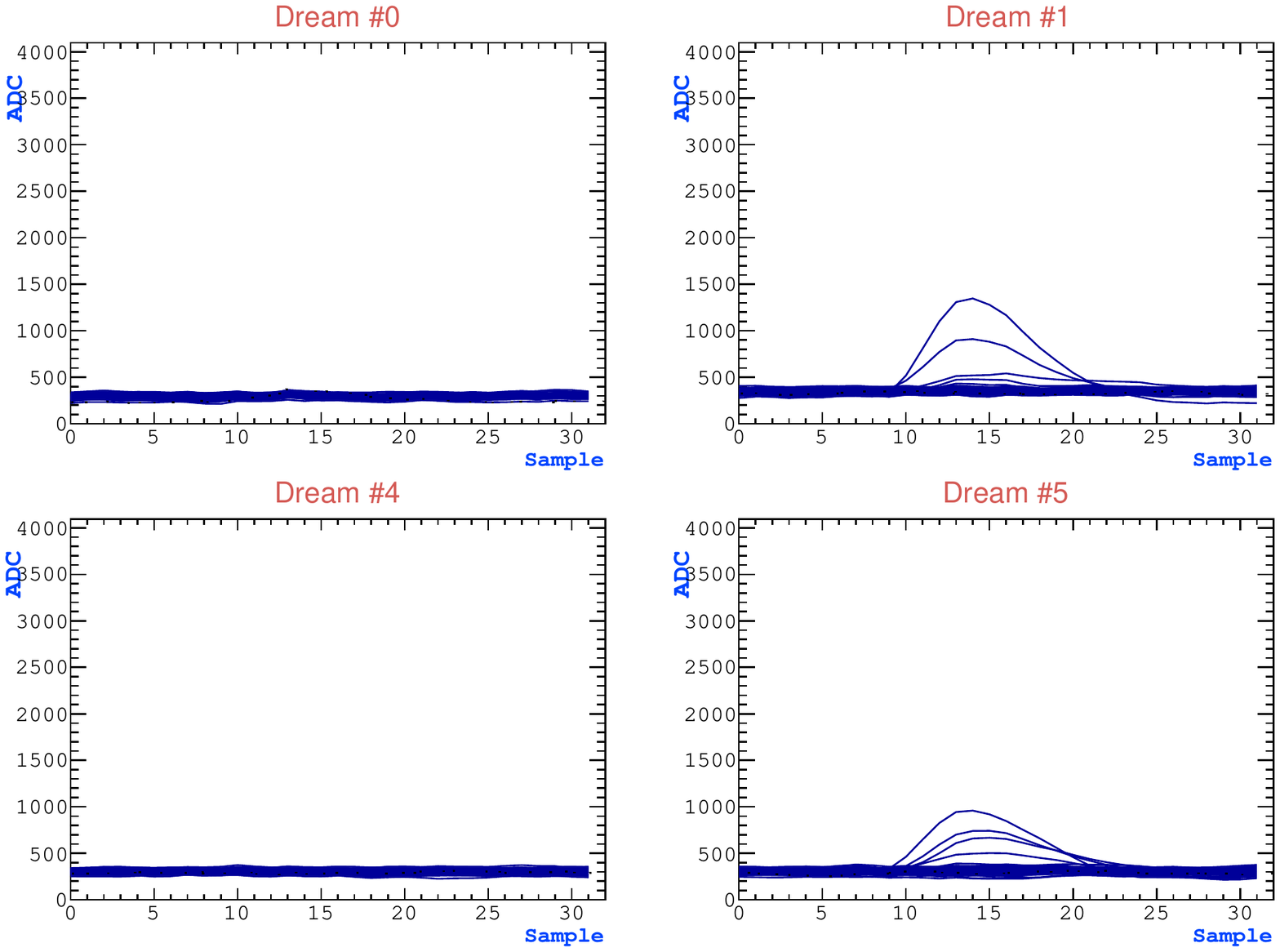}
	\includegraphics[width=0.7\textwidth]{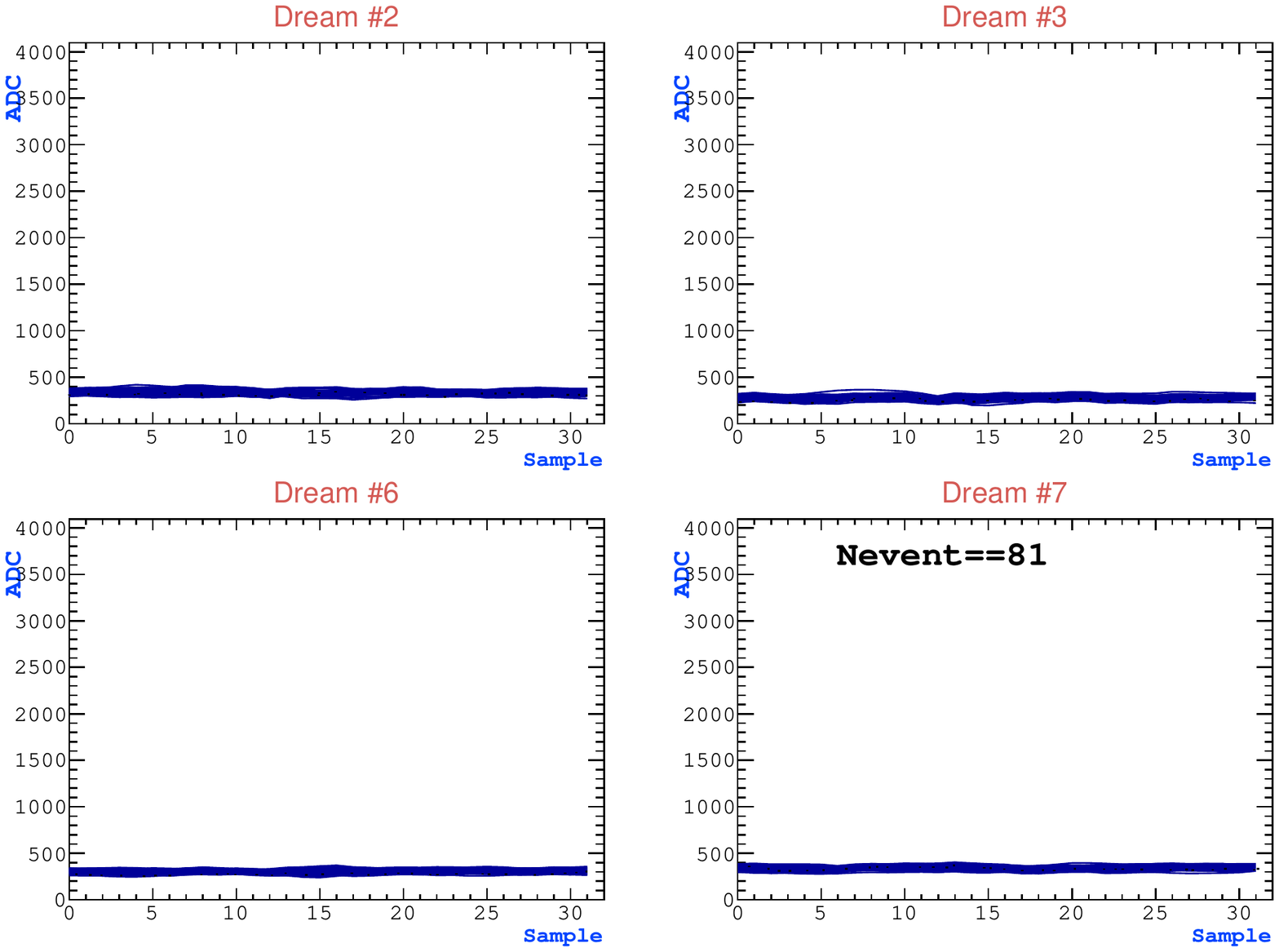}
	\caption{Typical cosmic ray event seen on the detector on one Front End Unit at the nominal voltage (\SI{570}{\volt}). Each graph shows the amplitudes of all the 64 channels of one DREAM ASIC for this event along the 32 samples. Each sample lasts \SI{60}{\nano\second}. The FEU is connected to two PCBs located at the same place on top of each other. ASICs labelled 0 to 3 are connected to L1 and those labelled 4 to 7 are connected to L3; therefore the signals seen on DREAM 1 and 5 are extremely likely to belong to the same cosmic ray particle.} 
	\label{fig:signals}
\end{figure}

The data analysis is performed in several stages. The code is based on the code developed for muon tomography studies~\cite{Bouteille:2017jty} and has been continuously adapted~\cite{chevalerias:tel-02268557}, developed and improved during LM1 characterisation. Starting from the electronics output signals, the data analysis is performed, as follows, in four steps that will be further detailed: 
\begin{itemize}
    \item \textbf{Pedestal substraction}: the amplitude of each strip  undergoes pedestal subtraction, together with common mode suppression described Subsection~\ref{subsubsection:pedestal}.
    \item \textbf{Hit finding}:  to remove electronic noise, only signals above five times the standard deviation of the strip pedestal are considered. The retained signals are called “hits" (see Fig.~\ref{fig:signals} and Subsection~\ref{subsubsection:hitfinding}).
    \item \textbf{Clustering}: nearby hits are gathered into “clusters" according to an algorithm described below. The typical number of strips forming a cluster is around three. If the module is fully efficient, a cosmic muon going through it is expected to leave a cluster in each of the four layers. The clustering is described in more detail in Subsection~\ref{subsubsection:clustering}.
    \item \textbf{Track to cluster matching}: After having aligned the module correctly, a matching algorithm is applied to associate the LM1 cluster to a reconstructed track in the \mcube. If the interpolation of the track position in the layer plane under study is in the module, a simple count of detected or undetected particles can provide the efficiency for each strip. This is described in Subsection~\ref{subsubsection:matching}
\end{itemize}

%
\subsubsection{Pedestal subtraction}
\label{subsubsection:pedestal}
\begin{figure}[!h]
    \centering
	\includegraphics[width=0.5\textwidth]{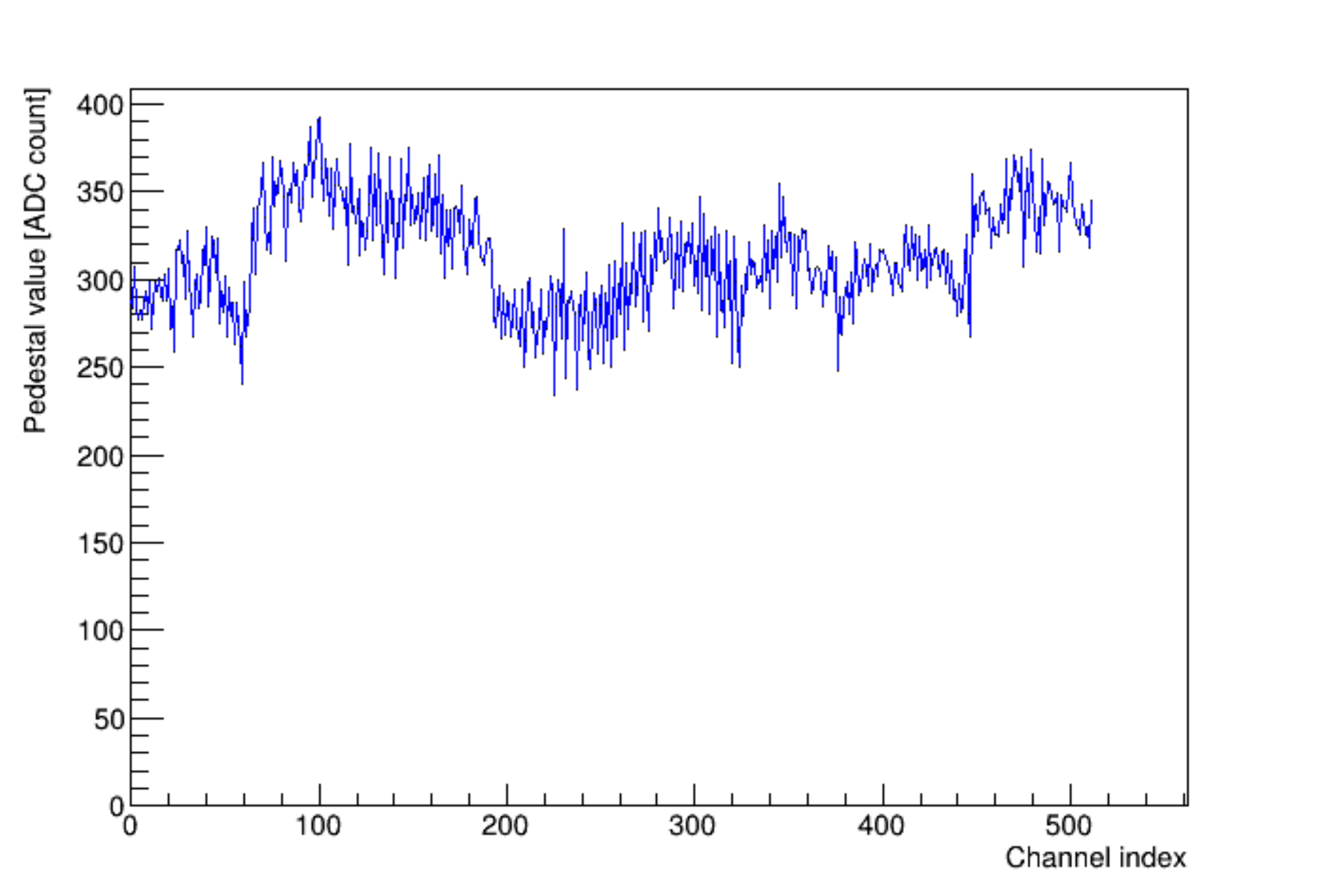}
	\caption{Example of values of pedestals as a function of the channel number in one Front End Unit.} 
	\label{fig:pedestalFEU}
\end{figure}

The pedestals are obtained for each strip, and are calculated as the mean values of the output amplitudes, using a random generator that exclude physics signals from cosmic ray, before starting to take cosmic data. The values of the pedestals for all channels of one Front End Unit are shown in Fig.~\ref{fig:pedestalFEU}.
%

Following pedestal subtraction, there is a remaining common mode noise to be suppressed. While the 50 samples are collected, the baseline amplitudes of all strips of a given Front End Unit card can move coherently. This is known as the “common mode".

%
\subsubsection{Hit selection}
\label{subsubsection:hitfinding}
\begin{figure}[!h]
    \centering
    \includegraphics[width=0.45\textwidth]{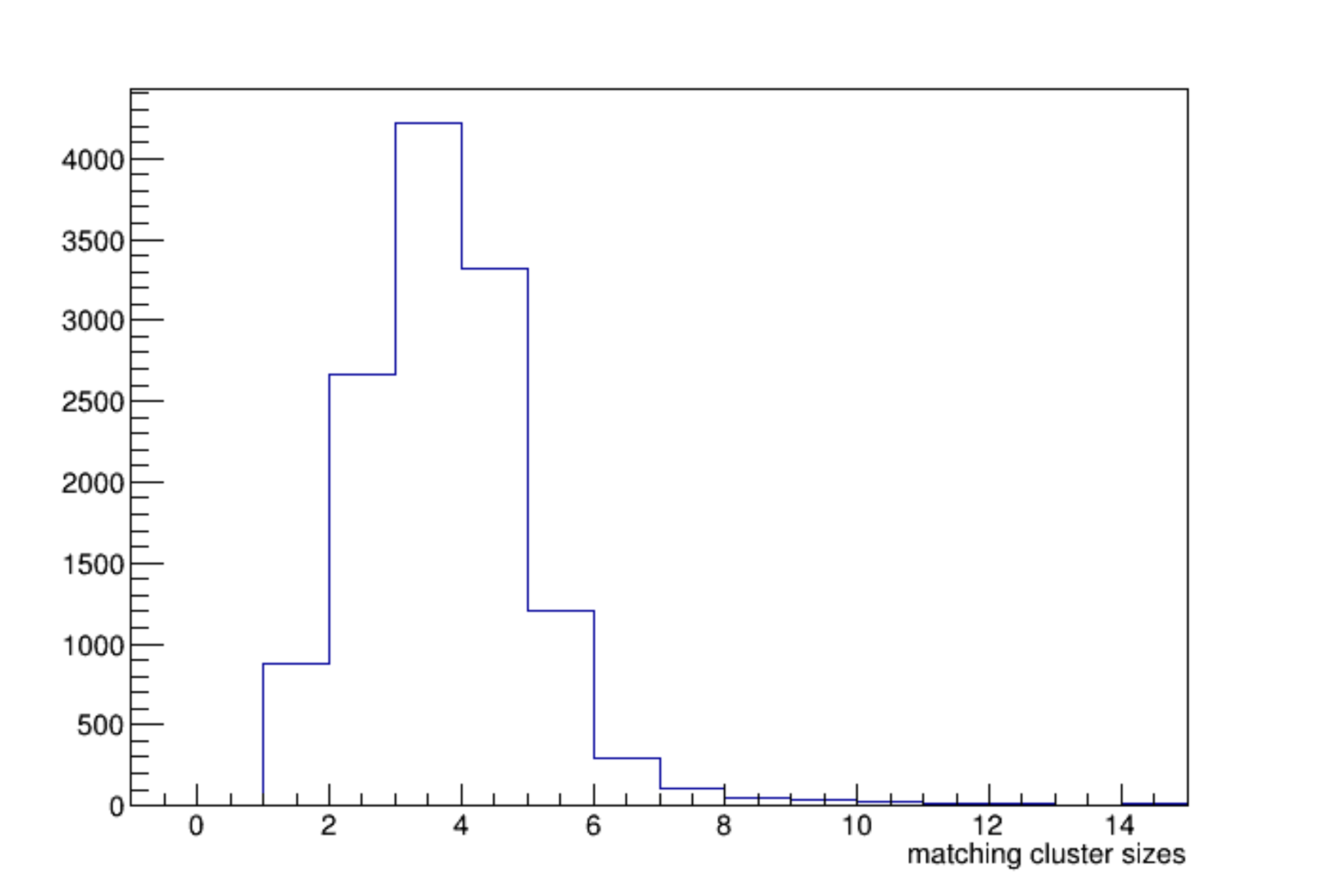}
	\includegraphics[width=0.45\textwidth]{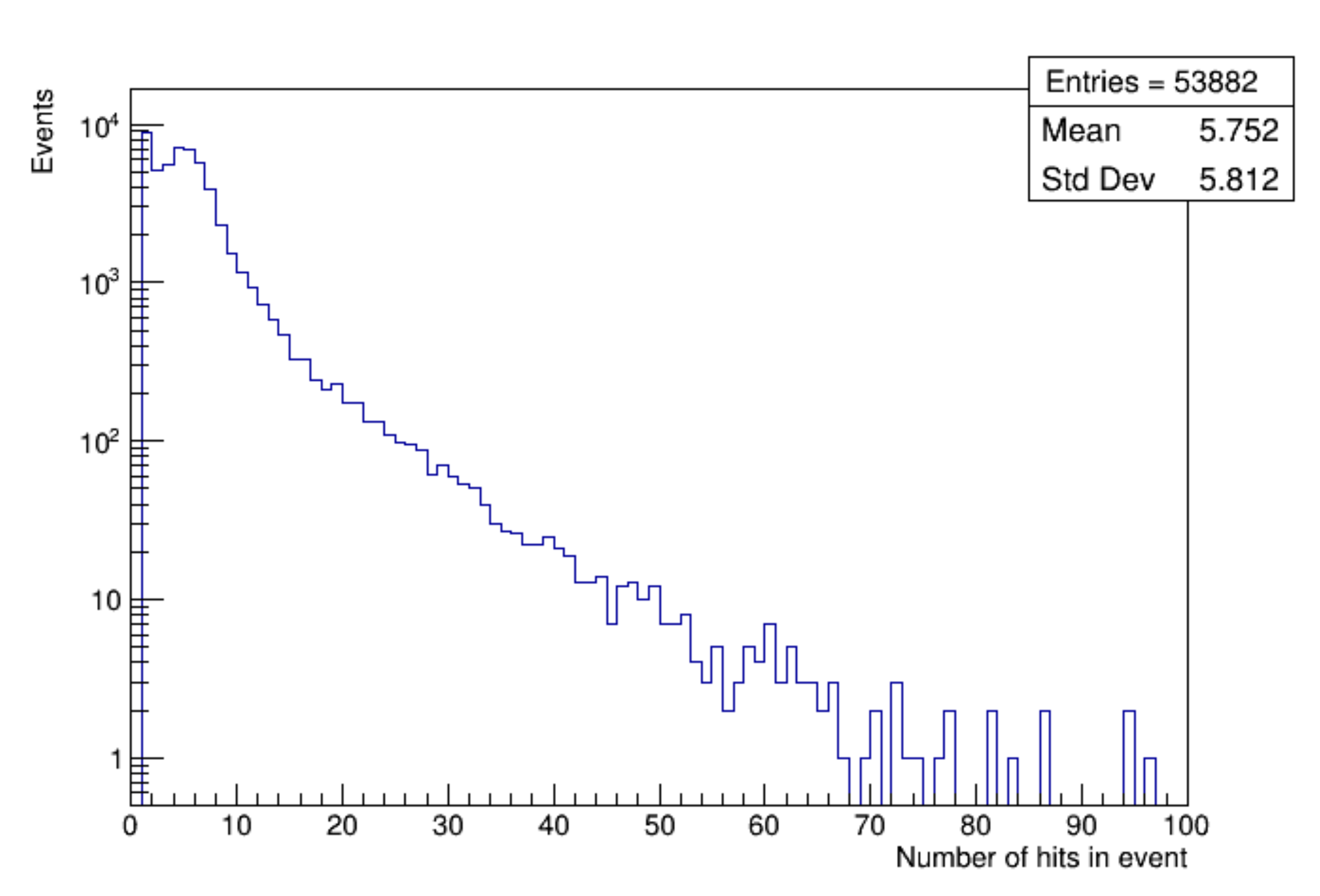}
	\caption{Left:Size of clusters matched to $good$ \mcube\ tracks for the layer 1 of Module 11 PCB1. Right: Distribution of hit multiplicity in one FEU for events recorded in the module.} 
	\label{fig:clusSize-and-HitsPerEvent}
\end{figure}

The standard deviation of a given strip's pedestal is typically around 10-15 ADC counts, as seen in Fig.~\ref{fig:pedestals}. Therefore, only signals exceeding 50-75 ADC counts are reconstructed as hits. The typical number of hits per event in one Front End Unit is around six, with a standard deviation of seven, as shown in Fig.~\ref{fig:clusSize-and-HitsPerEvent} right.

\begin{figure}[!h]
    \centering
    \includegraphics[width=0.45\textwidth]{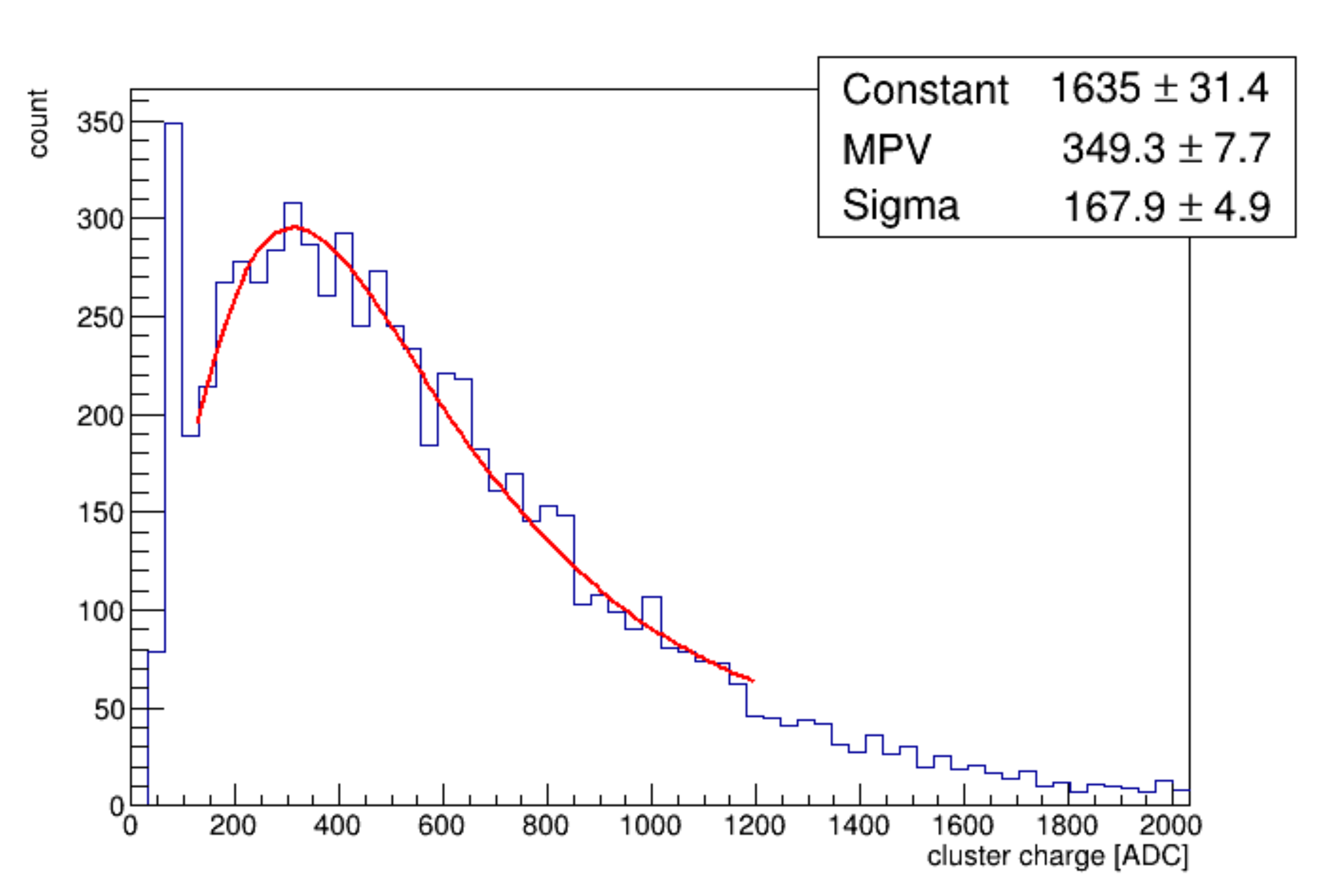}
    \caption{Charge distribution of clusters that matched \mcube\ tracks. }
	\label{fig:clusCharge}
\end{figure}
The distribution of cluster charges matched to \mcube\ is given in Fig.~\ref{fig:clusCharge}, where it is fitted by a Landau function. In data analysis, the most probable value is used to indicate the signal charge of each sector, while the sigma is treated as uncertainty. It was checked that the peak at low values of the cluster charge is mainly originating from a threshold effect partly removing the cluster components, and affecting size one clusters only.
\subsubsection{Clustering and demultiplexing}
\label{subsubsection:clustering}
 For each layer, the hits are gathered into clusters, using the following algorithm. 
Given that one electronic channel is linked to two strips in the detector due to multiplexing, when several electronic channels respond to a cosmic muon, all possible combinations of corresponding strips have to be considered.
For each event, the two associated strips from a hit are located through a map,  created for this run. Starting from the hit with the highest recorded amplitude, the program will try to find another hit that has an associated strip that is a neighbour of a strip associated to the first hit. Two strips are considered as neighbours if they are within \SI{0.9}{mm} of each other. Once the neighbours are found, the corresponding strips of both hits can be determined and the two hits are gathered into a cluster. This procedure is called  “demultiplexing”. Then the program loops over the remaining hits to merge other hits into this cluster, as long as one associated strip of the new hit is the neighbour of any determined strip in the cluster. If no neighbour strip can be found, then the clustering is finished. The clustering of another cluster continues in the same way, merging hits from the rest of them until all hits are gathered into clusters. However, chances are that sometimes neither associated strip of a hit has any neighbour strip. In this case, the clustering of this hit is performed by generating two unit-sized clusters in the positions of each associated strip. It has been checked that this duplication does not bias the final numbers.


The total charge of a given cluster is the sum over the constituent hits of their maximum amplitudes. The strip of the cluster is defined as the average strip number of the associated hits, weighted by their maximum amplitude. This strip number is therefore not a physical strip, but rather the centroid position calculated from the hit constituents.

\subsubsection{Matching tracks}
\label{subsubsection:matching}
\begin{figure}[!h]
    \centering
	\includegraphics[width=0.6\textwidth]{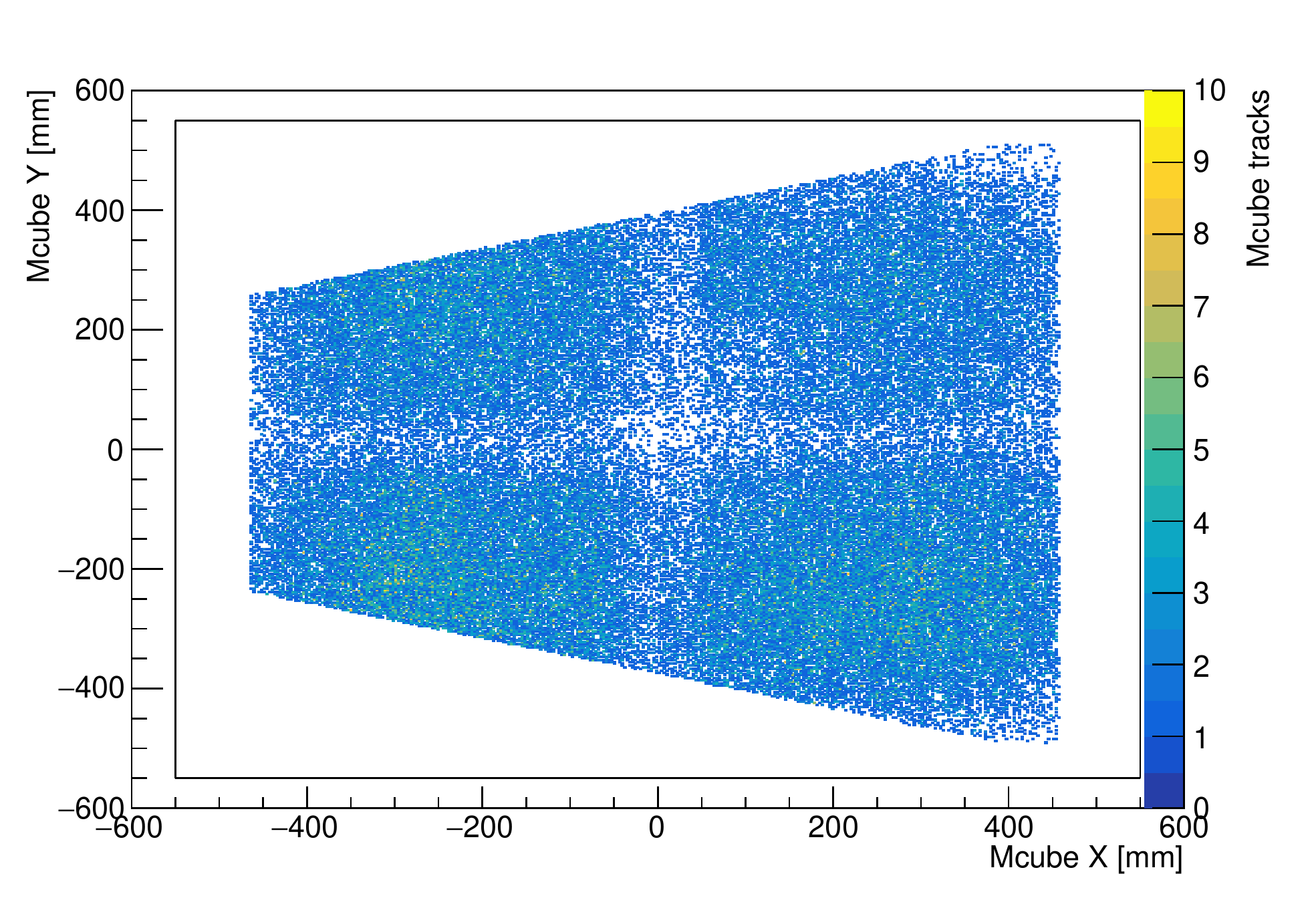}
	\caption{Distribution of \mcube\ tracks extrapolated to the module plane on Layer 1. Number of \mcube tracks during the data taking is binned per \SI{2.5}{\milli\meter}$\times$\SI{2.5}{\milli\meter}. The acceptances of each bulk detector result in a cross with lower statistics. The outline of the \mcube system is indicated by the solid lines.} 
	\label{fig:hitmapMcube}
\end{figure}

The possible \mcube tracks are reconstructed from the hits in all three \mcube layers. Because of the complexity of the \mcube setup (and in particular the factor 16 of multiplexing), a non-negligible fraction of the triggered events, a bit less than 40~\%, do not lead to a successful track reconstruction. Furthermore, considering the setup of the cosmic bench, only those tracks with $\chi^2_{X}$ less than 2.5, $\chi^2_{Y}$ less than 10 and incident angle less than 0.4\,rad are kept as good tracks. The degree of freedom is one, as a linear fit is performed across three \mcube layers, individually in the X and Y directions. X and Y are orthogonal axes defined by \mcube, and can be visualised in Fig.~\ref{fig:hitmapMcube},  where Y is is the direction of the readout strips on the Eta panels of the module. The information from cluster reconstruction is used when deciding which \mcube track to choose, since there might be more than one good tracks.
A match between a \mcube track and a reconstructed cluster is considered to be achieved if the residual between the $\vec{X}$ position of the cluster and the $\vec{X}$ position predicted by \mcube track is less than the residual cut (\SI{10}{mm}).
Among the good tracks that have the largest number of matching clusters, the one with minimum $\chi^2_{X}$ is chosen to be the best track of the event. In case of no cluster matches the \mcube tracks, the best track is simply assumed to be the good track with minimum $\chi^2_{X}$.
Once the best track of \mcube is selected, one starts to look at the clusters. For each layer of the detector, only the cluster with the best reconstructed position can represent the response in the layer. This cluster is the one with the smallest residual in $\vec{X}$ with respect to the position predicted by the  \mcube track. The residual cut will then be applied for the second time, in order to remove the best clusters that are too far away from the predicted $\vec{X}$ position.
A typical cut flow for cosmic muon events is shown in Table~\ref{tab:cut_flow}, showing that the different selection steps reduce the initial recorded data by a factor $\sim 12$.

\begin{table}[!h]
    \centering
    \footnotesize
    \begin{tabular}{|c|c|c|c|c|}
        \hline
        Cut & \multicolumn{4}{c|}{Entries}\\
        \hline
        \mcube triggered events & \multicolumn{4}{c|}{499949}\\
        \hline
        At least 1 reconstructed \mcube track & \multicolumn{4}{c|}{334791}\\
        \hline
        \makecell[c]{At least 1 good \mcube track ($\chi^2_{X}<2.5$, \\ $\chi^2_{Y}<10.0$ and incident angle $<0.4$\,rad)} & \multicolumn{4}{c|}{54583}\\
        \hline
         & Layer 1 & Layer 2 & Layer 3 & Layer 4\\
        \hline
        Geometric cut for \mcube track & 44447 & 44142 & 43319 & 45297\\
        \hline
        Residual cut for LM1 clusters (10\,mm) & 38831 & 38186 & 36818 & 39884\\
        \hline
    \end{tabular}
    \caption{Cut flow of cosmic muon analysis for Module 13 PCB4 and PCB5. Each event is triggered by the 3 \mcube layers in coincidence.}
    \label{tab:cut_flow}
\end{table}

The distribution of \mcube\ tracks extrapolated to the module plane in the eta top layer (L1) is shown in Fig.~\ref{fig:hitmapMcube}.

\begin{figure}[!h]
    \centering
    
	\includegraphics[width=0.45\textwidth]{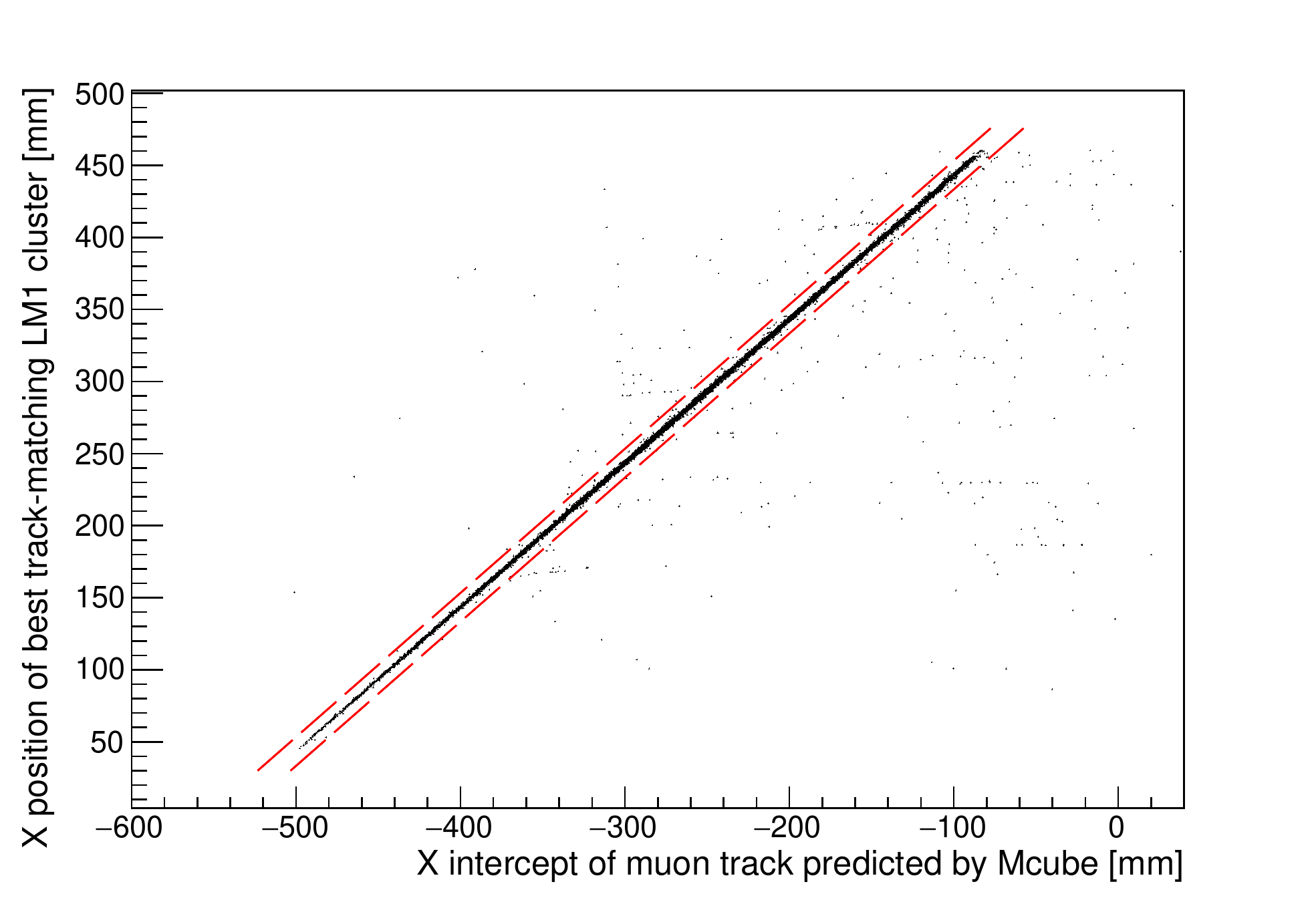}
	\includegraphics[width=0.5\textwidth]{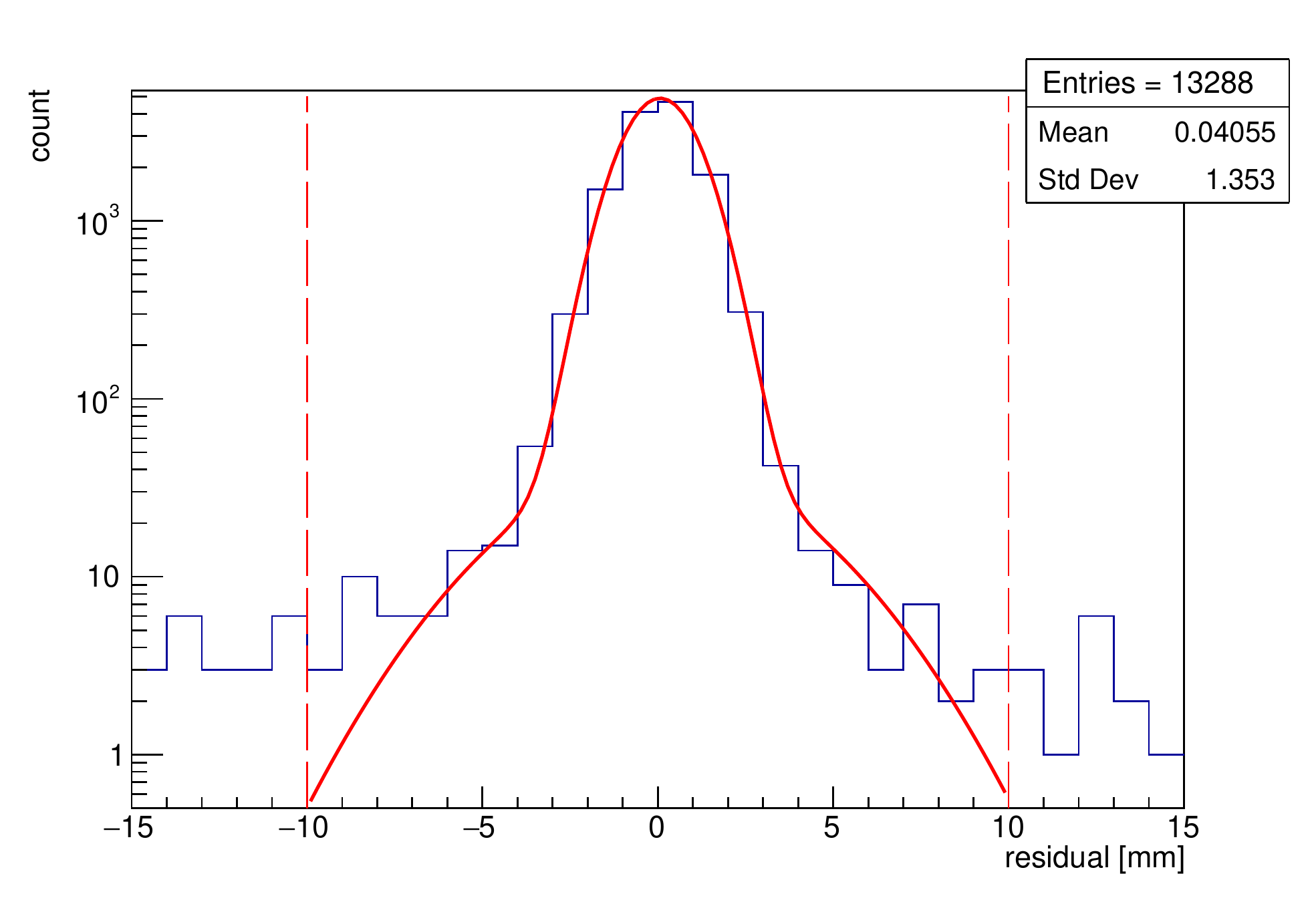}
	\caption{Left: Correlation between the best cluster position along $\vec{X}$ and the predicted position of its associated \mcube\ track in the corresponding cluster layer, before applying the tracking cut. Right: The blue histogram shows the difference between the best cluster position along $\vec{X}$ and the predicted position of its associated \mcube\ track in the corresponding cluster layer, before applying the tracking cut. The red curve shows the double-Gaussian fit. The red dashed lines stand for the 10\,mm residual cut in both plots.} 
	\label{fig:clusTrackCorr-clusRes}
\end{figure}

The linear correlation between the reconstructed best cluster position along $\vec{X}$ and the predicted position of the associated track is excellent, as shown in Fig~\ref{fig:clusTrackCorr-clusRes} left. Any significant deviation from the linear correlation is unlikely to be the response of the LM1 module to the cosmic muon. This is why the second residual cut is applied, to exclude these deviant clusters from the calculation of efficiency. The distribution of residuals for the correlation is shown in Fig.~\ref{fig:clusTrackCorr-clusRes} right, where the residual is defined to be the difference between the predicted position of the track along $\vec{X}$ and the best cluster. This distribution is not necessarily centered at 0 before module alignment, and indicates the value of the module translation to be applied such that the mean residual is 0. Once obtained, the mean value of the residual distribution is then subtracted from the position along $\vec{X}$ of the LM1 module, and another iteration of the track matching is performed. The standard deviation of the residual distribution, which is about 1.2\,mm, is dominated by the resolution of the \mcube\ tracks. Based on the distribution of cluster residuals, the value of the residual cut is chosen to be 10\,mm to suppress the remaining background while keeping the valid module clusters for the data analysis.

%
\subsection{Performance results}
\subsubsection{Gain homogeneity}

\begin{figure}[!h]
    \centering
	\includegraphics[width=0.6\textwidth]{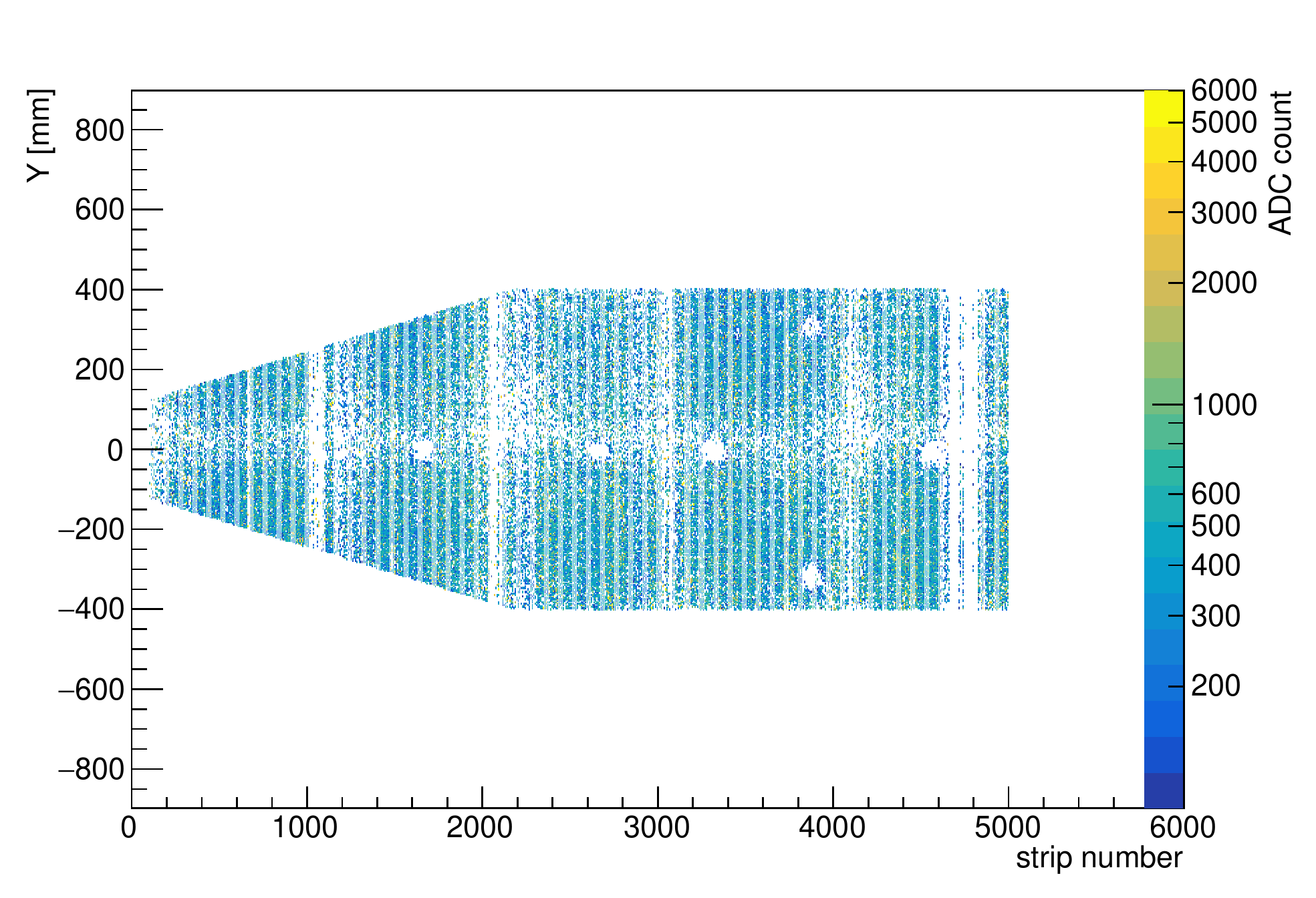}
	\caption{Cluster charge on Module 11 Layer 2 for all PCBs. 5 PCBs give 4 thin vertical gaps due to the passivation over the borders of PCBs. The empty gaps in addition to the interconnections and PCB borders (around strip number 4800) are caused by bad connections of electronics.} 
	\label{fig:clusterCharge2D}
\end{figure}

After matching clusters to \mcube\ tracks, it is possible to characterize the response of the module to cosmic muons. The absolute value of the gain  will depend on the transfer function of the electronics and the connection scheme. 
As can be seen in Fig.~\ref{fig:clusterCharge2D}, the gain is homogeneous over the full module. 
For this performance study, using a multiplexing of 2, the average number of ADC counts for one cluster is around 300, with an RMS of $\approx$30 corresponding to an absolute gain of $\approx$7000.

\begin{figure}[!h]
  \centering
  \begin{subfigure}[b]{0.47\textwidth}
    \includegraphics[width=\textwidth]{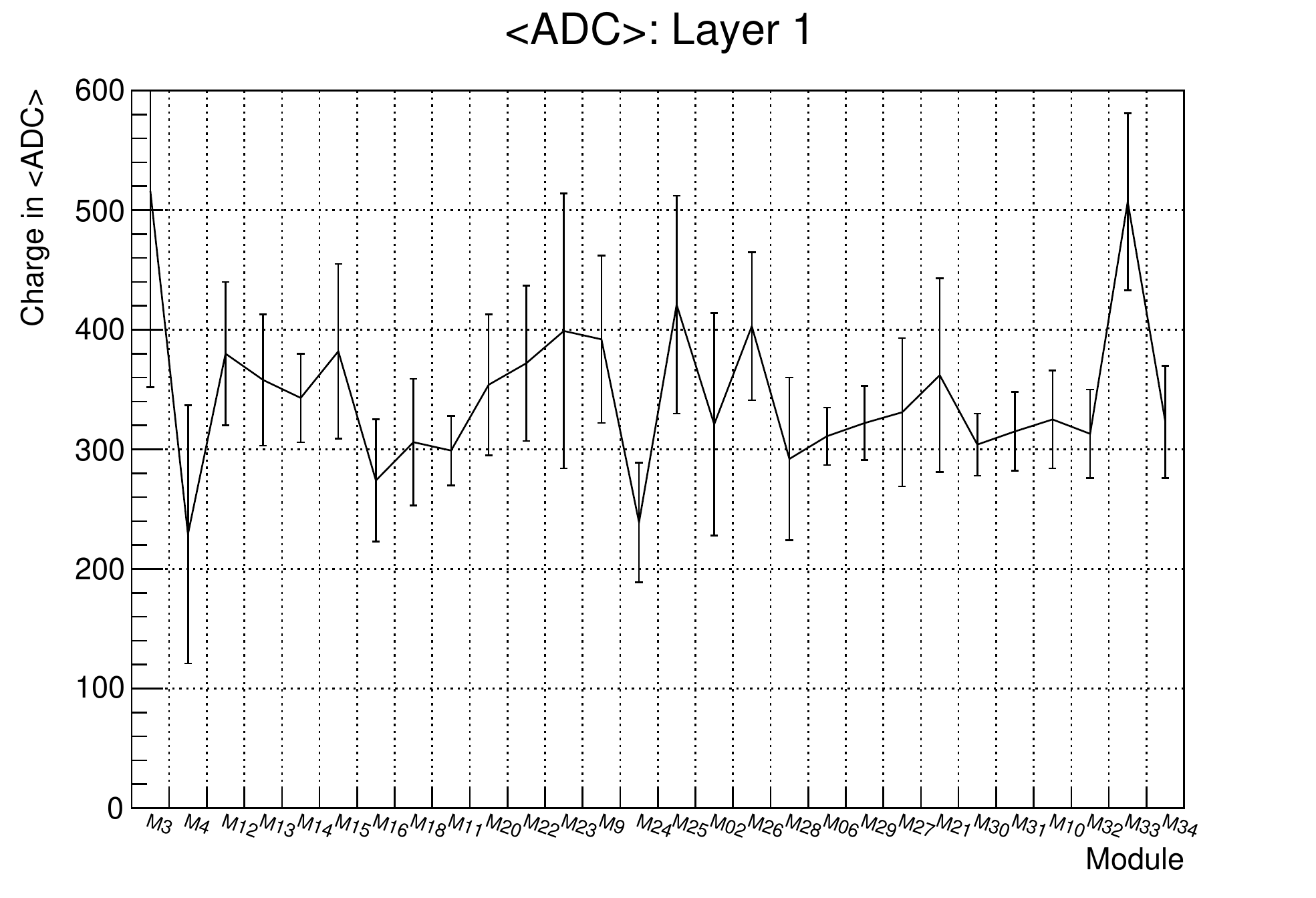}
    \caption{L1/Eta top}
    \label{fig:GainEvolL0}
  \end{subfigure}
  \begin{subfigure}[b]{0.47\textwidth}
    \includegraphics[width=\textwidth]{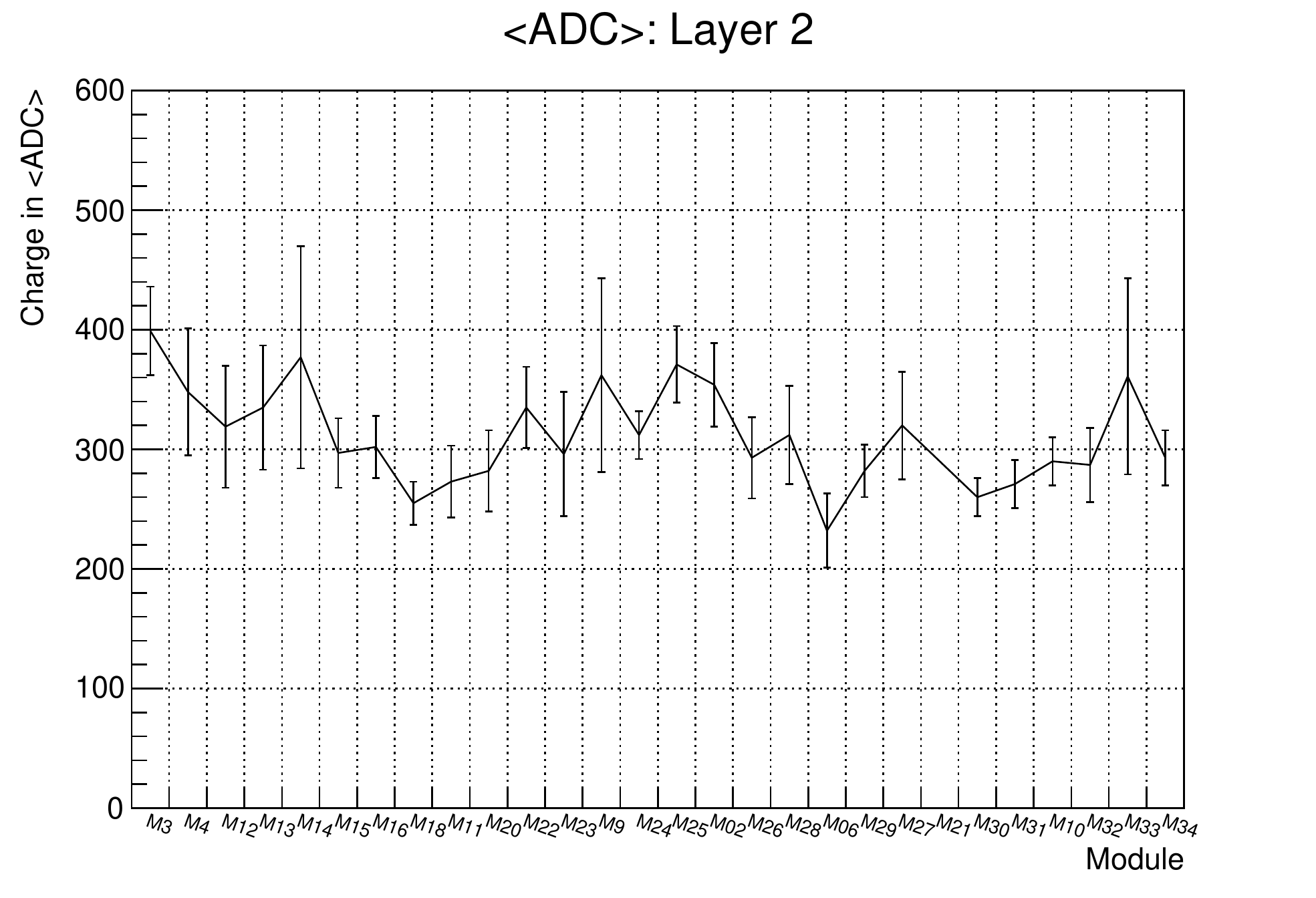}
    \caption{L2/Eta bottom}
    \label{fig:GainEvolL1}
  \end{subfigure}
  \\
  \begin{subfigure}[b]{0.47\textwidth}
    \includegraphics[width=\textwidth]{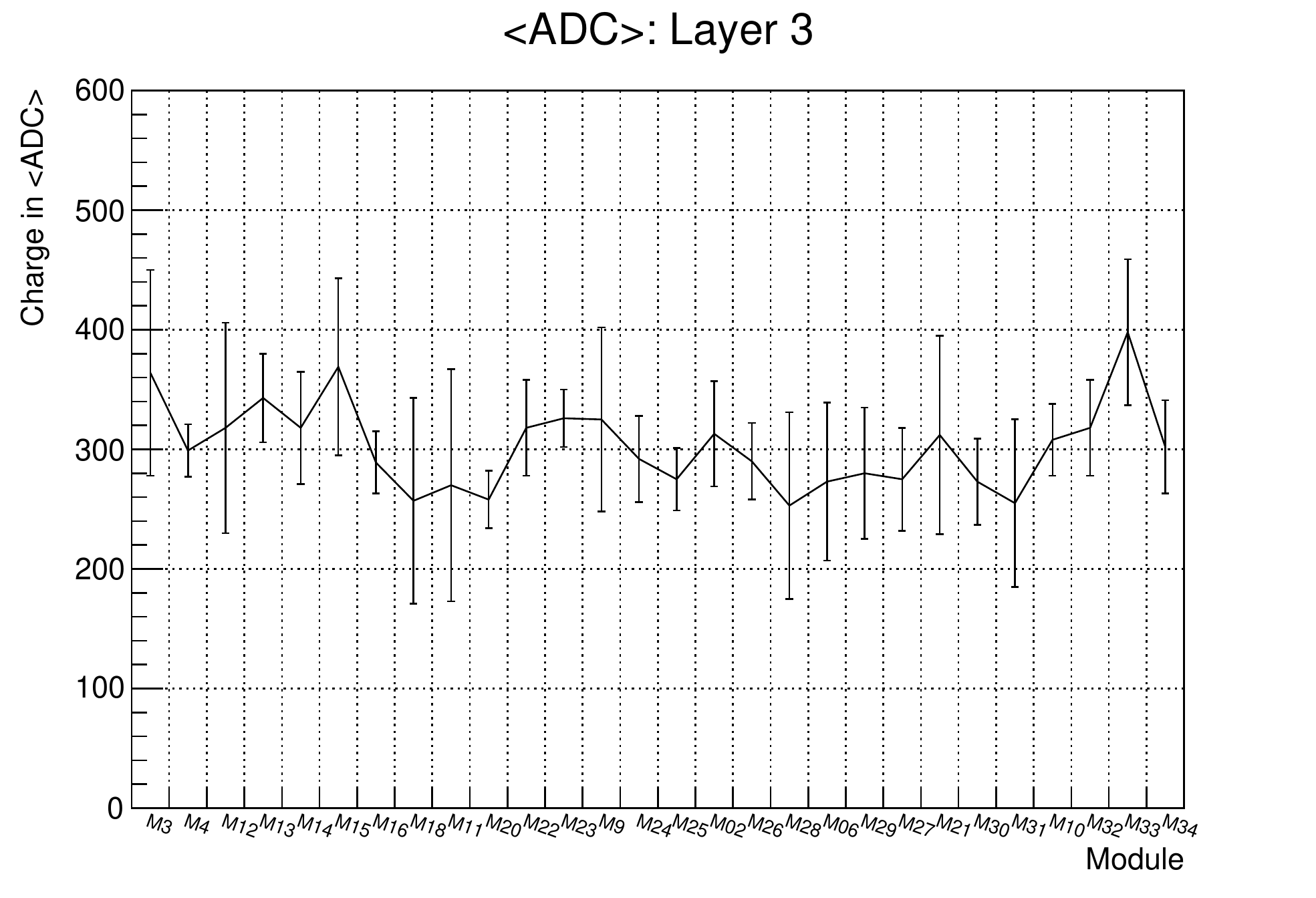}
    \caption{L3/Stereo top}
    \label{fig:GainEvolL2}
  \end{subfigure}
  \begin{subfigure}[b]{0.47\textwidth}
    \includegraphics[width=\textwidth]{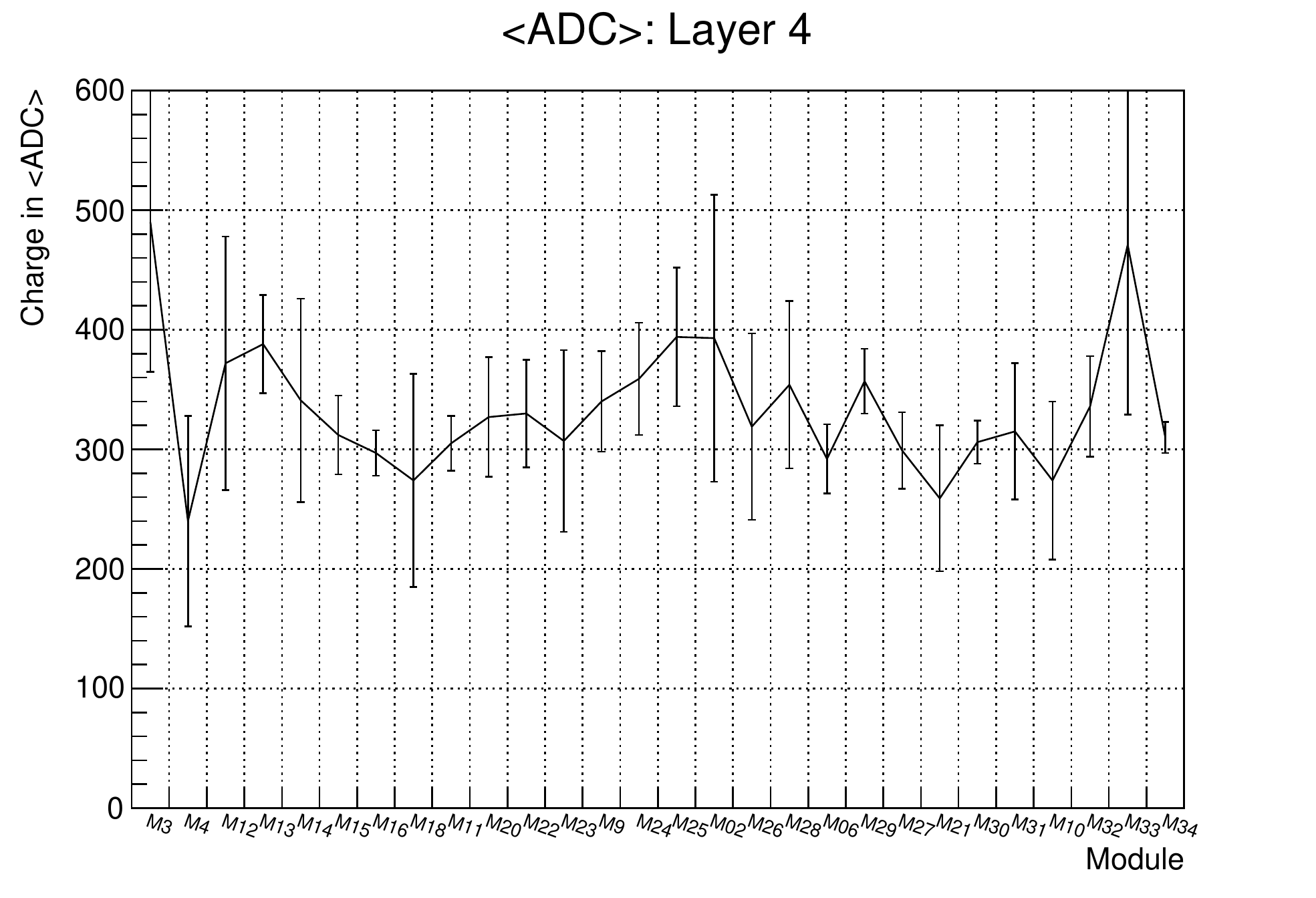}
    \caption{L4/Stereo bottom}
    \label{fig:GainEvolL3}
  \end{subfigure}
  \caption{Average number of ADC counts for each module validated on the cosmic bench. The uncertainty bars are the RMS evaluated from 10 different values, each one corresponding to a HV sector in the layer.
  }
  \label{fig:GainEvol}
\end{figure}

Figure~\ref{fig:GainEvol} shows the average gain per cluster for each layer as a function of the module ID number. The first point, for Module 3,  is significantly above the others, because multiplexing was not used when evaluating its performance.

\subsubsection{Efficiency}

\begin{figure}[!h]
    \centering
	\includegraphics[width=0.6\textwidth]{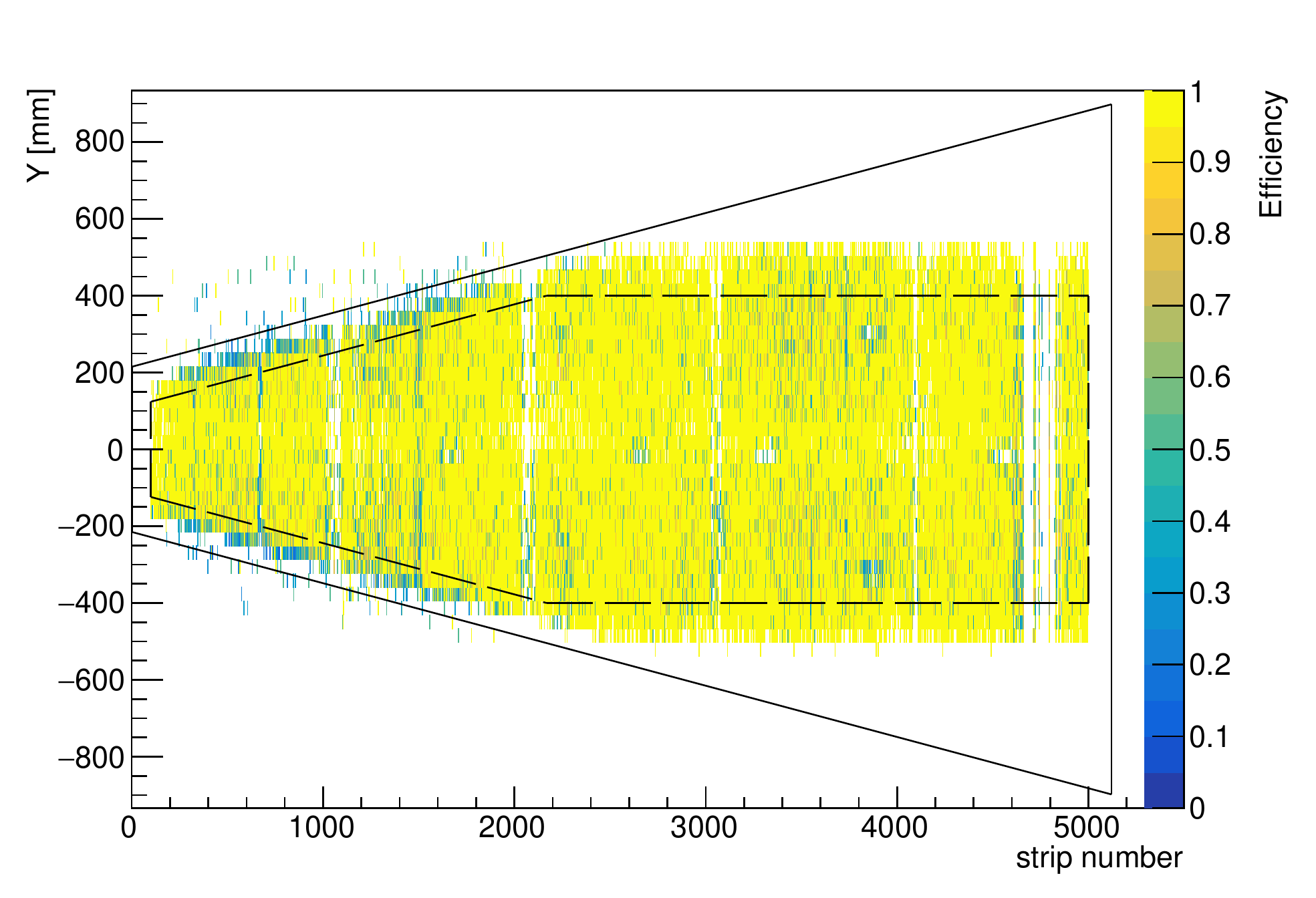}
	\caption{Efficiency map of Module 11, Layer 2. The solid lines show the real size of the sensitive area while the dashed lines indicate the geometric cut for efficiency. Given that the track matching is only limited to the X direction and that the Y positions of the LM1 clusters are assigned by the \mcube tracks, the entries beyond the sensitive area reflect the global noise level of the module. The contribution of global noise to the efficiency is below 1\%.} 
	\label{fig:efficiency2D}
\end{figure}

The efficiency is characterized in the 2D module plane, for each layer of the module under study. For each good track reconstructed by the \mcube\ detector, the presence of one cluster within 10\,mm of the extrapolated track within the plane, along the $\vec{X}$ axis, defines an efficient detection of the particle passing through. Since the acceptance of the \mcube\ detector is smaller than PCB 3, 4 and 5 along the $\vec{Y}$ direction, the efficiency does not take into the account the borders of these PCBs. 
The passivated areas on PCB 1 and 2 are also excluded for the computation of efficiency. Fig.~\ref{fig:efficiency2D} shows the efficiency as a function of the $\vec{Y}$ direction and strip number. The corresponding efficiency of each layer is summarized in Table~\ref{tab:efficiency_layers}, excluding passivated areas and non-connected channels. In Table~\ref{tab:efficiency_pcb}
the same information is shown broken out for each. The reduction of efficiency due to passivation is clearly affecting the smaller PCBs.

\begin{table}[!h]
    \centering
    
    \begin{tabular}{cccc}
        \toprule
        Layer & Raw  &Efficiency & Efficiency \\
          &  Efficiency &excluding bad connections& excluding passivated areas\\
        
        \midrule
        L1/Eta top & 0.86 $\pm$0.11 & 0.88$\pm$0.11 &0.89$\pm$0.03\\
        \midrule
        L2/Eta bottom & 0.85 $\pm$0.06 & 0.87$\pm$0.05 &0.88$\pm$0.03\\
        \midrule
        L3/Stereo top & 0.84 $\pm$0.06 & 0.86$\pm$0.05 &0.88$\pm$0.03\\
        \midrule
        L4/Stereo bottom & 0.87 $\pm$0.08 & 0.88$\pm$0.07 &0.89$\pm$0.03 \\
        \bottomrule
    \end{tabular}
    \caption{Mean efficiencies of the four layers of Module 13.}
    \label{tab:efficiency_layers}
\end{table}

\begin{table}[!h]
    \centering
   
    \begin{tabular}{cccc}
        \toprule
        Layer & Raw  &Efficiency & Efficiency \\
          &  Efficiency &excluding bad connections& excluding passivated areas\\
        
        \midrule
        PCB1 & 0.72 $\pm$0.06& 0.74$\pm$0.5&0.91$\pm$0.04\\
        \midrule
        PCB2 & 0.82 $\pm$0.03& 0.86$\pm$0.01&0.92$\pm$0.03\\
        \midrule
        PCB3 & 0.86 $\pm$0.03&0.91$\pm$0.02&0.91$\pm$0.02\\
        \midrule
        PCB4 & 0.87 $\pm$0.05&0.88$\pm$0.03&0.88$\pm$0.03\\
        \midrule
         PCB5 & 0.85$\pm$0.04&0.88$\pm$0.01&0.88$\pm$0.01 \\
        \bottomrule
    \end{tabular}
     \caption{Mean efficiencies per PCB for Module 13.}
     \label{tab:efficiency_pcb}
\end{table}

\begin{figure}[!h]
  \centering
  \begin{subfigure}[b]{0.47\textwidth}
    \includegraphics[width=\textwidth]{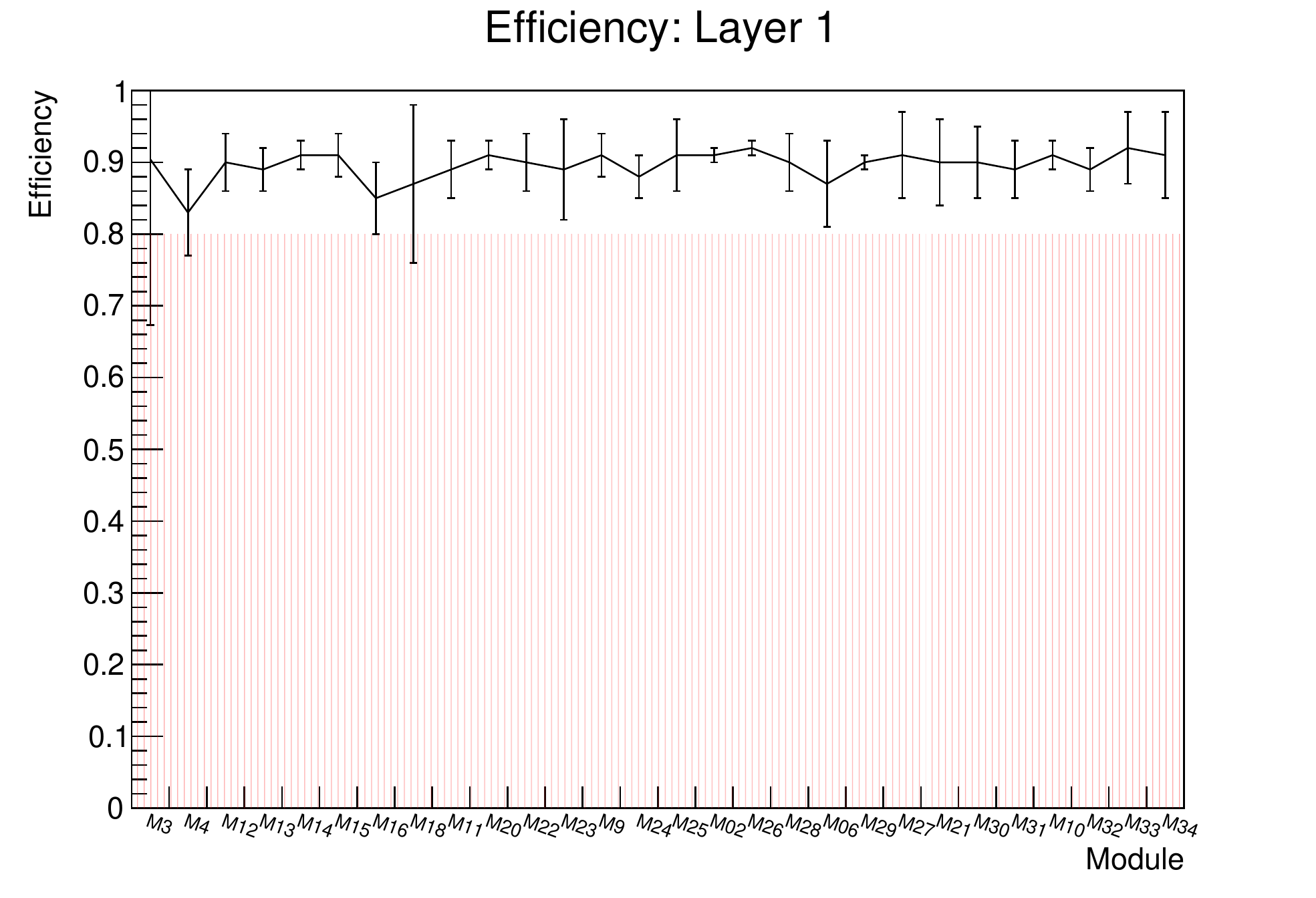}
    \caption{L1/Eta top}
    \label{fig:EffEvolL0}
  \end{subfigure}
  \begin{subfigure}[b]{0.47\textwidth}
    \includegraphics[width=\textwidth]{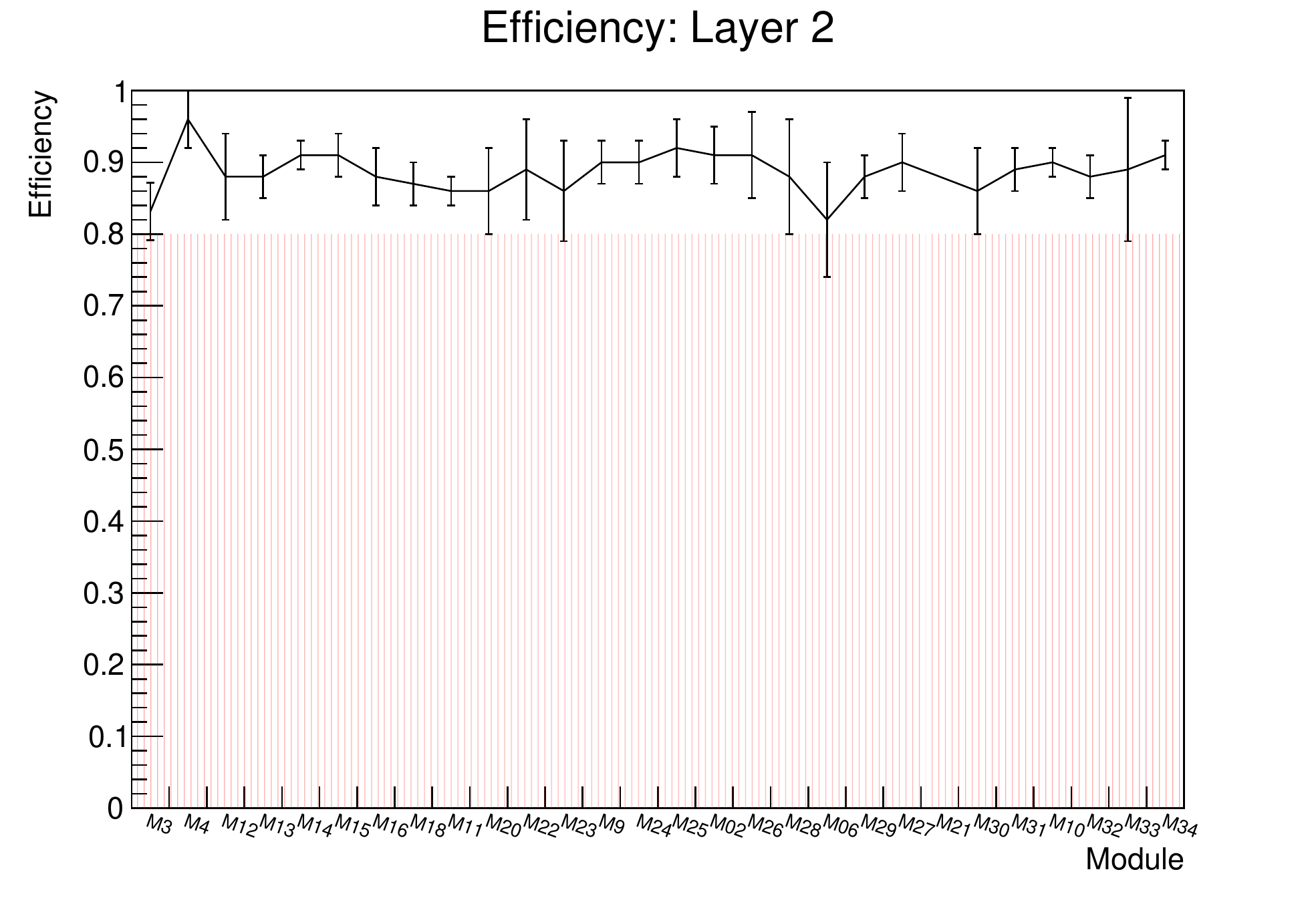}
    \caption{L2/Eta top}
    \label{fig:EffEvolL1}
  \end{subfigure}
  \\
  \begin{subfigure}[b]{0.47\textwidth}
    \includegraphics[width=\textwidth]{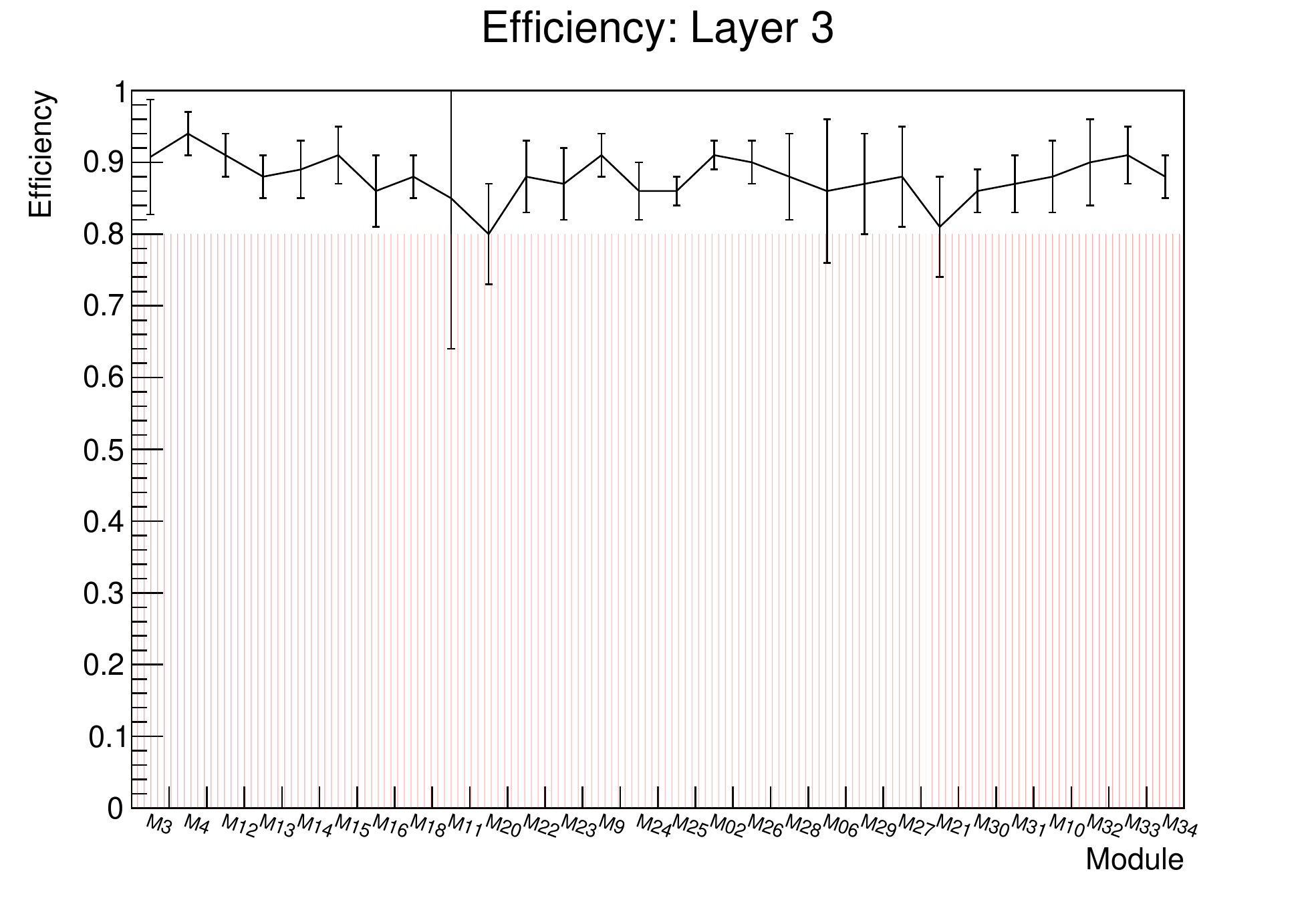}
    \caption{L3/Stereo top}
    \label{fig:EffEvolL2}
  \end{subfigure}
  \begin{subfigure}[b]{0.47\textwidth}
    \includegraphics[width=\textwidth]{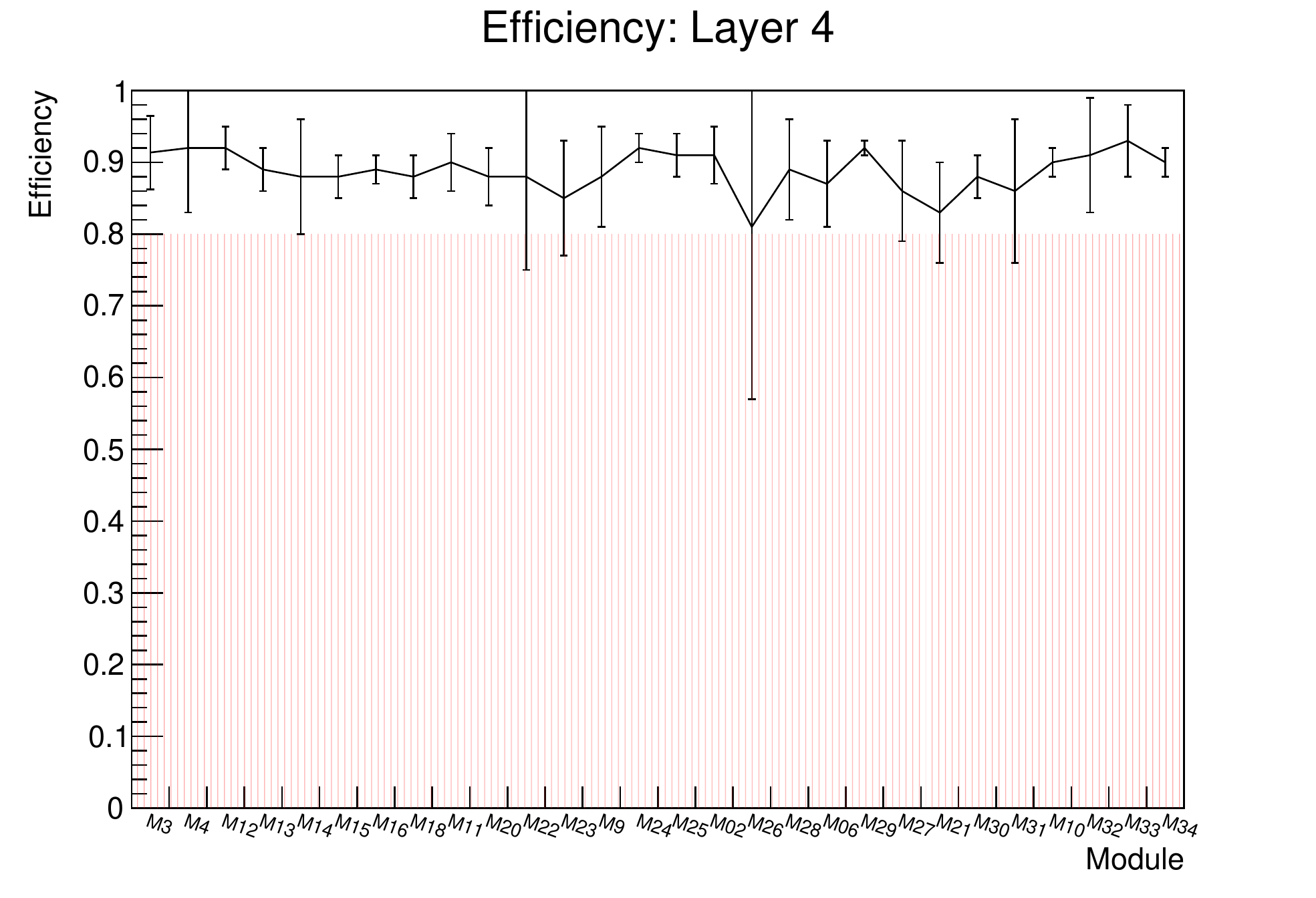}
    \caption{L4/stereo bottom}
    \label{fig:EffEvolL3}
  \end{subfigure}
  \caption{Average efficiency of each module validated on the cosmic bench. The uncertainty bars are the RMS evaluated from 10 different values, each one corresponding to a HV sector in the layer. Layer efficiencies falling below the red line do not fulfill the requirements.}
  \label{fig:EffEvol}
\end{figure}

Besides the relatively high efficiency per individual layer, an overall good homogeneity for all layers is also achieved. As shown in  Table~\ref{tab:efficiency_layers} and Table~\ref{tab:efficiency_pcb}, the final efficiency excluding bad connections and edge passivation is around $90\%$ with RMS smaller than $4\%$ by layer and by PCB.
As shown in Fig~\ref{fig:EffEvol}, efficiency has been monitored over all the modules production. In all cases, the detectors meet the requirement that all layers must reach $80\%$ efficiency. 

Module efficiency and HV stability can be enhanced when a small fraction of Isobutane is added to the Argon-CO$_2$ mixture as shown in~\cite{chevalerias:tel-02268557,mvd}. This option is currently being studied in the NSW Collaboration.

\clearpage
\subsubsection{Effect of planarity defects on performance results}
\begin{figure}[!h]
  \centering
  \begin{subfigure}[b]{0.46\textwidth}
    \includegraphics[width=\textwidth]{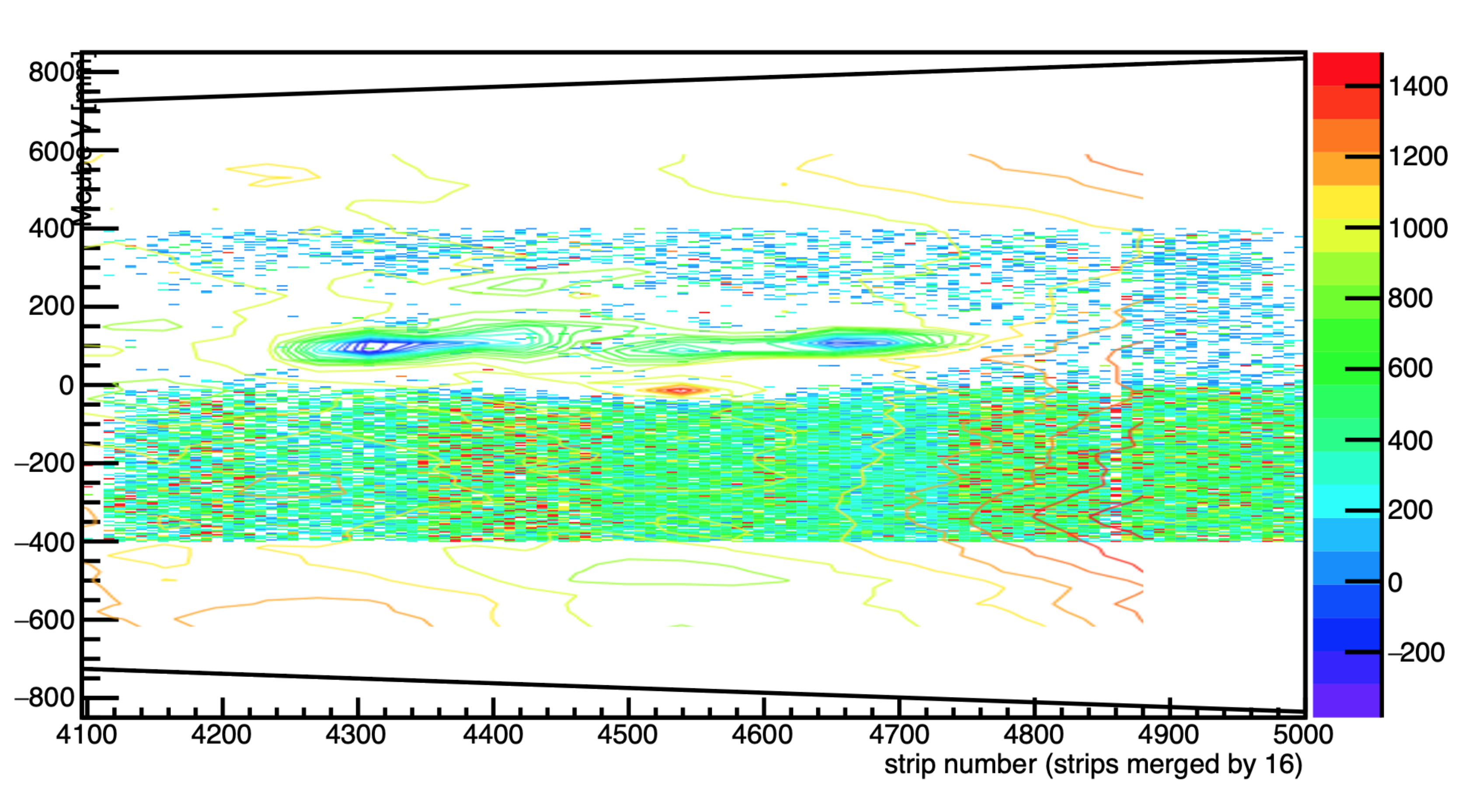}
    \caption{S16}
    \label{fig:CompS16}
  \end{subfigure}
  \begin{subfigure}[b]{0.47\textwidth}
    \includegraphics[width=\textwidth]{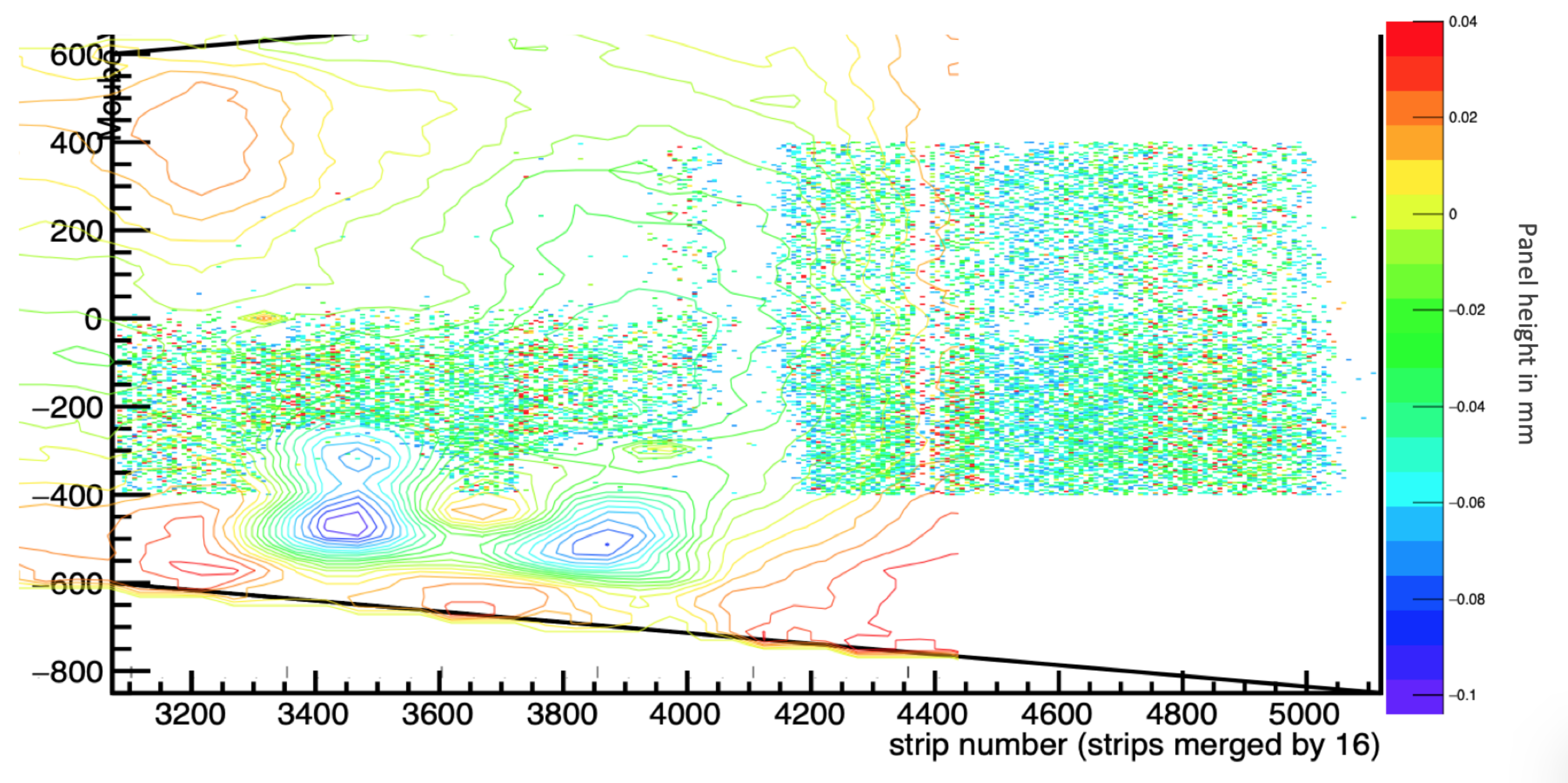}
    \caption{S14}
    \label{fig:CompS14}
  \end{subfigure}
  \caption{Comparison between signals observed on the cosmic test bench and the planarity scan made during construction. Contour lines correspond to height measurement from the planarity scans. The height is given by the vertical scales. The effect of holes is clearly seen in the inefficiency of detectors.
  }
    \label{fig:CompPlaEff}
\end{figure}

The planarity scans described in Section~\ref{section:planarity} reveal fine structure elements of the panels. Here the effect of these elements will be correlated to performance results for two specific panels that showed interesting features. During the planarity scans, depressions  of the order of 100\,\microns were observed for these two panels. These two panels were later assembled into quadruplets and tested in the cosmic test bench where the holes were seen as less efficient areas. Fig.~\ref{fig:CompPlaEff} shows that the effect of the lower areas propagates quite far from the depression areas. This effect probably comes from the non-attachment of the mesh electrostatically, especially for  S14 where there are higher areas that carry the mesh in between the holes. For instance in Fig.~\ref{fig:CompS14}, the right hole causes an inefficiency from the high point at strip=3700 to the edge of the PCB where the mesh is elevated by the inter-PCB passivation.

\section{Conclusion and outlook}
\label{section:conclusion}
The 32 LM1 quadruplets required for the NSWs have been produced. Two spares modules will also be produced. Quality control measurements show that the quadruplets meet the required specifications for planarity, alignment, efficiency and gain homogeneity.

While the four types of  MM quadruplets are assembled and qualified at the respective construction sites, their assembly into small and large sectors mounted on the NSW takes place at CERN. The sectors consist of a central mechanical spacer structure to which the detectors are attached  along with their respective services and read-out electronics. Each sector is built with four MM chambers (2 LM1 + 2 LM2 for large sectors) and two small-strip Thin Gap Chamber (sTGC) wedges, consisting of an assembly of three quadruplet modules kinematically mounted on the central structure.

Each chamber undergoes a first HV test at the CERN  to spot any HV instability or spiking behaviour. About 70\% of the production chambers are also HV-tested under irradiation to assess their behaviour in an environment comparable to the one foreseen for the ATLAS High-Luminosity operation. To reach this extremely- high radiation field, chambers are tested at the CERN Gamma Irradiation Facility (GIF++)\cite{Pfeiffer:2016hnl,Guida:2016ivc} that exploits a single 14\,TBq $^{137}$Cs radioactive source with a half-life of 30 years. The goal is to measure the behaviour of the chambers at different high voltages and incident fluxes by monitoring the current. Particular attention is given to current instabilities  and  spiking effects that might reduce the detector performance during the long-term operation of the NSW,  foreseen to be at least 15 years.

After these reception tests, the chambers are sent to mechanical and services integration. 
Then the MM double wedge and sTGC wedges are equipped with their read-out electronics. Before assembly with the sTGC wedges to form sectors, the MM double wedge is qualified in a  large cosmic ray facility. The MM double wedge and both sTGC wedges are assembled together before installation in the NSW. Finally the sector is connected to the services and electronics boxes installed on the NSW to proceed with the commissioning. The first NSW will be lowered to the ATLAS pit in July 2021. The second one will follow in October 2021.

\section*{Acknowledgments}
We want to dedicate this paper to the memory of  Stephanie Zimmerman, former leader of the ATLAS NSW project. The success of the project owes a lot to her dedication and her deep involvement. 

This work was performed within the ATLAS NSW Collaboration. We thank our colleagues for their commitment, the and useful discussions and the advices. We would like to thank in particular our colleagues from the National Laboratory of Frascati for their continuous support and help to keep a good production rate, in particular by implementing an additional integration station at Saclay.

We acknowledge support from the French Ile de France region for the construction of the CICLAD platform (Conception, Integration, and Characterization of Large Area Detectors).  We wish to thank Isabelle Trigger for the careful reading of our manuscript and  insightful comments and corrections.

.

\bibliography{mybibfile}
\bibliographystyle{elsarticle-num}
\end{document}